\documentclass[
 reprint,
 superscriptaddress,
 bibnotes,
 amsmath,
 amssymb,
 aps,
 prb,
 citeautoscript,
 floatfix,
]{revtex4-2}

\usepackage{graphicx}
\usepackage{dcolumn}
\usepackage{bm}
\usepackage[dvipsnames]{xcolor}
\usepackage[colorlinks=true, citecolor={blue!80!black}, urlcolor={blue!50!black}, linkcolor = {blue!80!black}]{hyperref}
\usepackage{lipsum}  
\usepackage{upgreek}
\usepackage{siunitx}
\usepackage{mathtools}
\usepackage{ulem}

\newcommand{\Vj}{$V_{\rm j}$}
\newcommand{\Vp}{$V_{\rm p}$}
\newcommand{\Vt}{$V_{\rm t}$}
\newcommand{\Vr}{$V_{\rm r}$}
\newcommand{\Vc}{$V_{\rm c}$}
\newcommand{\Vl}{$V_{\rm l}$}
\newcommand{\ft}{$f_{\rm 01}$}
\newcommand{\flux}{$\phi_{\rm ext}$}
\newcommand{\abs}[1]{\left|#1\right|}
\newcommand{\ket}[1]{\left|#1\right\rangle}

\begin{document}

\preprint{APS/123-QED}

\title{Singlet-doublet transitions of a quantum dot Josephson junction detected in a transmon circuit}

\author{Arno Bargerbos}
\thanks{These two authors contributed equally.}
\affiliation{QuTech and Kavli Institute of Nanoscience, Delft University of Technology, 2600 GA Delft, The Netherlands}

\author{Marta Pita-Vidal}
\thanks{These two authors contributed equally.}
\affiliation{QuTech and Kavli Institute of Nanoscience, Delft University of Technology, 2600 GA Delft, The Netherlands}

\author{Rok Žitko}
\affiliation{Jožef Stefan Institute, Jamova 39, SI-1000 Ljubljana, Slovenia}
\affiliation{Faculty of Mathematics and Physics, University of Ljubljana, Jadranska 19, SI-1000 Ljubljana, Slovenia}

\author{Jesús Ávila}
\affiliation{Instituto de Ciencia de Materiales de Madrid (ICMM),
Consejo Superior de Investigaciones Cientificas (CSIC), Sor Juana Ines de la Cruz 3, 28049 Madrid, Spain}

\author{Lukas J. Splitthoff}
\affiliation{QuTech and Kavli Institute of Nanoscience, Delft University of Technology, 2600 GA Delft, The Netherlands}

\author{Lukas Grünhaupt}
\affiliation{QuTech and Kavli Institute of Nanoscience, Delft University of Technology, 2600 GA Delft, The Netherlands}

\author{Jaap J. Wesdorp}
\affiliation{QuTech and Kavli Institute of Nanoscience, Delft University of Technology, 2600 GA Delft, The Netherlands}

\author{Christian K. Andersen}
\affiliation{QuTech and Kavli Institute of Nanoscience, Delft University of Technology, 2600 GA Delft, The Netherlands}

\author{Yu Liu}
\affiliation{Center for Quantum Devices, Niels Bohr Institute, University of Copenhagen, 2100 Copenhagen, Denmark}

\author{Leo P. Kouwenhoven}
\affiliation{QuTech and Kavli Institute of Nanoscience, Delft University of Technology, 2600 GA Delft, The Netherlands}

\author{Ramón Aguado}
\affiliation{Instituto de Ciencia de Materiales de Madrid (ICMM),
Consejo Superior de Investigaciones Cientificas (CSIC), Sor Juana Ines de la Cruz 3, 28049 Madrid, Spain}

\author{Angela Kou}
\affiliation{Department of Physics and Frederick Seitz Materials Research Laboratory,
University of Illinois Urbana-Champaign, Urbana, IL 61801, USA}

\author{Bernard van Heck}
\affiliation{Leiden Institute of Physics, Leiden University, Niels Bohrweg 2, 2333 CA Leiden, The Netherlands}

\date{\today}

\begin{abstract}
We realize a hybrid superconductor-semiconductor transmon device in which the Josephson effect is controlled by a gate-defined quantum dot in an InAs/Al nanowire. Microwave spectroscopy of the transmon's transition spectrum allows us to probe the ground state parity of the quantum dot as a function of gate voltages, external magnetic flux, and magnetic field applied parallel to the nanowire. The measured parity phase diagram is in agreement with that predicted by a single-impurity Anderson model with superconducting leads. Through continuous time monitoring of the circuit we furthermore resolve the quasiparticle dynamics of the quantum dot Josephson junction across the phase boundaries. Our results can facilitate the realization of semiconductor-based $0-\pi$ qubits and Andreev qubits.
\end{abstract}

\maketitle



\section{\label{sec:intro} Introduction}
Superconducting pairing and charging energy are two
 fundamental interactions that determine the behavior
of mesoscopic devices.
Notably, when a quantum dot (QD) is coupled to a superconductor, they compete to determine its ground state.
A large charging energy favors single-electron doublet occupancy of the dot and thus a spin-1/2 ground state, while a strong coupling to the superconducting leads favors double occupancy in a singlet configuration with zero spin.
A quantum phase transition between the singlet and doublet ground state can occur as system parameters such as the dot energy level and the coupling strength are varied.
The latter also controls the nature of the singlet ground state, which can be either of the Bardeen-Cooper-Schrieffer (BCS) type or of the Kondo type.
The rich phase diagram of the system, as well as its transport properties, are theoretically well captured by an Anderson model with superconducting leads \cite{Glazman1989, yoshioka2000, Choi2004, Oguri2004, Tanaka2007, MartinRodero2011, Karrasch2008, Luitz2012, Kadlecova2019, Meden2019}.

Quantum dots coupled to superconductors have been studied experimentally over the last two decades. Signatures of the singlet-doublet transition have been detected in tunneling spectroscopy measurements of N-QD-S devices (where N is a normal lead, and S is a superconducting one) via the observation of Fermi-level crossings \cite{Pillet2010, pillet2013, Deacon2010, Lee2014, Chang2013, Li2017, Lee2017, Valentini2021, Whiticar2021}.
Additionally, they have been detected in switching current measurements of S-QD-S devices via $\pi$-phase shifts in the current-phase relation of the resulting quantum dot Josephson junction \cite{VanDam2006a, Cleuziou2006, Jorgensen2007, Jorgenson2009, Eichler2009, Lee2012, Maurand2012, Kumar2014, Szombati2016, Delagrange2015, Delagrange2018, Corral2020, Whiticar2021}.

Recent experiments~\cite{Janvier2015,Hays2020,Hays2021} on Andreev pair and spin qubits~\cite{Zazunov2003a,Chtchelkatchev2003,Padurariu2010,pavesic2022} have renewed the interest in quantum dot junctions due to the possibility of tuning the ground state of the system to be in addressable spin states.
Knowledge of the phase diagram of the quantum dot junction is also beneficial for realizing proposals for a quantum-dot-based readout of topological qubits~\cite{Plugge_2017,Karzig2017,Smith2020}.

These developments have highlighted the need for a better fundamental understanding of the quantum dot junction and its dynamics, requiring tools which are not limited by the long integration times of low-frequency measurements nor by the invasiveness of transport probes.
To address this need, we have embedded a fully controllable quantum dot in a microwave superconducting circuit. This experimental choice is motivated by the success of circuit quantum electrodynamics (QED) techniques in the investigation of mesoscopic effects in Josephson junctions \cite{Janvier2015,deLange2015,Larsen2015,Tosi2018,Hays2018,Bargerbos2020,Kringhoj2020,Hays2020, Uilhoorn2021,Hays2021,canadas2021}, which stems from its enhanced energy and time resolution compared to low-frequency transport techniques.
In this context, the microwave response of a quantum dot junction has attracted recent theoretical~\cite{Kurilovich2021,Hermansen2022} and experimental~\cite{Fatemi2021} attention.

\begin{figure}
    \centering
    \includegraphics[scale=1.0]{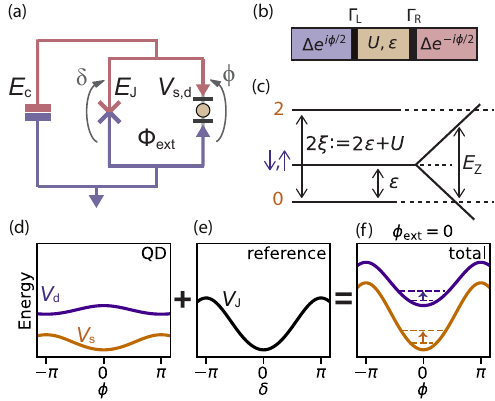}
    \caption{(a) Schematic diagram of a quantum dot junction incorporated into a transmon circuit. The transmon island  with charging energy  $E_{\rm c}$ is connected to ground by a SQUID formed by the parallel combination of a quantum dot junction and a reference junction. In this panel $\phi$ and $\delta$ denote the superconducting phase difference across the quantum dot and reference junctions respectively. $\Phi_\textrm{ext}$ is the externally applied magnetic flux through the SQUID loop. (b) Model diagram of the quantum dot junction in the excitation picture. Two $s$-wave superconductors are connected via tunnel barriers to a single level quantum dot. (c) Level diagram of the quantum dot hosting 0, 1, or 2 electrons when disconnected from the leads ($\Gamma_{\rm L}=\Gamma_{\rm R}=0$).  (d) Phase dependence of the Josephson potential of the quantum dot junction in the singlet (orange) and doublet (purple) state. 
    (e) Josephson potential of the reference junction. (f) Josephson potential of the DC SQUID for $\phi_\textrm{ext}=(2e/\hbar) \Phi_{\rm ext}$~=~0, with the quantum dot junction in the singlet (orange) and doublet (purple) state. The dashed lines represented the two lowest transmon energy levels in each branch of the Josephson potential, with the arrow denoting the resulting transition frequency, which can differ for the two quantum dot junction states (orange and purple arrows for singlet and doublet, respectively). }
    \label{fig:theorydiagram}
\end{figure}

The core of our experiment is a transmon circuit formed by an island with charging energy $E_{\rm c}$, coupled to ground via a superconducting quantum interference device (SQUID) formed by a junction with a known Josephson energy $E_{\rm J}$ and a quantum dot junction (Fig.~\ref{fig:theorydiagram}(a-c)).
The energy-phase relation of the quantum dot junction depends on whether it is in a singlet or doublet state, with a characteristic $\pi$-phase shift between the two relations (Fig.~\ref{fig:theorydiagram}(d)) \cite{spivak1991}.
The Josephson energies of the two junctions either add or subtract, depending on whether the quantum dot junction is in the singlet or doublet state, and on the value of the applied flux bias (Fig.~\ref{fig:theorydiagram}(f)).
The two branches of the spectrum give rise to two distinct transition frequencies of the transmon circuit, which can be detected and distinguished via standard circuit QED techniques \cite{Blais2004}.
Therefore, a transition from a singlet to a doublet state will appear as a discontinuous jump in a measurement of the transmon frequency spectrum.

Using this method, we have detected the singlet-doublet transition and reconstructed the phase diagram of a quantum dot junction as a function of all experimentally controlled parameters in a single device: the energy level of the dot, the tunnel couplings to the superconducting leads, the superconducting phase difference across the quantum dot junction, and also an external Zeeman field. The measured phase boundaries are in agreement with the single-impurity Anderson model with superconducting leads as calculated via the numerical renormalization group (NRG) \cite{wilson1975,satori1992,yoshioka2000,bulla2008} methods, and include parameter regimes that have experimentally not been explored before. Finally, we have investigated the rates at which the quantum dot switches between doublet and singlet occupation via real-time monitoring of the transmon circuit, allowing us to determine the switching time-scales of the quantum dot junction parity across the phase transition. 


\section{\label{sec:device} Device Overview}

\begin{figure}[t!]
    \centering
        \includegraphics[scale=1.0]{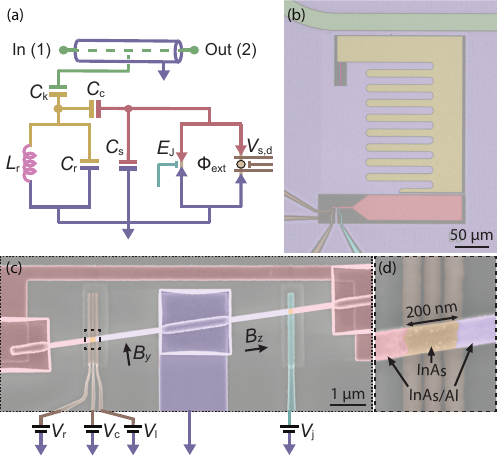}    \caption{{ \bf Device overview.}  (a) Diagram of the microwave circuit. A coplanar waveguide transmission line (green center conductor) is capacitively coupled to a grounded LC resonator. The resonator consists of an island (yellow) capacitively and inductively  (pink) shunted to ground (blue). The resonator is in turn capacitively coupled to a transmon island (red), which is shunted to ground capacitively as well as via two parallel Josephson junctions.
        (b) False-colored optical microscope image of device A showing the qubit island, the resonator island, the resonator inductor, the transmission line, the electrostatic gates and ground. (c) False-colored scanning electron micrograph (SEM) of the transmon's Josephson junctions, showing the InAs/Al nanowire into which the junctions are defined. The $B_y$ component of the magnetic field is used to tune $\Phi_{\rm ext}$ \cite{Note2}. $B_z$ is the magnetic field component parallel to the nanowire. (d) False-colored SEM of the quantum dot junction in which the quantum dot is gate defined. The three bottom gates have a width and spacing of \SI{40}{nm}, although this is obfuscated by the dielectric layer placed on top \cite{Note2}.}
    \label{fig:device}
\end{figure}

The quantum dot junction under investigation is formed in an epitaxial superconductor-semiconductor InAs/Al nanowire \cite{Krogstrup2015}. It is defined in an uncovered section of the nanowire where the Al has been etched away, and controlled by three electrostatic bottom gates  (Fig.~\ref{fig:device}(d)).
As shown in the circuit of Fig.~\ref{fig:device}(a), this quantum dot junction is placed in parallel to a second Josephson junction, hereafter referred to as the ``reference junction'', to form a SQUID.
The reference junction consists of a second uncovered segment of InAs on the same nanowire as the quantum dot junction.
Its Josephson energy $E_J$ can be tuned with a single electrostatic gate via the field effect. 

The SQUID connects a superconducting island to ground, resulting in a transmon circuit~\cite{Koch2007} governed by the Hamiltonian 
\begin{equation}
H = -4E_{\rm c}\partial_\phi^2 + V(\phi),
\label{eq:transmon_ham}
\end{equation} 
where $E_{\rm c} = e^2/2C_\Sigma$, $C_\Sigma$ is the total capacitance of the island to ground.
The Josephson potential $V(\phi)$ is determined by the phase-dependent energies of the reference junction, $V_{\rm J}(\delta)=E_{\rm J}(1-\cos\delta)$, and of the quantum dot junction, $V_{\rm s,d}(\phi)$:
\begin{equation}
V(\phi) = E_{\rm J}\left[1-\cos(\phi-\phi_{\rm ext})\right] + \begin{cases}
V_{\rm s}(\phi) & \textrm{singlet} \\
V_{\rm d}(\phi) & \textrm{doublet}\,.
\end{cases}
\label{eq:Josephson_potential}
\end{equation}
Here, the phase drops across the quantum dot junction ($\phi$) and across the reference junction ($\delta$) are connected according to $\phi-\delta=\phi_\textrm{ext}$, where $\phi_{\rm ext}=(2e/\hbar)\Phi_\textrm{ext}$ is the phase difference resulting from the externally applied magnetic flux through the SQUID loop, $\Phi_\textrm{ext}$.

The presence of the reference junction serves several purposes.
First, it allows us to tune the phase difference at the quantum dot junction by changing $\Phi_\textrm{ext}$ with the $B_y$ component of the magnetic field \cite{Note2}.
We operate the device in a regime where the reference junction has a Josephson energy that is substantially larger than that of the quantum dot; for most of the parameter regime explored in our measurements it is larger by more than an order of magnitude \cite{Note2}. This ensures that the phase drop across the reference junction is close to zero, while the phase difference across the quantum dot junction is close to~$\phi_{\rm ext}$  \cite{Della2007}.
Second, the ability to tune $E_{\rm J}$ independently of the quantum dot junction configuration ensures that the transition frequencies of the transmon circuit remain inside the measurement bandwidth for all parameter regimes of the quantum dot junction.
Finally, the Josephson energy of the reference junction is  such that $E_{\rm J}/E_{\rm c} > 25$, suppressing unwanted sensitivity to the offset charge of the superconducting island, justifying its absence in the Hamiltonian of Eq.~\eqref{eq:transmon_ham} \cite{Koch2007}.

In order to perform microwave spectroscopy measurements, the transmon is capacitively coupled to a readout resonator which is in turn coupled to a transmission line.
This allows us to measure the circuit's complex microwave transmission $S_{21}$ through the transmission line's input~(1) and output~(2) ports. 

We implement the circuit as shown in Fig.~\ref{fig:device}(b-d).
The device differs from conventional circuit QED geometries in several ways \cite{Blais2021}, in order to allow the application of magnetic fields in excess of \SI{100}{mT}.
Apart from the Josephson junctions, all circuit elements are made out of field compatible \SI{20}{nm}-thick NbTiN films \cite{Luthi2018}.
We additionally incorporate vortex pinning sites in the ground plane, the transmission line, the resonator island and the transmon island \cite{Kroll2019}.
We use a lumped element readout resonator, which has previously been successfully utilized in flux-sensitive devices up to \SI{1}{T} \cite{PitaVidal2020}.
Its capacitance is formed by an interdigitated capacitor to ground, while its inductance is formed by a \SI{200}{nm} wide NbTiN nanowire, which has a kinetic inductance of \SI{15}{pH\per\Box}.
This design localizes the regions of high current density at the narrow inductor where vortices are less likely to nucleate due to its reduced width \cite{Samkharadze2016}.
For the transmon circuit the SQUID loop area is chosen to be small, $\sim\SI{5}{\upmu m^2}$, in order to suppress flux noise from misalignment in large parallel magnetic fields.
Finally, InAs/Al nanowires, in which both junctions are defined, have been shown to support sizeable Josephson energies in fields in excess of \SI{1}{T} \cite{PitaVidal2020, Uilhoorn2021}. Further details about device fabrication as well as the cryogenic and room temperature measurement setup can be found in the Supplementary Information (Section II) \cite{Note2}.


\section{\label{sec:theory} Anderson model for a quantum dot junction }

As we will show, the quantum dot junction can be described by a single Anderson impurity tunnel-coupled to two superconducting leads (Fig.~\ref{fig:theorydiagram}(b)).
We review its most important properties to facilitate the discussion of the experimental results that follow.

The Hamiltonian of the model takes the form
\begin{equation}\label{eq:ham}
H=H_\textrm{dot}+H_\textrm{leads}+H_{\rm T}.
\end{equation}
The first term describes a single-level quantum dot,
\begin{equation}\label{eq:hdot}
H_\textrm{dot} = \sum_{\sigma=\uparrow,\downarrow} \epsilon_{\sigma}d^\dagger_\sigma d_\sigma  + U n_\uparrow n_\downarrow\,.
\end{equation}
Here, $\epsilon_{\uparrow,\downarrow} = \epsilon \pm E_{\rm Z}/2$ gives the single-particle energies: $\epsilon$ is the dot energy level measured with respect to the Fermi level in the leads, which can be controlled via the central electrostatic gate, and $E_{\rm Z} = g \mu_\mathrm{B} B$ is the Zeeman energy.
In the latter, $g$ is the effective g-factor of the level, $\mu_\mathrm{B}$ is the Bohr magneton, and $B$ is the magnetic field strength. In the experiment we choose the $B$-field direction to be parallel to the nanowire in order to maximize the magnetic field compatibility. Finally, $U>0$ is the repulsive Coulomb interaction between the electrons, which disfavors the double occupancy of the impurity, while $n_\sigma = d^\dagger_\sigma d_\sigma$ are number operators for the dot level, with $d_\sigma\,(d^\dagger_\sigma)$ the electron annihilation (creation) operators. The resulting energy diagram described by Eq.~\eqref{eq:hdot} is shown in Fig.~\ref{fig:theorydiagram}(c).

The many-particle energy levels of Eq.~\eqref{eq:hdot} are divided in two sectors, corresponding to their fermion parity, or equivalently, to their total spin $S$.
The singlet sector includes the states of even parity, which have $S=0$: the empty state $\ket{0}$ and the pair state $\ket{2}=d^\dagger_\uparrow d^\dagger_\downarrow \,\ket{0}$.
The doublet sector includes the states of odd parity, which have $S=1/2$:~$\ket{\uparrow}=~d^\dagger_\uparrow\ket{0}$ and $\ket{\downarrow}=d^\dagger_\downarrow\ket{0}$.
It is convenient to introduce the energy $\xi=\epsilon+U/2$, corresponding to half of the energy gap in the singlet sector, so that $\xi=0$ corresponds to the electron-hole symmetry point, where $\ket{0}$ and $\ket{2}$ are degenerate in energy.
The ground state of $H_\textrm{dot}$ belongs to the doublet sector for $\abs{\xi/U} < 1/2$.

The second term in Eq.~\eqref{eq:ham} describes two superconducting reservoirs,
\begin{equation}
H_\textrm{leads} = \sum_{i,k} \epsilon_k n_{i,k} + \sum_{i,k}\left(\Delta e^{-i\phi_i} c^\dagger_{i,k\uparrow}c^\dagger_{i,k\downarrow} + \textrm{h.c.}\right)
\end{equation}
where $i={\rm L},{\rm R}$ labels the left and right leads, $k$ labels spin-degenerate single-particle states, and $\Delta e^{-i\phi_i}$ is the $s$-wave pairing potential in each reservoir. The gauge-invariant phase difference between them is $\phi = \phi_1-\phi_2$. It is made experimentally tunable with magnetic flux as discussed in section \ref{sec:device}. We assume the reservoirs to have identical gap $\Delta$ and density of states $\rho$; this assumption should be reasonable since in the experiment the two leads are made out of a single hybrid nanowire. We further take the g-factor of the reservoirs to be zero, capturing the magnetic field dependence of the combined system in the effective quantum dot g-factor of Eq.~\eqref{eq:hdot}.

Finally, the third term is the tunneling Hamiltonian coupling the dot and the reservoirs,
\begin{equation}
H_{\rm T}=\sum_{i,k,\sigma} \left(t_i c^\dagger_{i,k,\sigma}d_\sigma + \textrm{h.c.}\right)\,,
\end{equation}
where $t_i$ are the dot-reservoir tunnel coupling strengths, which, for simplicity, we choose to be independent of $k$ and spin. The tunneling rate across each barrier is given by $\Gamma_i = \pi\rho \abs{t_i}^2$. In practice these rates can also be tuned with electrostatic gates. The tunneling terms in $H_{\rm T}$ break the conservation of the parity and spin in the quantum dot. Nevertheless, the notion of singlet and doublet sectors introduced for the dot Hamiltonian of Eq.~\eqref{eq:hdot} is inherited by the total Hamiltonian of Eq.~\eqref{eq:ham}, provided that the spin $S$ is now regarded as the total spin of the system, including that of quasi-particles in the reservoirs.

Over the years, the model of Eq.~\eqref{eq:ham} (or immediate extensions of it) has become paradigmatic to describe quantum dots coupled to superconducting leads.
It has been studied in different limits and using a variety of numerical methods, often requiring advanced many-body methods such as NRG and quantum Monte Carlo for full quantitative descriptions \cite{Meden2019}.

A salient feature of the model is that a quantum phase transition between doublet and singlet ground state can occur upon changing several experimentally-tunable parameters.
The dot energy level $\xi$, the coupling strengths $\Gamma_{\rm L,R}$, as well as the superconducting phase difference $\phi$ and the magnetic field $B$ all act to shift the relative positions of $V_{\mathrm{s}}$ and $V_{\mathrm{d}}$ and cause an energy crossing between the ground states of the two sectors.
In the measurements reported in Sec.~\ref{sec:measurement} and \ref{sec:2d-maps}, we vary all these parameters and compare the extracted phase boundaries to theory. 

For the theoretical comparison we use the NRG method \cite{wilson1975,zitko2009,zitko_rok_2021_4841076} to compute the lowest-lying eigenvalues in the singlet and doublet spin sectors for the Hamiltonian in Eq.~\eqref{eq:ham} as a function of the phase difference $\phi$.
This results in the Josephson potentials $V_\mathrm{s}(\phi)$ and $V_\mathrm{d}(\phi)$, which are then used as input to the model of Eq.~\eqref{eq:Josephson_potential} to calculate the transmon transition frequencies \cite{Note2}.
The projection onto the lowest-energy state of the Josephson junction in each sector is enough to capture the salient features of our experiment, although the inclusion of excited Andreev states of the quantum dot junction in the circuit model is theoretically possible~\cite{Zazunov2003a,Keselman2019,Avila2020,Avila2020a,Kurilovich2021}.

Experimentally, the observation of the phase transition is facilitated by the presence of a $\pi$-phase shift between $V_{\rm s}(\phi)$ and $V_{\rm d}(\phi)$.
The phase shift originates from the required permutation of a spin-up and a spin-down electron when a Cooper pair sequentially tunnels through a dot that initially is occupied by one quasiparticle~\cite{spivak1991}.
Thus, while $V_{\rm s}(\phi)$ has a minimum at $\phi=0$, as encountered for conventional Josephson junctions, $V_{\rm d}(\phi)$ has a minimum at $\phi=\pi$ (Fig.~\ref{fig:theorydiagram}(d)).
Therefore, a quantum dot junction in a doublet state is often denominated as a $\pi$-junction, and the singlet-doublet transition is also referred to as the 0-$\pi$ transition.
In the following sections we will use the presence or absence of such a $\pi$-phase shift to identify regions with a singlet or a doublet ground state \footnote{Our assumption that regimes with $0$-junction and $\pi$-junction behaviour correspond to the quantum dot junction being in a singlet or doublet state, respectively, is only valid in the single-level regime, where the level spacing of the quantum dot is significantly larger than $\Delta$ and $U$. In the multi-level regime, where excited states of the quantum dot are involved, the presence or absence of the $\pi$ offset also depends on the character of the orbital wavefunctions in addition to the fermion parity \cite{VanDam2006a}.}.


\section{\label{sec:measurement} Transmon spectroscopy of the quantum dot}

To perform spectroscopy of the resonator, we monitor the microwave transmission $S_{21}$ across the transmission line while varying the frequency of a single continuous microwave tone, $f_{\rm r}$. This results in a dip with Lorentzian lineshape around the resonance frequency of the lumped-element resonator.
Two-tone spectroscopy is subsequently performed by fixing the frequency of this first tone, $f_{\rm r}$, at the minimum of the transmission amplitude, $|S_{21}|$, while varying the frequency of a second tone, $f_{\rm t}$, also sent through the transmission line. 
When the second tone matches the frequency of the ground to first excited transmon transition, $f_{\rm t} = f_{01}$, a peak in  $|S_{21}|$  is observed due to the transmon-state-dependent dispersive shift of the resonator \cite{Blais2004}.
This gives us access to the transmon transition frequency. 

We are interested in the behavior of the device when a single level of the quantum dot provides the dominant contribution to the Josephson effect \cite{Note1}.
To find such a regime, we search for an isolated resonance in the gate dependence of the frequency spectrum.
Isolated resonances often occur when the gate voltages controlling the quantum dot are set close to their pinch-off values~\cite{Bargerbos2020,Kringhoj2020}, here operationally defined as the voltage values below which the quantum dot junction does not contribute appreciably to the transmon's transition frequency.
In order to identify the right gate configuration, we perform the following sequence of calibration measurements.
First, we characterize the reference junction with the quantum dot pinched-off; second, we explore the sizeable parameter space governed by the three quantum dot gates; third, we identify the relation between $B_y$ and \flux~through the transmon frequency's SQUID oscillations; and finally we define appropriate gate coordinates to account for cross-couplings.
These calibration measurements are detailed in the Supplementary Material (Section III) \cite{Note2}.
As a result of this procedure, the gate voltage of the reference junction \Vj~is fixed such that the transmon frequency when the quantum dot junction is pinched-off is $f_{\rm 01}^0\approx\SI{4.4}{GHz}$. 
Furthermore, we fix \Vl~= 470 mV and introduce virtual plunger (\Vp) and right tunnel (\Vt) gates as a linear combination of \Vc~and \Vr, such that in what follows $\xi$ is mostly independent of \Vt.

\begin{figure}[t!]
    \centering
    \includegraphics[scale=1.0]{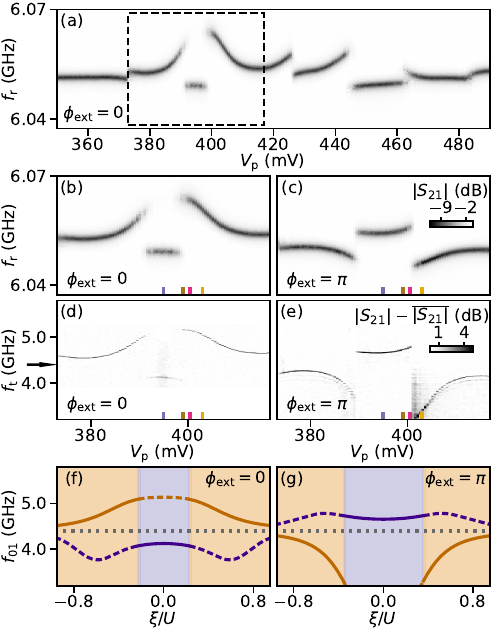} \caption{{\bf Resonator and transmon spectroscopy.} 
    (a) \Vp~dependence of single-tone spectroscopy for \flux~=~0, showing the resonator's transition frequency. \Vp~is a virtual gate voltage defined as a linear combination of \Vc~and \Vr~(see text). 
    (b) Zoom-in of (a) in the plunger gate range indicated with dashed lines in (a). (c) Same as (b) but for \flux~=~$\pi$. (d) \Vp~dependence of two-tone spectroscopy for \flux~=~0, showing the transmon's transition frequency. The black arrow indicates $f_{\rm 01}^0$, the transmon frequency set by the reference junction when the quantum dot is pinched off.  (e) Same as panel (d) but for \flux~=~$\pi$. For panels (a-e) \Vt~= 182 mV and \Vl~= 470 mV. (f) Theoretical estimates of the singlet (orange), doublet (purple) and reference junction-only (dotted, grey) transmon frequencies as $\xi$ is varied for \flux~=0. Solid (dashed) lines indicate which quantum dot occupation corresponds to the ground (excited) state. (g) Same as panel (f) but for \flux~=~$\pi$. For panels (f-g) $\Delta/h=\SI{46}{GHz}$, $U/\Delta=12.2$, $\Gamma_\mathrm{L}/\Delta=1.05$ and $\Gamma_\mathrm{R}/\Delta=1.12$. $f_{\rm t}$ and $f_{01}$ denote, respectively, the frequency of the second tone in two-tone spectroscopy and the first transmon transition frequency (see text). }
    \label{fig:plunger}
\end{figure}

We then move on to study the quantum dot junction. We first monitor the resonator frequency for \flux~$= 0 $ while the plunger gate voltage \Vp~is varied (Fig.~\ref{fig:plunger}(a)).
This reveals a resonant shape which is discontinuously interrupted near its peak at \Vp~=~395~mV, followed by other discontinuous jumps in the resonator frequency.
A zoom into the resonance is shown in Fig.~\ref{fig:plunger}(b) and the corresponding transmon transition frequency,  exhibiting the same discontinuity as the resonator, is shown in Fig.~\ref{fig:plunger}(d).
We identify  regions in \Vp~where the transmon frequency $f_{\rm 01}$ is larger and smaller than the reference frequency $f_{\rm 01}^0$.
This hierarchy is reversed upon changing the applied flux to \flux~$= \pi$, as shown in Figs.~\ref{fig:plunger}(c, e). 

These observed discontinuities in frequency are a signature of a singlet-doublet transition.
The change of the ground state of the quantum dot junction determines a sudden switch in the branch of the Josephson potential of Eq.~\eqref{eq:Josephson_potential} (from $V_\mathrm{s}$ to $V_\mathrm{d}$ or vice-versa) and, thus, a sudden change in the transmon frequency.
This is illustrated theoretically in Figs.~\ref{fig:plunger}(f-g), which show the expected evolution of the transmon frequencies as a function of $\xi$.
Here, the transition occurs as the single-particle energy level is tuned toward the electron-hole symmetry point $\xi=0$, where the doublet ground state is energetically favorable.
The resemblance of the dispersion of the transition frequencies in Figs.~\ref{fig:plunger}(f-g) to the experimental data in Figs.~\ref{fig:plunger}(d-e) confirms that \Vp~primarily tunes $\xi$, as was intended with the virtual gate voltage definition.

The occurrence of the singlet-doublet transition requires a change of the fermion parity of the quantum dot junction.
In the S-QD-S setup, this is possible in the presence of a population of excited quasiparticles in the superconducting leads, providing the required fermion parity reservoir.
The presence of these quasiparticles should further result in a finite occupation of both the singlet and doublet states when their energy difference is small compared to the effective temperature of the quasiparticle bath, namely in the vicinity of the transition.
Indeed, upon closer inspection of the data of Figs.~\ref{fig:plunger}(b-c), both branches of the spectrum are visible in a small frequency window surrounding each discontinuous jump.
This is because these transition spectra are obtained by averaging over many subsequent frequency sweeps, thus reflecting the occupation statistics of the junction. 
This feature is further discussed in the next Section.

In Figs.~\ref{fig:plunger}(d-e), the fact that the frequency shift of the transmon has the opposite sign for the singlet and doublet sectors is a consequence of the $\pi$-phase shift in the Josephson potential between the two sectors.
For the case \flux~=~0, the singlet potential interferes constructively with the reference junction potential, while the doublet potential interferes destructively, resulting in $f_{01} > f_{\rm 01}^0$ for the singlet and $f_{01} < f_{\rm 01}^0$ for the doublet.
This behaviour is reversed when \flux~=~$\pi$, and thus serves as a method for identifying the quantum dot junction state.


\section{\label{sec:2d-maps} Singlet-doublet transition boundaries}
Having established a method for identifying singlet and doublet states by transmon spectroscopy, we now experimentally investigate the phase diagrams of the quantum dot junction. We focus on the behaviour around \Vp~=~395~mV and monitor singlet-doublet transitions versus multiple different control parameters. 

\subsection{\label{ssec:plunger-flux} Plunger gate and Flux}

\begin{figure}[t!]
    \centering
    \includegraphics[scale=1.0]{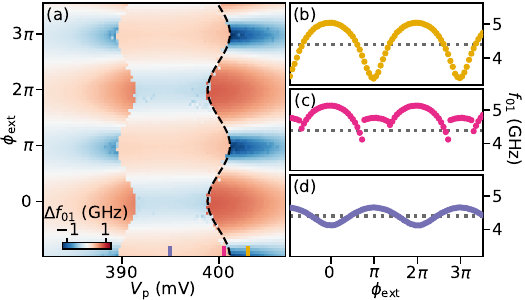} \caption{{\bf Flux and plunger gate dependence.} 
    (a)  $\Delta f_{\rm 01} = f_{\rm 01} - f_{\rm 01}^0$ versus \Vp~and \flux~as extracted from 
    two-tone spectroscopy. The dashed line is a sinusoidal guide for the eye, denoting the transition boundary in line with the theoretical expectation \cite{Note2}. (b)-(d) Three linecuts of \ft~versus \flux~at representative \Vp~values, indicated in panel (a) and Fig.~\ref{fig:plunger}(b-e). The dotted line indicates $f_{\rm 01}^0$. For all panels \Vt~=~182~mV.}
    \label{fig:plunger-flux}
\end{figure}

We first study the singlet-doublet phase map in \Vp~and \flux~space. Fig.~\ref{fig:plunger-flux}(a) shows the transmon frequency offset with respect to the frequency set by the reference junction, $\Delta f_{\rm 01} = f_{\rm 01} - f_{\rm 01}^0$, as a function of \Vp~and \flux.
As discussed in the previous Section, positive values of $\Delta f_{\rm 01}$ result from constructive interference between the two junctions, while negative $\Delta f_{\rm 01}$ values result from destructive interference.
Going from left to right, three distinct plunger regions can be observed, with a sudden flux offset of exactly $\pi$ between them (Fig.~\ref{fig:plunger-flux}(b,d)).
Based on the preceding discussion of Fig.~\ref{fig:plunger}, we identify the outer two regions as phases with a singlet ground state and the inner region as a doublet ground state.
We note that the change in contrast between the two singlet regions suggests that \Vp~also weakly tunes $\Gamma_{\rm L,R}$ in addition to $\xi$. 

For values of \Vp~close to the singlet-doublet transition we also observe a sinusoidal dependence of the transition boundary on the external flux, resulting in an enhanced region of doublet occupation around \flux~=~$\pi$ with respect to \flux~=~0.
This comes about from interference between tunneling processes involving the two superconducting leads of the quantum dot junction \cite{Oguri2004, Delagrange2015}, as further discussed in Sec.~\ref{ssec:tunnel}.
At a value of \Vp~fixed near this boundary one thus also observes a singlet-doublet transition versus the external flux (Fig.~\ref{fig:plunger-flux}(c)).

In Fig.~\ref{fig:plunger-flux} and subsequent figures, the transition boundary between the singlet and doublet phase appears to be sharp and not affected by the thermal broadening typical of transport experiments~\cite{VanDam2006a, Cleuziou2006, Jorgensen2007, Szombati2016, Delagrange2015, Delagrange2018}.
The sharpness is a result of a selective spectroscopy technique.
As detailed in the Supplementary Information (Section III.E) \cite{Note2}, in the vicinity of a transition two resonant dips appear in single-tone spectroscopy, one for the singlet and one for the doublet.
In this circumstance, the center frequency of either dip can be chosen as the readout frequency for the subsequent two-tone spectroscopy measurement.
This binary choice selects the transmon transition frequency belonging to the corresponding quantum dot junction state.
It is reasonable to assume that the most prominent dip corresponds to the state of the quantum dot junction which is more prominently occupied, and thus lower in energy.
If this is the case, the extracted phase boundaries are a close approximation of the zero-temperature phase diagram of the quantum dot junction.
When the occupations of singlet and doublet states are almost equally probable, the selective spectroscopy method is affected by selection errors, which leads to the pixelation effects visible in Fig.~\ref{fig:plunger-flux} near the phase boundaries.
In Sec.~\ref{sec:qpp}, we will explicitly measure the lifetimes of the quantum dot in the singlet and doublet states, substantiating the latter statements.

\subsection{\label{ssec:tunnel}  Tunnel gate}

\begin{figure}
    \centering
    \includegraphics[scale=1.0]{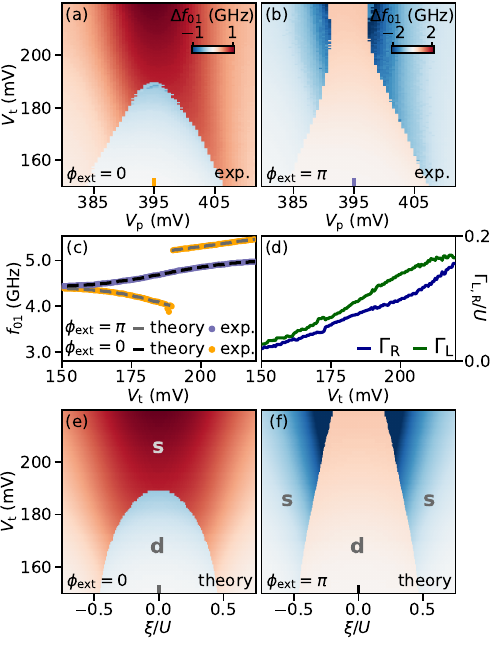}  \caption{ {\bf Tunnel gate dependence.} (a) $\Delta f_{\rm 01}$  versus \Vp~and \Vt~at \flux~=~0, where \Vt~is a virtual gate voltage defined as a linear combination of \Vc and \Vr (see text). The blue region corresponds to a negative supercurrent contribution from the quantum dot junction, while the red region corresponds to a positive contribution. (b) The same measurement as (a) repeated for \flux~=~$\pi$. 
    (c) Linecuts of (a) and (b) at \Vp~=~\SI{395}{mV} overlayed with best-fits based on NRG calculations. (d) Extracted dependence of $\Gamma_{\rm L,R}$ on \Vt.
    (e) Calculated transmon frequencies based on NRG calculations at \flux~=~0 as matched to the measured data, with the \Vt~axis as given in figure (d). The color bar is shared with panel (a). (f) Same as (e) but for \flux~=~$\pi$, with the same color bar as (b). For the NRG calculations in panels (c-f) we fix $\Delta/h=\SI{46}{GHz}$ and $U/\Delta=12.2$.} 
    \label{fig:tunnel}
\end{figure}

\begin{figure*}[ht!]
    \centering
    \includegraphics[scale=1.0]{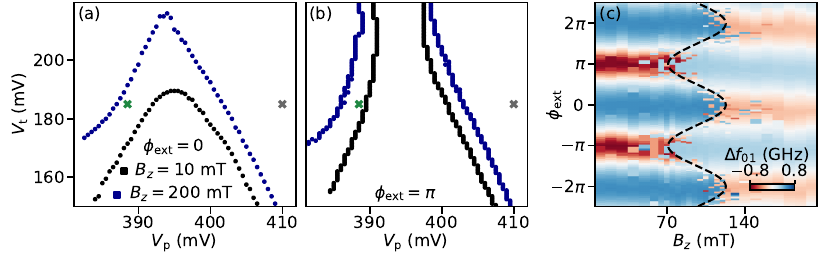}  \caption{{\bf Parallel magnetic field dependence.}
    (a)  Borders between singlet and doublet regions for $B_z$~=~10~mT (black) and $B_z$~=~200~mT (blue).  \flux~=~0 (b) Same as (a) but for \flux~=~$\pi$. The grey marker denotes the gate point used to re-calibrate the flux axis after varying $B_z$ \cite{Note2}. (c) $\Delta f_{\rm 01}$ versus $B_z$ and \flux, measured at the gate point indicated in (a) and (b) with a green marker. The sinusoidal dashed line serves as a guide for the eye, in line with the transition boundary expect from theory \cite{Note2}.}
    \label{fig:Zeeman}
\end{figure*}

Next, we explore the singlet-doublet transition in plunger and tunnel gate space, where the tunnel gate is expected to control $\Gamma_{\rm L,R}$. Fig.~\ref{fig:tunnel}(a) shows $\Delta f_{\rm 01}$ versus plunger and tunnel gates at \flux~=~0.
We can identify the region where $\Delta f_{\rm 01} > 0$ as the singlet phase and the region where $\Delta f_{\rm 01} < 0$ as the doublet phase.
The region of doublet occupancy takes the shape of a dome, similar to the one coarsely seen in flux-insensitive tunneling spectroscopy experiments \cite{Lee2014, Lee2017}.
This shape is in accordance with theoretical expectations for the boundary in the $\xi-\Gamma$ plane. 
Its physical origin depends on the parameter regime~\cite{Kadlecova2019}.
For $U\ll\Delta$ it arises due to an increase in induced superconductivity on the dot with increasing values of $\Gamma$, favoring BCS-like singlet occupation.
For $U\gg\Delta$ it instead comes about from increased anti-ferromagnetic Kondo exchange interactions between the spin on the dot and the quasiparticles in the leads, favoring a Yu-Shiba-Rusinov (YSR)-like singlet occupation.
In both regimes the singlets compete with doublets, ultimately determining the transition to a singlet ground state at large enough $\Gamma = \Gamma_{\mathrm{L}} + \Gamma_{\mathrm{R}}$.

We investigate the same plunger and tunnel gate dependence at an external flux \flux~=~$\pi$, see Fig.~\ref{fig:tunnel}(b). 
We find that the doublet phase is enhanced considerably compared to \flux~=~0, due to the previously mentioned interference between tunneling processes to the superconducting leads.
Notably, rather than a dome-like shape, the phase boundary takes a characteristic ``chimney'' shape that was theoretically predicted~\cite{Oguri2004} but, to our knowledge, not yet confirmed experimentally before these measurements. 
Unlike the dome, the chimney does not close for any \Vt. 
In an extended gate range, it is seen to connect to another doublet region of the parameter space which was disconnected from the dome of Fig.~\ref{fig:tunnel}(a) at \flux~=~0 (Supplementary Information Section III.D \cite{Note2}).

The chimney at \flux~=~$\pi$ is much less thoroughly researched than the dome at \flux~=~0.
The open questions include that of the exact nature of the doublet states as a function of the $U/\Delta$ and $\Gamma/U$ ratios, and the role of the flux bias \cite{kirsanskas2015,kadlecova2017,zalom2021,escribano2022}.
In particular, when $U\gg\Delta$, the doublet state for small $\Gamma$ is a decoupled doublet state with a single local moment in the quantum dot. On the other hand, in the same limit but at large $\Gamma$ (i.e. in the neck of the chimney), the strong exchange interaction with both superconductors is expected to lead to some mixing with the doublet states that involve one Bogoliubov quasiparticle from each lead~\cite{zitko2017}, causing an overscreening of the local moment in the quantum dot.
The role of the exchange interaction is more pronounced at \flux~=~$\pi$ also because the anomalous component of the hybridisation (describing the proximity effect) is suppressed due to the cancellation of contributions from the left and right leads~\cite{zalom2021}, where the cancellation is exact when $\Gamma_L = \Gamma_R$. This further stabilizes the spin-doublet states. The experimental observation of the chimney calls for more thorough theoretical studies of this parameter regime of the model.

We compare the results at both values of external flux to the expected transition frequencies obtained from NRG calculations using Eq.~\eqref{eq:ham}.
We assume that $\xi=0$ at \Vp~=~\SI{395}{mV} since this is the symmetry point of the experimental data.
At this point, by requiring simultaneous agreement between experiment and theory for both values of external flux (Fig.~\ref{fig:tunnel}(c)), we are able to extract several of the model parameters. We find that $\Delta/h = \SI{46}{GHz}$ (\SI{190}{\upmu eV}), close to the bulk value of Al.
We furthermore extract $U/\Delta = 12.2$, corresponding to a sizeable charging energy of $\SI{2.3}{meV}$.
It places the nature of the singlets near $\xi=0$ in the strongly correlated regime, with a YSR-like character rather than a BCS-like one.
By matching values of $\Gamma_{\rm L,R}$ to \Vt~we then find that $\Gamma/U$ varies between 0.05 and 0.4, while $\Gamma_{\rm R}/\Gamma_{\rm L} \approx 0.75-1$ in the range of gates explored (Fig.~\ref{fig:tunnel}(d)).
The details of the numerical procedure as well as error estimation can be found in the Supplementary Information (Section I.C), including estimates based on an alternative potential shape for the reference junction~\cite{Note2}.

The extracted set of parameters is consistent with the observed dome shape at \flux~=~0, as shown in  Fig.~\ref{fig:tunnel}(e).
Additionally, as a result of the ratio $\Gamma_{\rm R}/\Gamma_{\rm L}$ remaining close to 1, the extracted parameters also match the observed diverging behaviour at \flux~=~$\pi$ (Fig.~\ref{fig:tunnel}(f)), which was not enforced in the parameter extraction.
We note that in these panels we did not map \Vp~to $\xi$ beyond identifying \Vp~=~\SI{395}{mV} with $\xi=0$.
A unique mapping could not be constructed due to the unintended dependence of $\Gamma$ on \Vp.
We speculate that this causes the remaining discrepancies between the measured and calculated boundaries in the horizontal direction. 

\subsection{\label{ssec:field}  Parallel magnetic field}

Finally, we investigate the effect of a parallel magnetic field on the phase transition boundaries. Here, we expect a magnetic-field induced singlet-doublet transition to occur \cite{Lee2014,Valentini2021,Whiticar2021}.
As $B_z$ increases, the doublet sector separates into spin species that are aligned and anti-aligned with respect to the magnetic field, dispersing in opposite energy directions. The singlet ground state energy, on the other hand, is approximately independent of magnetic field.
Given an appropriate zero-field energy level configuration, for some $B_z$ value the energy of one of the two doublet states will thus become lower than that of the singlet, and become the ground state instead (see Fig.~\ref{fig:theorydiagram}(c)).

Such a transition will only occur for specific configurations of~\Vp~and~\Vt~in the experimentally accessible range of magnetic fields. We therefore start by applying $B_z =$~\SI{200}{mT} parallel to the nanowire axis, a sizeable magnetic field, yet one for which the $E_{\rm J}$ of the reference junction is not yet substantially suppressed. At this field we investigate the effect on the \Vp~and \Vt~phase map.
The result, shown in Figs.~\ref{fig:Zeeman}(a-b), reveals an expansion of the doublet region for both \flux~=~0 and \flux~$=\pi$.
We can classify different regions in the parameter space by comparing the phase boundaries at $B_z =$~\SI{10}{mT} and $B_z =$~\SI{200}{mT}.
There are regions in which a singlet ground state remains a singlet ground state, independent of the flux and the magnetic field, as well as regions where a singlet-doublet transition occurs depending on the value of the flux.
There is also a region that starts off as a singlet ground state and ends up as a doublet ground state at high field,  for all values of the flux.
Thus, fixing \Vp~and \Vt~in this region, we expect to observe a transition with $B_z$ for any value of \flux.
A measurement of $\Delta f_{\rm 01} $ versus \flux~and $B_z$ (Fig.~\ref{fig:Zeeman}(c)) indeed reveals such a transition, occurring at a different magnetic field depending on the flux value.
For details about the data analysis and identification of the flux axis we refer to the Supplementary Information (Section IV) \cite{Note2}.


\section{\label{sec:qpp} Dynamics of the singlet-doublet  transition}
In the preceding sections we made use of selective spectroscopy to reconstruct the phase transition boundaries. We now turn to time-resolved spectroscopy techniques to study the parity dynamics of the quantum dot junction close to the transition, aiming to characterize the lifetimes of singlet (even parity) and doublet (odd parity) states.
These methods have previously been used to study quasiparticle dynamics in superconducting qubits~\cite{Serniak2019, Uilhoorn2021}, and recently also applied to a nanowire junction to study the poisoning of Andreev bound states~\cite{Hays2018, Hays2020, Wesdorp2021}.

\begin{figure}[t!]
    \centering
    \includegraphics[scale=1.0]{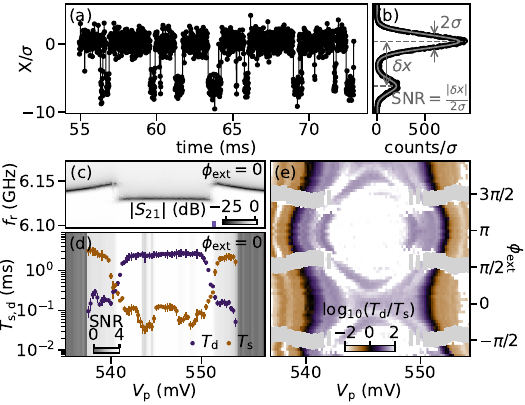}  \caption{{\bf Dependence of parity lifetimes on \Vp~and \flux~ for device B.}  (a)  A \SI{18}{\milli s} cut of a continuously measured time trace integrated in time bins of $t_{\rm int} =$ 11.4~$\upmu$s, revealing jumps between two distinct states. $X$ is the common axis onto which the quadratures of the outgoing microwave field are rotated to obtain the highest SNR, which takes a value of 3.3 in this panel. (b) 1D histogram of the response in (a) (black) and the best fit of a double Gaussian line-shape (gray). The separation of their centers $\delta x$ and their width $\sigma$ together define the SNR. The ratio of their amplitudes determines the ratio of the lifetimes. For panels (a-b) \Vp~=~551.4 mV  and \flux~=~0. (c) \Vp~dependence of $|S_{21}|$ at \flux~=~0. (d) \Vp~dependence of the extracted lifetimes at \flux~=~0. Markers indicate the mean while error bars indicate the maximum and minimum values of 10 consecutive \SI{2}{s} time traces. The SNR is shown in greyscale in the background. For points where ${\rm SNR}<1$, the extracted lifetimes are discarded. (e) 2D map of ${\rm log_{10}}(T_{\rm d}/T_{\rm s})$ versus \Vp~and \flux, extracted from a \SI{2}{s} time trace for each pixel.
    White pixels indicate points at which SNR~$<1$, while grey regions indicate where the resonator frequencies of singlet and doublet states overlap and thus cannot be distinguished.}
    \label{fig:QPP_dependence}
\end{figure}

To resolve individual switching events we use a second device (device B) with a larger signal-to-noise ratio (SNR) than the device used for the preceding sections (device A), enabling the use of short acquisition times.
Device B is nearly identical to device A, except for two features meant to increase the SNR: (1) a stronger coupling between the resonator and the transmission line; (2) an additional capacitor at its input port, which increases the directionality of the outgoing signal~\cite{Heinsoo2018}.
On device B we perform measurements on microsecond timescales by directly monitoring changes in the outgoing signal at a fixed readout frequency.
A continuous measurement of the outgoing microwave field then reveals a random telegraph signal between two different levels, a consequence of the switches in the quantum dot junction parity (Fig.~\ref{fig:QPP_dependence}(a-b)).
Owing to the increased temporal resolution of the detection method, even short-lived excited state occupation of a few $\upmu$s can now be detected.
The characteristic time scales of the telegraph signal reflect the underlying lifetimes of the singlet and doublet states, $T_{\rm s}$ and $T_{\rm d}$, or equivalently their decay rates,  $\Gamma_{\rm s}=1/T_{\rm s}$ and $\Gamma_{\rm d}=1/T_{\rm d}$.
These quantities can be extracted via a spectral analysis of the time traces, as described in detail in the Supplementary Information (Section V) \cite{Note2}.

To investigate the switching dynamics we tune device B to a regime similar to that of  Sec.~\ref{ssec:plunger-flux} studied in device A. By measuring $S_{21}$ with single-tone spectroscopy we once-more find ground state transitions between singlet and doublet as a function of \Vp~(Fig.~\ref{fig:QPP_dependence}(c)).
The discontinuous resonant shape, akin to that of Fig.~\ref{fig:plunger}(b), is symmetric around \Vp~=~546 mV, which we identify with $\xi=0$.
The singlet and doublet resonant frequencies are simultaneously visible close to the discontinuity at the transition.
The time-resolved measurements over the same gate voltage range reveal a smooth but strong evolution of the parity lifetimes with \Vp~(Fig.~\ref{fig:QPP_dependence}(d)). The hierarchy of lifetimes inverts as \Vp~is tuned across the phase transition, reflecting the change in the ground state parity.
Away from the transition, in either the singlet or doublet phase, we observe the lifetime in the ground state sector to be on the order of several milliseconds, exceeding that of the excited state by more than an order of magnitude.
These numbers are very favorable for the implementation of Andreev pair qubits~\cite{Janvier2015} as well as Andreev spin qubits~\cite{Hays2018, Hays2020, Hays2021}, whose control has so far been limited by microsecond parity lifetimes . 

We further explore the evolution of the relative lifetimes versus \Vp~and \flux. Fig.~\ref{fig:QPP_dependence}(e) shows a two-dimensional map of ${\rm log_{10}}(T_{\rm d}/T_{\rm s})$, which is a measure of the lifetime asymmetry.
We find behaviour similar to that previously seen in Fig.~\ref{fig:plunger-flux}, with a sinusoidal boundary of equal rates, indicative of the singlet-doublet transition.
Furthermore, we observe a strong polarization of the junction parity inside the doublet phase ($T_{\rm d}\gg T_{\rm s}$), where the signal-to-noise ratio (SNR) eventually becomes limited by our ability to resolve the rare and short-lived switches out of the ground state.
Additionally, we find a modulation of $T_{\rm d}$ with flux, with longer lifetimes at $\phi_\textrm{ext}=\pi$ \cite{Note2}.
This flux dependence likely originates from the oscillation of the singlet-doublet energy gap with flux, but might also be indicative of a coherent suppression of the tunneling rates, as previously observed in Al tunnel junctions~\cite{Pop2014}.
The polarization of the junction parity also occurs inside the singlet phase, where $T_{\rm s}\gg T_{\rm d}$ for \Vp~values away from the transition (Fig.~\ref{fig:QPP_dependence}d).

Strong parity polarization may not be surprising for a system in thermal equilibrium at temperatures below \SI{100}{mK}, typical of these experiments, corresponding to a thermal energy small compared to the singlet-doublet energy difference away from the transition.
However, parity lifetimes in Josephson junctions are seldom determined by thermal fluctuations, but rather by highly energetic non-equilibrium quasiparticles often observed in superconducting circuits \cite{Glazman2021}.
While non-equilibrium quasiparticles are most likely also present in our device, we believe that their influence is suppressed by the large charging energy of the quantum dot junction.

Finally, we observe a non-monotonic variation of the rate asymmetry inside both the singlet and doublet phase, forming apparent contours of fixed lifetimes.
We hypothesize two possible reasons, distinct in origin but possibly co-existing, behind this structure in the data: it could be caused by parity pumping mechanisms where the readout tone is resonant with the energy difference between singlet and doublet \cite{Wesdorp2021}, as well as by the spectral density of the non-equilibrium quasiparticles present in the environment \cite{Serniak2018}.
Further investigation of the tunnel gate, power, and temperature dependence of the rate asymmetry can be found in the Supplementary Information (Section VI) \cite{Note2}; we leave a more detailed study for future work.


\section{\label{sec:conclusion} Conclusions}

We have demonstrated the use of a transmon circuit to sensitively detect the ground state parity of a quantum dot Josephson junction. The transition frequency of the transmon exhibits a discontinuity if the ground state of the device changes from a singlet to a doublet, due to the presence of a $\pi$-phase shift in the Josephson potential of the junction.
This allowed us to accurately reconstruct the occurrence of the singlet-doublet transition as a function of all control parameters available in a single device, matching them to those expected from NRG calculations of an Anderson impurity model.
In particular, we have observed the flux-induced enhancement of the doublet phase, in the form of the striking transformation of a dome-shaped phase boundary at $\phi_\textrm{ext}=0$ into a chimney-shaped phase boundary at $\phi_\textrm{ext}=\pi$ (Fig.~\ref{fig:tunnel}).

In future research, this singlet-doublet tuning capability could become beneficial for several applications.
First, it can be used to define and control Andreev pair and spin qubits, and to couple them to conventional superconducting qubits.
Second, tuning the dot to the doublet phase is a robust way to induce a $\pi$ phase shift, which could be exploited to define a hybrid $0-\pi$ qubit that does not rely on the fine-tuning of the applied flux \cite{Larsen2020}.
Third, it can facilitate the bottom-up realization of a topological superconductor from a chain of proximitized quantum dots \cite{Sau2012,Fulga2013,Stenger2018}.
Finally, fast gate or flux-based switching between the 0 and $\pi$ shift of the dot can also be of interest for applications in Josephson magnetic random access memory (JMRAM) technologies \cite{Dayton2018}.

We have subsequently used continuous time-domain monitoring of the transmon resonant frequency to determine the lifetimes of singlet and doublet states.
We find that the time between switching events is strongly enhanced when the quantum dot is tuned away from the phase transition. 
Since our estimates indicate that $U\gg\Delta$ in our devices, we attribute this effect to the large energy difference associated with charging the quantum dot.
These findings are encouraging for Andreev qubits, which benefit from long parity lifetimes, and suggest that large-$U$ quantum dots could be effective as filters for high-energy quasiparticles.
However, further work is required to understand the full dependence of parity lifetimes on $U$.

In this work we have focused on the study of a single-level quantum dot by tuning our junction very close to pinch-off.
Looking forward, there is much left to explore in the parameter space of such a device.
To begin with, it would be interesting to understand whether the crossover from the BCS-like to the YSR-like singlet has any signature in the microwave response of the system.
Second, opening the junction further brings the quantum dot into a multi-level regime, not captured by the single impurity Anderson model, and still largely unexplored.
Finally, while we have primarily studied the ground state properties of the quantum dot junction, microwave spectroscopy should allow to study its excitations, as e.g. recently demonstrated in Refs.~\cite{Fatemi2021,canadas2021}, particularly at $\phi_\textrm{ext}=\pi$.

Further work will also aim at elucidating the role of spin-orbit coupling in the quantum dot junction.
It is well known that, when time-reversal invariance is broken, spin-orbit coupling can induce a spin-splitting of energy levels in the doublet sector~\cite{Chtchelkatchev2003,Padurariu2010,Tosi2018}, essential for Andreev spin qubits.
While this effect could have been expected to occur in the measurements presented here, it was not detected; we speculate that the level spacing in the dot was too large to result in a significant splitting \cite{Padurariu2010}.

Important extensions of our work could arise if the hybrid nanowire in our microwave circuit was driven into the Majorana topological phase~\cite{Ginossar2014,Keselman2019,Avila2020,Avila2020a}, which is currently challenging because of a large parameter space~\cite{Pikulin2021} and because of demanding disorder requirements~\cite{Stanescu2021}.
Including a quantum dot in a Josephson junction between two topological superconductors could be beneficial for the detection of the $4\pi$ Josephson effect: as we have seen, it mitigates quasiparticle poisoning, although it would not resolve~\cite{Schulenborg2020} the problem of distinguishing Majorana zero modes from trivial zero-energy Andreev bound states~\cite{prada2020}.
Finally, the manipulation of quantum dots coupled to superconducting leads is an essential ingredient of scalable proposals for topological quantum computation~\cite{Karzig2017}.

\begin{acknowledgments}

We acknowledge fruitful discussion with Gijs de Lange, and we thank Valla Fatemi, Pavel Kurilovich, Max Hays, Spencer Diamond, Nick Frattini, and Vladislav Kurilovich for sharing their related work \cite{Fatemi2021} prior to its publication, and for productive discussions. We thank Valla Fatemi and Pavel Kurilovich also for their feedback on this manuscript. We further thank Peter Krogstrup for guidance in the material growth, Daniël Bouman for nanowire placement, and James Kroll for help with nanofabrication. This research is co-funded by the allowance for Top consortia for Knowledge and Innovation (TKI’s) from the Dutch Ministry of Economic Affairs, research project {\it Scalable circuits of Majorana qubits with topological protection} (i39, SCMQ) with project number 14SCMQ02, from the Dutch Research Council (NWO), and the Microsoft Quantum initiative. R. \v{Z}. acknowledges the support of the Slovenian Research agency (ARRS) under P1-0044, P1-0416, and J1-3008. R. A. acknowledges support from the Spanish Ministry of Science and Innovation through Grant PGC2018-097018-B-I00 and from the CSIC Research Platform on Quantum Technologies PTI-001. B.v.H. was partially supported by the Dutch Research Council (NWO).
\end{acknowledgments}

\section*{Author contributions}
A.B., M.P.V., B.v.H. and A.K. conceived the experiment.
Y.L. developed and provided the nanowire materials.
A.B., M.P.V., L.S., L.G. and J.J.W prepared the experimental setup and data acquisition tools.
L.S. deposited the nanowires.
A.B. and M.P.V. designed and fabricated the device, performed the measurements and analysed the data, with continuous feedback from L.S., L.G., J.J.W, C.K.A, A.K. and B.v.H.
R.A., J.A., B.v.H. and R.Z. provided theory support during and after the measurements, and formulated the theoretical framework to analyze the experiment.
R.Z. performed the NRG calculations.
A.B., M.P.V. and B.v.H. wrote the code to compute the circuit energy levels and extract experimental parameters.
L.P.K., R.A., A.K. and B.v.H. supervised the work.
A.B., M.P.V., R.Z. and B.v.H. wrote the manuscript with feedback from all authors.

\footnotetext[2]{See Supplemental Material at [URL], which contains further details about theoretical modeling, device fabrication, experimental setup, device tune-up,  analysis for in-field data, processing of time-domain data, as well as power, tunnel gate and temperature dependence of the parity lifetimes. It includes Refs. \cite{kadlecova2017,rok_zitko_2022_5874832, Schroer2011, Kringhoj2018, Antipov2018, Winkler2019, Kringhoj2020b, Jong2021, Spanton2017, Hart2019, Splitthoff2022, Welch1967, Wesdorp2021b}}. 
\bibliography{ms.bib}

\begin{thebibliography}{105}%
\makeatletter
\providecommand \@ifxundefined [1]{%
 \@ifx{#1\undefined}
}%
\providecommand \@ifnum [1]{%
 \ifnum #1\expandafter \@firstoftwo
 \else \expandafter \@secondoftwo
 \fi
}%
\providecommand \@ifx [1]{%
 \ifx #1\expandafter \@firstoftwo
 \else \expandafter \@secondoftwo
 \fi
}%
\providecommand \natexlab [1]{#1}%
\providecommand \enquote  [1]{``#1''}%
\providecommand \bibnamefont  [1]{#1}%
\providecommand \bibfnamefont [1]{#1}%
\providecommand \citenamefont [1]{#1}%
\providecommand \href@noop [0]{\@secondoftwo}%
\providecommand \href [0]{\begingroup \@sanitize@url \@href}%
\providecommand \@href[1]{\@@startlink{#1}\@@href}%
\providecommand \@@href[1]{\endgroup#1\@@endlink}%
\providecommand \@sanitize@url [0]{\catcode `\\12\catcode `\$12\catcode
  `\&12\catcode `\#12\catcode `\^12\catcode `\_12\catcode `\%12\relax}%
\providecommand \@@startlink[1]{}%
\providecommand \@@endlink[0]{}%
\providecommand \url  [0]{\begingroup\@sanitize@url \@url }%
\providecommand \@url [1]{\endgroup\@href {#1}{\urlprefix }}%
\providecommand \urlprefix  [0]{URL }%
\providecommand \Eprint [0]{\href }%
\providecommand \doibase [0]{https://doi.org/}%
\providecommand \selectlanguage [0]{\@gobble}%
\providecommand \bibinfo  [0]{\@secondoftwo}%
\providecommand \bibfield  [0]{\@secondoftwo}%
\providecommand \translation [1]{[#1]}%
\providecommand \BibitemOpen [0]{}%
\providecommand \bibitemStop [0]{}%
\providecommand \bibitemNoStop [0]{.\EOS\space}%
\providecommand \EOS [0]{\spacefactor3000\relax}%
\providecommand \BibitemShut  [1]{\csname bibitem#1\endcsname}%
\let\auto@bib@innerbib\@empty
\bibitem [{\citenamefont {Glazman}\ and\ \citenamefont
  {Matveev}(1989)}]{Glazman1989}%
  \BibitemOpen
  \bibfield  {author} {\bibinfo {author} {\bibfnamefont {L.~I.}\ \bibnamefont
  {Glazman}}\ and\ \bibinfo {author} {\bibfnamefont {K.~A.}\ \bibnamefont
  {Matveev}},\ }\bibfield  {title} {\bibinfo {title} {Resonant {J}osephson
  current through {K}ondo impurities in a tunnel barrier},\ }\href
  {http://jetpletters.ru/ps/1121/article_16988.shtml} {\bibfield  {journal}
  {\bibinfo  {journal} {JETP Lett.}\ }\textbf {\bibinfo {volume} {49}},\
  \bibinfo {pages} {570} (\bibinfo {year} {1989})}\BibitemShut {NoStop}%
\bibitem [{\citenamefont {Yoshioka}\ and\ \citenamefont
  {Ohashi}(2000)}]{yoshioka2000}%
  \BibitemOpen
  \bibfield  {author} {\bibinfo {author} {\bibfnamefont {T.}~\bibnamefont
  {Yoshioka}}\ and\ \bibinfo {author} {\bibfnamefont {Y.}~\bibnamefont
  {Ohashi}},\ }\bibfield  {title} {\bibinfo {title} {Numerical renormalization
  group studies on single impurity {{A}nderson} model in superconductivity: a
  unified treatment of magnetic, nonmagnetic impurities, and resonance
  scattering},\ }\href {https://doi.org/10.1143/jpsj.69.1812} {\bibfield
  {journal} {\bibinfo  {journal} {J. Phys. Soc. Japan}\ }\textbf {\bibinfo
  {volume} {69}},\ \bibinfo {pages} {1812} (\bibinfo {year}
  {2000})}\BibitemShut {NoStop}%
\bibitem [{\citenamefont {Choi}\ \emph {et~al.}(2004)\citenamefont {Choi},
  \citenamefont {Lee}, \citenamefont {Kang},\ and\ \citenamefont
  {Belzig}}]{Choi2004}%
  \BibitemOpen
  \bibfield  {author} {\bibinfo {author} {\bibfnamefont {M.-S.}\ \bibnamefont
  {Choi}}, \bibinfo {author} {\bibfnamefont {M.}~\bibnamefont {Lee}}, \bibinfo
  {author} {\bibfnamefont {K.}~\bibnamefont {Kang}},\ and\ \bibinfo {author}
  {\bibfnamefont {W.}~\bibnamefont {Belzig}},\ }\bibfield  {title} {\bibinfo
  {title} {{K}ondo effect and {J}osephson current through a quantum dot between
  two superconductors},\ }\href {https://doi.org/10.1103/PhysRevB.70.020502}
  {\bibfield  {journal} {\bibinfo  {journal} {Phys. Rev. B}\ }\textbf {\bibinfo
  {volume} {70}},\ \bibinfo {pages} {020502} (\bibinfo {year}
  {2004})}\BibitemShut {NoStop}%
\bibitem [{\citenamefont {Oguri}\ \emph {et~al.}(2004)\citenamefont {Oguri},
  \citenamefont {Tanaka},\ and\ \citenamefont {Hewson}}]{Oguri2004}%
  \BibitemOpen
  \bibfield  {author} {\bibinfo {author} {\bibfnamefont {A.}~\bibnamefont
  {Oguri}}, \bibinfo {author} {\bibfnamefont {Y.}~\bibnamefont {Tanaka}},\ and\
  \bibinfo {author} {\bibfnamefont {A.~C.}\ \bibnamefont {Hewson}},\ }\bibfield
   {title} {\bibinfo {title} {Quantum phase transition in a minimal model for
  the {K}ondo effect in a {J}osephson junction},\ }\href
  {https://doi.org/10.1143/JPSJ.73.2494} {\bibfield  {journal} {\bibinfo
  {journal} {J. Phys. Soc. Japan}\ }\textbf {\bibinfo {volume} {73}},\ \bibinfo
  {pages} {2494} (\bibinfo {year} {2004})}\BibitemShut {NoStop}%
\bibitem [{\citenamefont {Tanaka}\ \emph {et~al.}(2007)\citenamefont {Tanaka},
  \citenamefont {Oguri},\ and\ \citenamefont {Hewson}}]{Tanaka2007}%
  \BibitemOpen
  \bibfield  {author} {\bibinfo {author} {\bibfnamefont {Y.}~\bibnamefont
  {Tanaka}}, \bibinfo {author} {\bibfnamefont {A.}~\bibnamefont {Oguri}},\ and\
  \bibinfo {author} {\bibfnamefont {A.~C.}\ \bibnamefont {Hewson}},\ }\bibfield
   {title} {\bibinfo {title} {{K}ondo effect in asymmetric {J}osephson
  couplings through a quantum dot},\ }\href
  {https://doi.org/10.1088/1367-2630/10/2/029801} {\bibfield  {journal}
  {\bibinfo  {journal} {New J. Phys.}\ }\textbf {\bibinfo {volume} {9}},\
  \bibinfo {pages} {115} (\bibinfo {year} {2007})}\BibitemShut {NoStop}%
\bibitem [{\citenamefont {Mart{\'i}n-Rodero}\ and\ \citenamefont
  {Yeyati}(2011)}]{MartinRodero2011}%
  \BibitemOpen
  \bibfield  {author} {\bibinfo {author} {\bibfnamefont {A.}~\bibnamefont
  {Mart{\'i}n-Rodero}}\ and\ \bibinfo {author} {\bibfnamefont {A.~L.}\
  \bibnamefont {Yeyati}},\ }\bibfield  {title} {\bibinfo {title} {{J}osephson
  and {A}ndreev transport through quantum dots},\ }\href
  {https://doi.org/10.1080/00018732.2011.624266} {\bibfield  {journal}
  {\bibinfo  {journal} {Adv. Phys.}\ }\textbf {\bibinfo {volume} {60}},\
  \bibinfo {pages} {899} (\bibinfo {year} {2011})}\BibitemShut {NoStop}%
\bibitem [{\citenamefont {Karrasch}\ \emph {et~al.}(2008)\citenamefont
  {Karrasch}, \citenamefont {Oguri},\ and\ \citenamefont
  {Meden}}]{Karrasch2008}%
  \BibitemOpen
  \bibfield  {author} {\bibinfo {author} {\bibfnamefont {C.}~\bibnamefont
  {Karrasch}}, \bibinfo {author} {\bibfnamefont {A.}~\bibnamefont {Oguri}},\
  and\ \bibinfo {author} {\bibfnamefont {V.}~\bibnamefont {Meden}},\ }\bibfield
   {title} {\bibinfo {title} {{J}osephson current through a single {{A}nderson}
  impurity coupled to {BCS} leads},\ }\href
  {https://doi.org/10.1103/physrevb.77.024517} {\bibfield  {journal} {\bibinfo
  {journal} {Phys. Rev. B}\ }\textbf {\bibinfo {volume} {77}},\ \bibinfo
  {pages} {024517} (\bibinfo {year} {2008})}\BibitemShut {NoStop}%
\bibitem [{\citenamefont {Luitz}\ \emph {et~al.}(2012)\citenamefont {Luitz},
  \citenamefont {Assaad}, \citenamefont {Novotn\'y}, \citenamefont {Karrasch},\
  and\ \citenamefont {Meden}}]{Luitz2012}%
  \BibitemOpen
  \bibfield  {author} {\bibinfo {author} {\bibfnamefont {D.~J.}\ \bibnamefont
  {Luitz}}, \bibinfo {author} {\bibfnamefont {F.~F.}\ \bibnamefont {Assaad}},
  \bibinfo {author} {\bibfnamefont {T.}~\bibnamefont {Novotn\'y}}, \bibinfo
  {author} {\bibfnamefont {C.}~\bibnamefont {Karrasch}},\ and\ \bibinfo
  {author} {\bibfnamefont {V.}~\bibnamefont {Meden}},\ }\bibfield  {title}
  {\bibinfo {title} {Understanding the {J}osephson current through a
  {K}ondo-correlated quantum dot},\ }\href
  {https://doi.org/10.1103/PhysRevLett.108.227001} {\bibfield  {journal}
  {\bibinfo  {journal} {Phys. Rev. Lett.}\ }\textbf {\bibinfo {volume} {108}},\
  \bibinfo {pages} {227001} (\bibinfo {year} {2012})}\BibitemShut {NoStop}%
\bibitem [{\citenamefont {Kadlecov\'a}\ \emph {et~al.}(2019)\citenamefont
  {Kadlecov\'a}, \citenamefont {\ifmmode~\check{Z}\else \v{Z}\fi{}onda},
  \citenamefont {Pokorn\'y},\ and\ \citenamefont {Novotn\'y}}]{Kadlecova2019}%
  \BibitemOpen
  \bibfield  {author} {\bibinfo {author} {\bibfnamefont {A.}~\bibnamefont
  {Kadlecov\'a}}, \bibinfo {author} {\bibfnamefont {M.}~\bibnamefont
  {\ifmmode~\check{Z}\else \v{Z}\fi{}onda}}, \bibinfo {author} {\bibfnamefont
  {V.}~\bibnamefont {Pokorn\'y}},\ and\ \bibinfo {author} {\bibfnamefont
  {T.}~\bibnamefont {Novotn\'y}},\ }\bibfield  {title} {\bibinfo {title}
  {Practical guide to quantum phase transitions in quantum-dot-based tunable
  {J}osephson junctions},\ }\href
  {https://doi.org/10.1103/PhysRevApplied.11.044094} {\bibfield  {journal}
  {\bibinfo  {journal} {Phys. Rev. Applied}\ }\textbf {\bibinfo {volume}
  {11}},\ \bibinfo {pages} {044094} (\bibinfo {year} {2019})}\BibitemShut
  {NoStop}%
\bibitem [{\citenamefont {Meden}(2019)}]{Meden2019}%
  \BibitemOpen
  \bibfield  {author} {\bibinfo {author} {\bibfnamefont {V.}~\bibnamefont
  {Meden}},\ }\bibfield  {title} {\bibinfo {title} {The
  {A}nderson{\textendash}{J}osephson quantum dot{\textemdash}a theory
  perspective},\ }\href {https://doi.org/10.1088/1361-648x/aafd6a} {\bibfield
  {journal} {\bibinfo  {journal} {J. Phys. Condens. Matter}\ }\textbf {\bibinfo
  {volume} {31}},\ \bibinfo {pages} {163001} (\bibinfo {year}
  {2019})}\BibitemShut {NoStop}%
\bibitem [{\citenamefont {Pillet}\ \emph {et~al.}(2010)\citenamefont {Pillet},
  \citenamefont {Quay}, \citenamefont {Morfin}, \citenamefont {Bena},
  \citenamefont {Yeyati},\ and\ \citenamefont {Joyez}}]{Pillet2010}%
  \BibitemOpen
  \bibfield  {author} {\bibinfo {author} {\bibfnamefont {J.~D.}\ \bibnamefont
  {Pillet}}, \bibinfo {author} {\bibfnamefont {C.~H.~L.}\ \bibnamefont {Quay}},
  \bibinfo {author} {\bibfnamefont {P.}~\bibnamefont {Morfin}}, \bibinfo
  {author} {\bibfnamefont {C.}~\bibnamefont {Bena}}, \bibinfo {author}
  {\bibfnamefont {A.~L.}\ \bibnamefont {Yeyati}},\ and\ \bibinfo {author}
  {\bibfnamefont {P.}~\bibnamefont {Joyez}},\ }\bibfield  {title} {\bibinfo
  {title} {{{A}ndreev bound states in supercurrent-carrying carbon nanotubes
  revealed}},\ }\href {https://doi.org/10.1038/nphys1811} {\bibfield  {journal}
  {\bibinfo  {journal} {Nat. Phys.}\ }\textbf {\bibinfo {volume} {6}},\
  \bibinfo {pages} {965} (\bibinfo {year} {2010})}\BibitemShut {NoStop}%
\bibitem [{\citenamefont {Pillet}\ \emph {et~al.}(2013)\citenamefont {Pillet},
  \citenamefont {Joyez}, \citenamefont {\v{Z}itko},\ and\ \citenamefont
  {Goffman}}]{pillet2013}%
  \BibitemOpen
  \bibfield  {author} {\bibinfo {author} {\bibfnamefont {J.-D.}\ \bibnamefont
  {Pillet}}, \bibinfo {author} {\bibfnamefont {P.}~\bibnamefont {Joyez}},
  \bibinfo {author} {\bibfnamefont {R.}~\bibnamefont {\v{Z}itko}},\ and\
  \bibinfo {author} {\bibfnamefont {M.~F.}\ \bibnamefont {Goffman}},\
  }\bibfield  {title} {\bibinfo {title} {Tunneling spectroscopy of a single
  quantum dot coupled to a superconductor: From {{K}ondo} ridge to {{A}ndreev}
  bound states},\ }\href {https://doi.org/10.1103/physrevb.88.045101}
  {\bibfield  {journal} {\bibinfo  {journal} {Phys. Rev. B}\ }\textbf {\bibinfo
  {volume} {88}},\ \bibinfo {pages} {045101} (\bibinfo {year}
  {2013})}\BibitemShut {NoStop}%
\bibitem [{\citenamefont {Deacon}\ \emph {et~al.}(2010)\citenamefont {Deacon},
  \citenamefont {Tanaka}, \citenamefont {Oiwa}, \citenamefont {Sakano},
  \citenamefont {Yoshida}, \citenamefont {Shibata}, \citenamefont {Hirakawa},\
  and\ \citenamefont {Tarucha}}]{Deacon2010}%
  \BibitemOpen
  \bibfield  {author} {\bibinfo {author} {\bibfnamefont {R.~S.}\ \bibnamefont
  {Deacon}}, \bibinfo {author} {\bibfnamefont {Y.}~\bibnamefont {Tanaka}},
  \bibinfo {author} {\bibfnamefont {A.}~\bibnamefont {Oiwa}}, \bibinfo {author}
  {\bibfnamefont {R.}~\bibnamefont {Sakano}}, \bibinfo {author} {\bibfnamefont
  {K.}~\bibnamefont {Yoshida}}, \bibinfo {author} {\bibfnamefont
  {K.}~\bibnamefont {Shibata}}, \bibinfo {author} {\bibfnamefont
  {K.}~\bibnamefont {Hirakawa}},\ and\ \bibinfo {author} {\bibfnamefont
  {S.}~\bibnamefont {Tarucha}},\ }\bibfield  {title} {\bibinfo {title}
  {{Tunneling spectroscopy of {A}ndreev energy levels in a quantum dot coupled
  to a superconductor}},\ }\href
  {https://doi.org/10.1103/PhysRevLett.104.076805} {\bibfield  {journal}
  {\bibinfo  {journal} {Phys. Rev. Lett.}\ }\textbf {\bibinfo {volume} {104}},\
  \bibinfo {pages} {076805} (\bibinfo {year} {2010})}\BibitemShut {NoStop}%
\bibitem [{\citenamefont {Lee}\ \emph {et~al.}(2014)\citenamefont {Lee},
  \citenamefont {Jiang}, \citenamefont {Houzet}, \citenamefont {Aguado},
  \citenamefont {Lieber},\ and\ \citenamefont {Franceschi}}]{Lee2014}%
  \BibitemOpen
  \bibfield  {author} {\bibinfo {author} {\bibfnamefont {E.~J.~H.}\
  \bibnamefont {Lee}}, \bibinfo {author} {\bibfnamefont {X.}~\bibnamefont
  {Jiang}}, \bibinfo {author} {\bibfnamefont {M.}~\bibnamefont {Houzet}},
  \bibinfo {author} {\bibfnamefont {R.}~\bibnamefont {Aguado}}, \bibinfo
  {author} {\bibfnamefont {C.~M.}\ \bibnamefont {Lieber}},\ and\ \bibinfo
  {author} {\bibfnamefont {S.~D.}\ \bibnamefont {Franceschi}},\ }\bibfield
  {title} {\bibinfo {title} {{Spin-resolved {A}ndreev levels and parity
  crossings in hybrid superconductor-semiconductor nanostructures}},\ }\href
  {https://doi.org/10.1038/nnano.2013.267} {\bibfield  {journal} {\bibinfo
  {journal} {Nat. Nanotechnol.}\ }\textbf {\bibinfo {volume} {9}},\ \bibinfo
  {pages} {79} (\bibinfo {year} {2014})}\BibitemShut {NoStop}%
\bibitem [{\citenamefont {Chang}\ \emph {et~al.}(2013)\citenamefont {Chang},
  \citenamefont {Manucharyan}, \citenamefont {Jespersen}, \citenamefont
  {Nyg{\aa}rd},\ and\ \citenamefont {Marcus}}]{Chang2013}%
  \BibitemOpen
  \bibfield  {author} {\bibinfo {author} {\bibfnamefont {W.}~\bibnamefont
  {Chang}}, \bibinfo {author} {\bibfnamefont {V.~E.}\ \bibnamefont
  {Manucharyan}}, \bibinfo {author} {\bibfnamefont {T.~S.}\ \bibnamefont
  {Jespersen}}, \bibinfo {author} {\bibfnamefont {J.}~\bibnamefont
  {Nyg{\aa}rd}},\ and\ \bibinfo {author} {\bibfnamefont {C.~M.}\ \bibnamefont
  {Marcus}},\ }\bibfield  {title} {\bibinfo {title} {{Tunneling spectroscopy of
  quasiparticle bound states in a spinful {J}osephson junction}},\ }\href
  {https://doi.org/10.1103/PhysRevLett.110.217005} {\bibfield  {journal}
  {\bibinfo  {journal} {Phys. Rev. Lett.}\ }\textbf {\bibinfo {volume} {110}},\
  \bibinfo {pages} {217005} (\bibinfo {year} {2013})}\BibitemShut {NoStop}%
\bibitem [{\citenamefont {Li}\ \emph {et~al.}(2017)\citenamefont {Li},
  \citenamefont {Kang}, \citenamefont {Caroff},\ and\ \citenamefont
  {Xu}}]{Li2017}%
  \BibitemOpen
  \bibfield  {author} {\bibinfo {author} {\bibfnamefont {S.}~\bibnamefont
  {Li}}, \bibinfo {author} {\bibfnamefont {N.}~\bibnamefont {Kang}}, \bibinfo
  {author} {\bibfnamefont {P.}~\bibnamefont {Caroff}},\ and\ \bibinfo {author}
  {\bibfnamefont {H.~Q.}\ \bibnamefont {Xu}},\ }\bibfield  {title} {\bibinfo
  {title} {$0\text{\ensuremath{-}}\ensuremath{\pi}$ phase transition in hybrid
  superconductor--{I}n{S}b nanowire quantum dot devices},\ }\href
  {https://doi.org/10.1103/PhysRevB.95.014515} {\bibfield  {journal} {\bibinfo
  {journal} {Phys. Rev. B}\ }\textbf {\bibinfo {volume} {95}},\ \bibinfo
  {pages} {014515} (\bibinfo {year} {2017})}\BibitemShut {NoStop}%
\bibitem [{\citenamefont {Lee}\ \emph {et~al.}(2017)\citenamefont {Lee},
  \citenamefont {Jiang}, \citenamefont {{\v{Z}}itko}, \citenamefont {Aguado},
  \citenamefont {Lieber},\ and\ \citenamefont {Franceschi}}]{Lee2017}%
  \BibitemOpen
  \bibfield  {author} {\bibinfo {author} {\bibfnamefont {E.~J.~H.}\
  \bibnamefont {Lee}}, \bibinfo {author} {\bibfnamefont {X.}~\bibnamefont
  {Jiang}}, \bibinfo {author} {\bibfnamefont {R.}~\bibnamefont {{\v{Z}}itko}},
  \bibinfo {author} {\bibfnamefont {R.}~\bibnamefont {Aguado}}, \bibinfo
  {author} {\bibfnamefont {C.~M.}\ \bibnamefont {Lieber}},\ and\ \bibinfo
  {author} {\bibfnamefont {S.~D.}\ \bibnamefont {Franceschi}},\ }\bibfield
  {title} {\bibinfo {title} {{Scaling of subgap excitations in a
  superconductor-semiconductor nanowire quantum dot}},\ }\href
  {https://doi.org/10.1103/PhysRevB.95.180502} {\bibfield  {journal} {\bibinfo
  {journal} {Phys. Rev. B}\ }\textbf {\bibinfo {volume} {95}},\ \bibinfo
  {pages} {180502} (\bibinfo {year} {2017})}\BibitemShut {NoStop}%
\bibitem [{\citenamefont {Valentini}\ \emph {et~al.}(2021)\citenamefont
  {Valentini}, \citenamefont {Pe{\~n}aranda}, \citenamefont {Hofmann},
  \citenamefont {Brauns}, \citenamefont {Hauschild}, \citenamefont {Krogstrup},
  \citenamefont {San-Jose}, \citenamefont {Prada}, \citenamefont {Aguado},\
  and\ \citenamefont {Katsaros}}]{Valentini2021}%
  \BibitemOpen
  \bibfield  {author} {\bibinfo {author} {\bibfnamefont {M.}~\bibnamefont
  {Valentini}}, \bibinfo {author} {\bibfnamefont {F.}~\bibnamefont
  {Pe{\~n}aranda}}, \bibinfo {author} {\bibfnamefont {A.}~\bibnamefont
  {Hofmann}}, \bibinfo {author} {\bibfnamefont {M.}~\bibnamefont {Brauns}},
  \bibinfo {author} {\bibfnamefont {R.}~\bibnamefont {Hauschild}}, \bibinfo
  {author} {\bibfnamefont {P.}~\bibnamefont {Krogstrup}}, \bibinfo {author}
  {\bibfnamefont {P.}~\bibnamefont {San-Jose}}, \bibinfo {author}
  {\bibfnamefont {E.}~\bibnamefont {Prada}}, \bibinfo {author} {\bibfnamefont
  {R.}~\bibnamefont {Aguado}},\ and\ \bibinfo {author} {\bibfnamefont
  {G.}~\bibnamefont {Katsaros}},\ }\bibfield  {title} {\bibinfo {title}
  {Nontopological zero-bias peaks in full-shell nanowires induced by
  flux-tunable {A}ndreev states},\ }\href
  {https://doi.org/10.1126/science.abf1513} {\bibfield  {journal} {\bibinfo
  {journal} {Science}\ }\textbf {\bibinfo {volume} {373}},\ \bibinfo {pages}
  {82} (\bibinfo {year} {2021})}\BibitemShut {NoStop}%
\bibitem [{\citenamefont {Whiticar}\ \emph {et~al.}(2021)\citenamefont
  {Whiticar}, \citenamefont {Fornieri}, \citenamefont {Banerjee}, \citenamefont
  {Drachmann}, \citenamefont {Gronin}, \citenamefont {Gardner}, \citenamefont
  {Lindemann}, \citenamefont {Manfra},\ and\ \citenamefont
  {Marcus}}]{Whiticar2021}%
  \BibitemOpen
  \bibfield  {author} {\bibinfo {author} {\bibfnamefont {A.~M.}\ \bibnamefont
  {Whiticar}}, \bibinfo {author} {\bibfnamefont {A.}~\bibnamefont {Fornieri}},
  \bibinfo {author} {\bibfnamefont {A.}~\bibnamefont {Banerjee}}, \bibinfo
  {author} {\bibfnamefont {A.~C.~C.}\ \bibnamefont {Drachmann}}, \bibinfo
  {author} {\bibfnamefont {S.}~\bibnamefont {Gronin}}, \bibinfo {author}
  {\bibfnamefont {G.~C.}\ \bibnamefont {Gardner}}, \bibinfo {author}
  {\bibfnamefont {T.}~\bibnamefont {Lindemann}}, \bibinfo {author}
  {\bibfnamefont {M.~J.}\ \bibnamefont {Manfra}},\ and\ \bibinfo {author}
  {\bibfnamefont {C.~M.}\ \bibnamefont {Marcus}},\ }\bibfield  {title}
  {\bibinfo {title} {{Zeeman-driven parity transitions in an {A}ndreev quantum
  dot}},\ }\href@noop {} {\bibfield  {journal} {\bibinfo  {journal} {arXiv
  e-prints}\ } (\bibinfo {year} {2021})},\ \Eprint
  {https://arxiv.org/abs/2101.09706} {arXiv:2101.09706} \BibitemShut {NoStop}%
\bibitem [{\citenamefont {van Dam}\ \emph {et~al.}(2006)\citenamefont {van
  Dam}, \citenamefont {Nazarov}, \citenamefont {Bakkers}, \citenamefont
  {Franceschi},\ and\ \citenamefont {Kouwenhoven}}]{VanDam2006a}%
  \BibitemOpen
  \bibfield  {author} {\bibinfo {author} {\bibfnamefont {J.~A.}\ \bibnamefont
  {van Dam}}, \bibinfo {author} {\bibfnamefont {Y.~V.}\ \bibnamefont
  {Nazarov}}, \bibinfo {author} {\bibfnamefont {E.~P. A.~M.}\ \bibnamefont
  {Bakkers}}, \bibinfo {author} {\bibfnamefont {S.~D.}\ \bibnamefont
  {Franceschi}},\ and\ \bibinfo {author} {\bibfnamefont {L.~P.}\ \bibnamefont
  {Kouwenhoven}},\ }\bibfield  {title} {\bibinfo {title} {{Supercurrent
  reversal in quantum dots}},\ }\href {https://doi.org/10.1038/nature05018}
  {\bibfield  {journal} {\bibinfo  {journal} {Nature}\ }\textbf {\bibinfo
  {volume} {442}},\ \bibinfo {pages} {667} (\bibinfo {year}
  {2006})}\BibitemShut {NoStop}%
\bibitem [{\citenamefont {Cleuziou}\ \emph {et~al.}(2006)\citenamefont
  {Cleuziou}, \citenamefont {Wernsdorfer}, \citenamefont {Bouchiat},
  \citenamefont {Ondarcuhu},\ and\ \citenamefont {Monthioux}}]{Cleuziou2006}%
  \BibitemOpen
  \bibfield  {author} {\bibinfo {author} {\bibfnamefont {J.~P.}\ \bibnamefont
  {Cleuziou}}, \bibinfo {author} {\bibfnamefont {W.}~\bibnamefont
  {Wernsdorfer}}, \bibinfo {author} {\bibfnamefont {V.}~\bibnamefont
  {Bouchiat}}, \bibinfo {author} {\bibfnamefont {T.}~\bibnamefont
  {Ondarcuhu}},\ and\ \bibinfo {author} {\bibfnamefont {M.}~\bibnamefont
  {Monthioux}},\ }\bibfield  {title} {\bibinfo {title} {{Carbon nanotube
  superconducting quantum interference device}},\ }\href
  {https://doi.org/10.1038/nnano.2006.54} {\bibfield  {journal} {\bibinfo
  {journal} {Nat. Nanotechnol.}\ }\textbf {\bibinfo {volume} {1}},\ \bibinfo
  {pages} {53} (\bibinfo {year} {2006})}\BibitemShut {NoStop}%
\bibitem [{\citenamefont {J{\o}rgensen}\ \emph {et~al.}(2007)\citenamefont
  {J{\o}rgensen}, \citenamefont {Novotn{\'{y}}}, \citenamefont
  {Grove-Rasmussen}, \citenamefont {Flensberg},\ and\ \citenamefont
  {Lindelof}}]{Jorgensen2007}%
  \BibitemOpen
  \bibfield  {author} {\bibinfo {author} {\bibfnamefont {H.~I.}\ \bibnamefont
  {J{\o}rgensen}}, \bibinfo {author} {\bibfnamefont {T.}~\bibnamefont
  {Novotn{\'{y}}}}, \bibinfo {author} {\bibfnamefont {K.}~\bibnamefont
  {Grove-Rasmussen}}, \bibinfo {author} {\bibfnamefont {K.}~\bibnamefont
  {Flensberg}},\ and\ \bibinfo {author} {\bibfnamefont {P.~E.}\ \bibnamefont
  {Lindelof}},\ }\bibfield  {title} {\bibinfo {title} {{Critical current
  0-$\pi$ transition in designed {J}osephson quantum dot junctions}},\ }\href
  {https://doi.org/10.1021/nl071152w} {\bibfield  {journal} {\bibinfo
  {journal} {Nano Lett.}\ }\textbf {\bibinfo {volume} {7}},\ \bibinfo {pages}
  {2441} (\bibinfo {year} {2007})}\BibitemShut {NoStop}%
\bibitem [{\citenamefont {J\o{}rgensen}\ \emph {et~al.}(2009)\citenamefont
  {J\o{}rgensen}, \citenamefont {Grove-Rasmussen}, \citenamefont {Flensberg},\
  and\ \citenamefont {Lindelof}}]{Jorgenson2009}%
  \BibitemOpen
  \bibfield  {author} {\bibinfo {author} {\bibfnamefont {H.~I.}\ \bibnamefont
  {J\o{}rgensen}}, \bibinfo {author} {\bibfnamefont {K.}~\bibnamefont
  {Grove-Rasmussen}}, \bibinfo {author} {\bibfnamefont {K.}~\bibnamefont
  {Flensberg}},\ and\ \bibinfo {author} {\bibfnamefont {P.~E.}\ \bibnamefont
  {Lindelof}},\ }\bibfield  {title} {\bibinfo {title} {Critical and excess
  current through an open quantum dot: Temperature and magnetic-field
  dependence},\ }\href {https://doi.org/10.1103/PhysRevB.79.155441} {\bibfield
  {journal} {\bibinfo  {journal} {Phys. Rev. B}\ }\textbf {\bibinfo {volume}
  {79}},\ \bibinfo {pages} {155441} (\bibinfo {year} {2009})}\BibitemShut
  {NoStop}%
\bibitem [{\citenamefont {Eichler}\ \emph {et~al.}(2009)\citenamefont
  {Eichler}, \citenamefont {Deblock}, \citenamefont {Weiss}, \citenamefont
  {Karrasch}, \citenamefont {Meden}, \citenamefont {Sch\"onenberger},\ and\
  \citenamefont {Bouchiat}}]{Eichler2009}%
  \BibitemOpen
  \bibfield  {author} {\bibinfo {author} {\bibfnamefont {A.}~\bibnamefont
  {Eichler}}, \bibinfo {author} {\bibfnamefont {R.}~\bibnamefont {Deblock}},
  \bibinfo {author} {\bibfnamefont {M.}~\bibnamefont {Weiss}}, \bibinfo
  {author} {\bibfnamefont {C.}~\bibnamefont {Karrasch}}, \bibinfo {author}
  {\bibfnamefont {V.}~\bibnamefont {Meden}}, \bibinfo {author} {\bibfnamefont
  {C.}~\bibnamefont {Sch\"onenberger}},\ and\ \bibinfo {author} {\bibfnamefont
  {H.}~\bibnamefont {Bouchiat}},\ }\bibfield  {title} {\bibinfo {title} {Tuning
  the {J}osephson current in carbon nanotubes with the {K}ondo effect},\ }\href
  {https://doi.org/10.1103/PhysRevB.79.161407} {\bibfield  {journal} {\bibinfo
  {journal} {Phys. Rev. B}\ }\textbf {\bibinfo {volume} {79}},\ \bibinfo
  {pages} {161407} (\bibinfo {year} {2009})}\BibitemShut {NoStop}%
\bibitem [{\citenamefont {Lee}\ \emph {et~al.}(2012)\citenamefont {Lee},
  \citenamefont {Jiang}, \citenamefont {Aguado}, \citenamefont {Katsaros},
  \citenamefont {Lieber},\ and\ \citenamefont {De~Franceschi}}]{Lee2012}%
  \BibitemOpen
  \bibfield  {author} {\bibinfo {author} {\bibfnamefont {E.~J.~H.}\
  \bibnamefont {Lee}}, \bibinfo {author} {\bibfnamefont {X.}~\bibnamefont
  {Jiang}}, \bibinfo {author} {\bibfnamefont {R.}~\bibnamefont {Aguado}},
  \bibinfo {author} {\bibfnamefont {G.}~\bibnamefont {Katsaros}}, \bibinfo
  {author} {\bibfnamefont {C.~M.}\ \bibnamefont {Lieber}},\ and\ \bibinfo
  {author} {\bibfnamefont {S.}~\bibnamefont {De~Franceschi}},\ }\bibfield
  {title} {\bibinfo {title} {Zero-bias anomaly in a nanowire quantum dot
  coupled to superconductors},\ }\href
  {https://doi.org/10.1103/PhysRevLett.109.186802} {\bibfield  {journal}
  {\bibinfo  {journal} {Phys. Rev. Lett.}\ }\textbf {\bibinfo {volume} {109}},\
  \bibinfo {pages} {186802} (\bibinfo {year} {2012})}\BibitemShut {NoStop}%
\bibitem [{\citenamefont {Maurand}\ \emph {et~al.}(2012)\citenamefont
  {Maurand}, \citenamefont {Meng}, \citenamefont {Bonet}, \citenamefont
  {Florens}, \citenamefont {Marty},\ and\ \citenamefont
  {Wernsdorfer}}]{Maurand2012}%
  \BibitemOpen
  \bibfield  {author} {\bibinfo {author} {\bibfnamefont {R.}~\bibnamefont
  {Maurand}}, \bibinfo {author} {\bibfnamefont {T.}~\bibnamefont {Meng}},
  \bibinfo {author} {\bibfnamefont {E.}~\bibnamefont {Bonet}}, \bibinfo
  {author} {\bibfnamefont {S.}~\bibnamefont {Florens}}, \bibinfo {author}
  {\bibfnamefont {L.}~\bibnamefont {Marty}},\ and\ \bibinfo {author}
  {\bibfnamefont {W.}~\bibnamefont {Wernsdorfer}},\ }\bibfield  {title}
  {\bibinfo {title} {First-order
  $0\mathrm{\text{\ensuremath{-}}}\ensuremath{\pi}$ quantum phase transition in
  the {K}ondo regime of a superconducting carbon-nanotube quantum dot},\ }\href
  {https://doi.org/10.1103/PhysRevX.2.011009} {\bibfield  {journal} {\bibinfo
  {journal} {Phys. Rev. X}\ }\textbf {\bibinfo {volume} {2}},\ \bibinfo {pages}
  {011009} (\bibinfo {year} {2012})}\BibitemShut {NoStop}%
\bibitem [{\citenamefont {Kumar}\ \emph {et~al.}(2014)\citenamefont {Kumar},
  \citenamefont {Gaim}, \citenamefont {Steininger}, \citenamefont {Yeyati},
  \citenamefont {Mart\'{\i}n-Rodero}, \citenamefont {H\"uttel},\ and\
  \citenamefont {Strunk}}]{Kumar2014}%
  \BibitemOpen
  \bibfield  {author} {\bibinfo {author} {\bibfnamefont {A.}~\bibnamefont
  {Kumar}}, \bibinfo {author} {\bibfnamefont {M.}~\bibnamefont {Gaim}},
  \bibinfo {author} {\bibfnamefont {D.}~\bibnamefont {Steininger}}, \bibinfo
  {author} {\bibfnamefont {A.~L.}\ \bibnamefont {Yeyati}}, \bibinfo {author}
  {\bibfnamefont {A.}~\bibnamefont {Mart\'{\i}n-Rodero}}, \bibinfo {author}
  {\bibfnamefont {A.~K.}\ \bibnamefont {H\"uttel}},\ and\ \bibinfo {author}
  {\bibfnamefont {C.}~\bibnamefont {Strunk}},\ }\bibfield  {title} {\bibinfo
  {title} {Temperature dependence of {A}ndreev spectra in a superconducting
  carbon nanotube quantum dot},\ }\href
  {https://doi.org/10.1103/PhysRevB.89.075428} {\bibfield  {journal} {\bibinfo
  {journal} {Phys. Rev. B}\ }\textbf {\bibinfo {volume} {89}},\ \bibinfo
  {pages} {075428} (\bibinfo {year} {2014})}\BibitemShut {NoStop}%
\bibitem [{\citenamefont {Szombati}\ \emph {et~al.}(2016)\citenamefont
  {Szombati}, \citenamefont {Nadj-Perge}, \citenamefont {Car}, \citenamefont
  {Plissard}, \citenamefont {Bakkers},\ and\ \citenamefont
  {Kouwenhoven}}]{Szombati2016}%
  \BibitemOpen
  \bibfield  {author} {\bibinfo {author} {\bibfnamefont {D.~B.}\ \bibnamefont
  {Szombati}}, \bibinfo {author} {\bibfnamefont {S.}~\bibnamefont
  {Nadj-Perge}}, \bibinfo {author} {\bibfnamefont {D.}~\bibnamefont {Car}},
  \bibinfo {author} {\bibfnamefont {S.~R.}\ \bibnamefont {Plissard}}, \bibinfo
  {author} {\bibfnamefont {E.~P. A.~M.}\ \bibnamefont {Bakkers}},\ and\
  \bibinfo {author} {\bibfnamefont {L.~P.}\ \bibnamefont {Kouwenhoven}},\
  }\bibfield  {title} {\bibinfo {title} {{J}osephson $\phi_0$-junction in
  nanowire quantum dots},\ }\href {https://doi.org/10.1038/nphys3742}
  {\bibfield  {journal} {\bibinfo  {journal} {Nat. Phys.}\ }\textbf {\bibinfo
  {volume} {12}},\ \bibinfo {pages} {568} (\bibinfo {year} {2016})}\BibitemShut
  {NoStop}%
\bibitem [{\citenamefont {Delagrange}\ \emph {et~al.}(2015)\citenamefont
  {Delagrange}, \citenamefont {Luitz}, \citenamefont {Weil}, \citenamefont
  {Kasumov}, \citenamefont {Meden}, \citenamefont {Bouchiat},\ and\
  \citenamefont {Deblock}}]{Delagrange2015}%
  \BibitemOpen
  \bibfield  {author} {\bibinfo {author} {\bibfnamefont {R.}~\bibnamefont
  {Delagrange}}, \bibinfo {author} {\bibfnamefont {D.~J.}\ \bibnamefont
  {Luitz}}, \bibinfo {author} {\bibfnamefont {R.}~\bibnamefont {Weil}},
  \bibinfo {author} {\bibfnamefont {A.}~\bibnamefont {Kasumov}}, \bibinfo
  {author} {\bibfnamefont {V.}~\bibnamefont {Meden}}, \bibinfo {author}
  {\bibfnamefont {H.}~\bibnamefont {Bouchiat}},\ and\ \bibinfo {author}
  {\bibfnamefont {R.}~\bibnamefont {Deblock}},\ }\bibfield  {title} {\bibinfo
  {title} {{Manipulating the magnetic state of a carbon nanotube {J}osephson
  junction using the superconducting phase}},\ }\href
  {https://doi.org/10.1103/physrevb.91.241401} {\bibfield  {journal} {\bibinfo
  {journal} {Phys. Rev. B}\ }\textbf {\bibinfo {volume} {91}},\ \bibinfo
  {pages} {241401} (\bibinfo {year} {2015})}\BibitemShut {NoStop}%
\bibitem [{\citenamefont {Delagrange}\ \emph {et~al.}(2018)\citenamefont
  {Delagrange}, \citenamefont {Weil}, \citenamefont {Kasumov}, \citenamefont
  {Ferrier}, \citenamefont {Bouchiat},\ and\ \citenamefont
  {Deblock}}]{Delagrange2018}%
  \BibitemOpen
  \bibfield  {author} {\bibinfo {author} {\bibfnamefont {R.}~\bibnamefont
  {Delagrange}}, \bibinfo {author} {\bibfnamefont {R.}~\bibnamefont {Weil}},
  \bibinfo {author} {\bibfnamefont {A.}~\bibnamefont {Kasumov}}, \bibinfo
  {author} {\bibfnamefont {M.}~\bibnamefont {Ferrier}}, \bibinfo {author}
  {\bibfnamefont {H.}~\bibnamefont {Bouchiat}},\ and\ \bibinfo {author}
  {\bibfnamefont {R.}~\bibnamefont {Deblock}},\ }\bibfield  {title} {\bibinfo
  {title} {{$0\mathrm{\text{\ensuremath{-}}}\ensuremath{\pi}$ quantum
  transition in a carbon nanotube {J}osephson junction: Universal phase
  dependence and orbital degeneracy}},\ }\href
  {https://doi.org/10.1016/j.physb.2017.09.034} {\bibfield  {journal} {\bibinfo
   {journal} {Phys. Rev. B}\ }\textbf {\bibinfo {volume} {536}},\ \bibinfo
  {pages} {211} (\bibinfo {year} {2018})}\BibitemShut {NoStop}%
\bibitem [{\citenamefont {Garc\'{\i}a~Corral}\ \emph
  {et~al.}(2020)\citenamefont {Garc\'{\i}a~Corral}, \citenamefont {van Zanten},
  \citenamefont {Franke}, \citenamefont {Courtois}, \citenamefont {Florens},\
  and\ \citenamefont {Winkelmann}}]{Corral2020}%
  \BibitemOpen
  \bibfield  {author} {\bibinfo {author} {\bibfnamefont {A.}~\bibnamefont
  {Garc\'{\i}a~Corral}}, \bibinfo {author} {\bibfnamefont {D.~M.~T.}\
  \bibnamefont {van Zanten}}, \bibinfo {author} {\bibfnamefont {K.~J.}\
  \bibnamefont {Franke}}, \bibinfo {author} {\bibfnamefont {H.}~\bibnamefont
  {Courtois}}, \bibinfo {author} {\bibfnamefont {S.}~\bibnamefont {Florens}},\
  and\ \bibinfo {author} {\bibfnamefont {C.~B.}\ \bibnamefont {Winkelmann}},\
  }\bibfield  {title} {\bibinfo {title} {Magnetic-field-induced transition in a
  quantum dot coupled to a superconductor},\ }\href
  {https://doi.org/10.1103/PhysRevResearch.2.012065} {\bibfield  {journal}
  {\bibinfo  {journal} {Phys. Rev. Research}\ }\textbf {\bibinfo {volume}
  {2}},\ \bibinfo {pages} {012065} (\bibinfo {year} {2020})}\BibitemShut
  {NoStop}%
\bibitem [{\citenamefont {Janvier}\ \emph {et~al.}(2015)\citenamefont
  {Janvier}, \citenamefont {Tosi}, \citenamefont {Bretheau}, \citenamefont
  {Girit}, \citenamefont {Stern}, \citenamefont {Bertet}, \citenamefont
  {Joyez}, \citenamefont {Vion}, \citenamefont {Esteve}, \citenamefont
  {Goffman}, \citenamefont {Pothier},\ and\ \citenamefont
  {Urbina}}]{Janvier2015}%
  \BibitemOpen
  \bibfield  {author} {\bibinfo {author} {\bibfnamefont {C.}~\bibnamefont
  {Janvier}}, \bibinfo {author} {\bibfnamefont {L.}~\bibnamefont {Tosi}},
  \bibinfo {author} {\bibfnamefont {L.}~\bibnamefont {Bretheau}}, \bibinfo
  {author} {\bibfnamefont {{\c{C}}.}~\bibnamefont {Girit}}, \bibinfo {author}
  {\bibfnamefont {M.}~\bibnamefont {Stern}}, \bibinfo {author} {\bibfnamefont
  {P.}~\bibnamefont {Bertet}}, \bibinfo {author} {\bibfnamefont
  {P.}~\bibnamefont {Joyez}}, \bibinfo {author} {\bibfnamefont
  {D.}~\bibnamefont {Vion}}, \bibinfo {author} {\bibfnamefont {D.}~\bibnamefont
  {Esteve}}, \bibinfo {author} {\bibfnamefont {M.~F.}\ \bibnamefont {Goffman}},
  \bibinfo {author} {\bibfnamefont {H.}~\bibnamefont {Pothier}},\ and\ \bibinfo
  {author} {\bibfnamefont {C.}~\bibnamefont {Urbina}},\ }\bibfield  {title}
  {\bibinfo {title} {Coherent manipulation of {A}ndreev states in
  superconducting atomic contacts},\ }\href
  {https://doi.org/10.1126/science.aab2179} {\bibfield  {journal} {\bibinfo
  {journal} {Science}\ }\textbf {\bibinfo {volume} {349}},\ \bibinfo {pages}
  {1199} (\bibinfo {year} {2015})}\BibitemShut {NoStop}%
\bibitem [{\citenamefont {Hays}\ \emph {et~al.}(2020)\citenamefont {Hays},
  \citenamefont {Fatemi}, \citenamefont {Serniak}, \citenamefont {Bouman},
  \citenamefont {Diamond}, \citenamefont {de~Lange}, \citenamefont {Krogstrup},
  \citenamefont {Nyg{\aa}rd}, \citenamefont {Geresdi},\ and\ \citenamefont
  {Devoret}}]{Hays2020}%
  \BibitemOpen
  \bibfield  {author} {\bibinfo {author} {\bibfnamefont {M.}~\bibnamefont
  {Hays}}, \bibinfo {author} {\bibfnamefont {V.}~\bibnamefont {Fatemi}},
  \bibinfo {author} {\bibfnamefont {K.}~\bibnamefont {Serniak}}, \bibinfo
  {author} {\bibfnamefont {D.}~\bibnamefont {Bouman}}, \bibinfo {author}
  {\bibfnamefont {S.}~\bibnamefont {Diamond}}, \bibinfo {author} {\bibfnamefont
  {G.}~\bibnamefont {de~Lange}}, \bibinfo {author} {\bibfnamefont
  {P.}~\bibnamefont {Krogstrup}}, \bibinfo {author} {\bibfnamefont
  {J.}~\bibnamefont {Nyg{\aa}rd}}, \bibinfo {author} {\bibfnamefont
  {A.}~\bibnamefont {Geresdi}},\ and\ \bibinfo {author} {\bibfnamefont {M.~H.}\
  \bibnamefont {Devoret}},\ }\bibfield  {title} {\bibinfo {title} {Continuous
  monitoring of a trapped superconducting spin},\ }\href
  {https://doi.org/10.1038/s41567-020-0952-3} {\bibfield  {journal} {\bibinfo
  {journal} {Nat. Phys.}\ }\textbf {\bibinfo {volume} {16}},\ \bibinfo {pages}
  {1103} (\bibinfo {year} {2020})}\BibitemShut {NoStop}%
\bibitem [{\citenamefont {Hays}\ \emph {et~al.}(2021)\citenamefont {Hays},
  \citenamefont {Fatemi}, \citenamefont {Bouman}, \citenamefont {Cerrillo},
  \citenamefont {Diamond}, \citenamefont {Serniak}, \citenamefont {Connolly},
  \citenamefont {Krogstrup}, \citenamefont {Nyg{\aa}rd}, \citenamefont
  {Yeyati}, \citenamefont {Geresdi},\ and\ \citenamefont {Devoret}}]{Hays2021}%
  \BibitemOpen
  \bibfield  {author} {\bibinfo {author} {\bibfnamefont {M.}~\bibnamefont
  {Hays}}, \bibinfo {author} {\bibfnamefont {V.}~\bibnamefont {Fatemi}},
  \bibinfo {author} {\bibfnamefont {D.}~\bibnamefont {Bouman}}, \bibinfo
  {author} {\bibfnamefont {J.}~\bibnamefont {Cerrillo}}, \bibinfo {author}
  {\bibfnamefont {S.}~\bibnamefont {Diamond}}, \bibinfo {author} {\bibfnamefont
  {K.}~\bibnamefont {Serniak}}, \bibinfo {author} {\bibfnamefont
  {T.}~\bibnamefont {Connolly}}, \bibinfo {author} {\bibfnamefont
  {P.}~\bibnamefont {Krogstrup}}, \bibinfo {author} {\bibfnamefont
  {J.}~\bibnamefont {Nyg{\aa}rd}}, \bibinfo {author} {\bibfnamefont {A.~L.}\
  \bibnamefont {Yeyati}}, \bibinfo {author} {\bibfnamefont {A.}~\bibnamefont
  {Geresdi}},\ and\ \bibinfo {author} {\bibfnamefont {M.~H.}\ \bibnamefont
  {Devoret}},\ }\bibfield  {title} {\bibinfo {title} {Coherent manipulation of
  an {A}ndreev spin qubit},\ }\href {https://doi.org/10.1126/science.abf0345}
  {\bibfield  {journal} {\bibinfo  {journal} {Science}\ }\textbf {\bibinfo
  {volume} {373}},\ \bibinfo {pages} {430} (\bibinfo {year}
  {2021})}\BibitemShut {NoStop}%
\bibitem [{\citenamefont {Zazunov}\ \emph {et~al.}(2003)\citenamefont
  {Zazunov}, \citenamefont {Shumeiko}, \citenamefont {Bratus'}, \citenamefont
  {Lantz},\ and\ \citenamefont {Wendin}}]{Zazunov2003a}%
  \BibitemOpen
  \bibfield  {author} {\bibinfo {author} {\bibfnamefont {A.}~\bibnamefont
  {Zazunov}}, \bibinfo {author} {\bibfnamefont {V.~S.}\ \bibnamefont
  {Shumeiko}}, \bibinfo {author} {\bibfnamefont {E.~N.}\ \bibnamefont
  {Bratus'}}, \bibinfo {author} {\bibfnamefont {J.}~\bibnamefont {Lantz}},\
  and\ \bibinfo {author} {\bibfnamefont {G.}~\bibnamefont {Wendin}},\
  }\bibfield  {title} {\bibinfo {title} {{{A}ndreev Level Qubit}},\ }\href
  {https://doi.org/10.1103/PhysRevLett.90.087003} {\bibfield  {journal}
  {\bibinfo  {journal} {Phys. Rev. Lett.}\ }\textbf {\bibinfo {volume} {90}},\
  \bibinfo {pages} {4} (\bibinfo {year} {2003})}\BibitemShut {NoStop}%
\bibitem [{\citenamefont {Chtchelkatchev}\ and\ \citenamefont
  {Nazarov}(2003)}]{Chtchelkatchev2003}%
  \BibitemOpen
  \bibfield  {author} {\bibinfo {author} {\bibfnamefont {N.~M.}\ \bibnamefont
  {Chtchelkatchev}}\ and\ \bibinfo {author} {\bibfnamefont {Y.~V.}\
  \bibnamefont {Nazarov}},\ }\bibfield  {title} {\bibinfo {title} {{A}ndreev
  quantum dots for spin manipulation},\ }\href
  {https://doi.org/10.1103/PhysRevLett.90.226806} {\bibfield  {journal}
  {\bibinfo  {journal} {Phys. Rev. Lett.}\ }\textbf {\bibinfo {volume} {90}},\
  \bibinfo {pages} {226806} (\bibinfo {year} {2003})}\BibitemShut {NoStop}%
\bibitem [{\citenamefont {Padurariu}\ and\ \citenamefont
  {Nazarov}(2010)}]{Padurariu2010}%
  \BibitemOpen
  \bibfield  {author} {\bibinfo {author} {\bibfnamefont {C.}~\bibnamefont
  {Padurariu}}\ and\ \bibinfo {author} {\bibfnamefont {Y.~V.}\ \bibnamefont
  {Nazarov}},\ }\bibfield  {title} {\bibinfo {title} {Theoretical proposal for
  superconducting spin qubits},\ }\href
  {https://doi.org/10.1103/PhysRevB.81.144519} {\bibfield  {journal} {\bibinfo
  {journal} {Phys. Rev. B}\ }\textbf {\bibinfo {volume} {81}},\ \bibinfo
  {pages} {144519} (\bibinfo {year} {2010})}\BibitemShut {NoStop}%
\bibitem [{\citenamefont {Pave\ifmmode \check{s}\else
  \v{s}\fi{}i\ifmmode~\acute{c}\else \'{c}\fi{}}\ and\ \citenamefont
  {\ifmmode~\check{Z}\else \v{Z}\fi{}itko}(2022)}]{pavesic2022}%
  \BibitemOpen
  \bibfield  {author} {\bibinfo {author} {\bibfnamefont {L.}~\bibnamefont
  {Pave\ifmmode \check{s}\else \v{s}\fi{}i\ifmmode~\acute{c}\else \'{c}\fi{}}}\
  and\ \bibinfo {author} {\bibfnamefont {R.}~\bibnamefont
  {\ifmmode~\check{Z}\else \v{Z}\fi{}itko}},\ }\bibfield  {title} {\bibinfo
  {title} {Qubit based on spin-singlet {Yu-Shiba-Rusinov} states},\ }\href
  {https://doi.org/10.1103/PhysRevB.105.075129} {\bibfield  {journal} {\bibinfo
   {journal} {Phys. Rev. B}\ }\textbf {\bibinfo {volume} {105}},\ \bibinfo
  {pages} {075129} (\bibinfo {year} {2022})}\BibitemShut {NoStop}%
\bibitem [{\citenamefont {Plugge}\ \emph {et~al.}(2017)\citenamefont {Plugge},
  \citenamefont {Rasmussen}, \citenamefont {Egger},\ and\ \citenamefont
  {Flensberg}}]{Plugge_2017}%
  \BibitemOpen
  \bibfield  {author} {\bibinfo {author} {\bibfnamefont {S.}~\bibnamefont
  {Plugge}}, \bibinfo {author} {\bibfnamefont {A.}~\bibnamefont {Rasmussen}},
  \bibinfo {author} {\bibfnamefont {R.}~\bibnamefont {Egger}},\ and\ \bibinfo
  {author} {\bibfnamefont {K.}~\bibnamefont {Flensberg}},\ }\bibfield  {title}
  {\bibinfo {title} {{M}ajorana box qubits},\ }\href
  {https://doi.org/10.1088/1367-2630/aa54e1} {\bibfield  {journal} {\bibinfo
  {journal} {New J. Phys.}\ }\textbf {\bibinfo {volume} {19}},\ \bibinfo
  {pages} {012001} (\bibinfo {year} {2017})}\BibitemShut {NoStop}%
\bibitem [{\citenamefont {Karzig}\ \emph {et~al.}(2017)\citenamefont {Karzig},
  \citenamefont {Knapp}, \citenamefont {Lutchyn}, \citenamefont {Bonderson},
  \citenamefont {Hastings}, \citenamefont {Nayak}, \citenamefont {Alicea},
  \citenamefont {Flensberg}, \citenamefont {Plugge}, \citenamefont {Oreg},
  \citenamefont {Marcus},\ and\ \citenamefont {Freedman}}]{Karzig2017}%
  \BibitemOpen
  \bibfield  {author} {\bibinfo {author} {\bibfnamefont {T.}~\bibnamefont
  {Karzig}}, \bibinfo {author} {\bibfnamefont {C.}~\bibnamefont {Knapp}},
  \bibinfo {author} {\bibfnamefont {R.~M.}\ \bibnamefont {Lutchyn}}, \bibinfo
  {author} {\bibfnamefont {P.}~\bibnamefont {Bonderson}}, \bibinfo {author}
  {\bibfnamefont {M.~B.}\ \bibnamefont {Hastings}}, \bibinfo {author}
  {\bibfnamefont {C.}~\bibnamefont {Nayak}}, \bibinfo {author} {\bibfnamefont
  {J.}~\bibnamefont {Alicea}}, \bibinfo {author} {\bibfnamefont
  {K.}~\bibnamefont {Flensberg}}, \bibinfo {author} {\bibfnamefont
  {S.}~\bibnamefont {Plugge}}, \bibinfo {author} {\bibfnamefont
  {Y.}~\bibnamefont {Oreg}}, \bibinfo {author} {\bibfnamefont {C.~M.}\
  \bibnamefont {Marcus}},\ and\ \bibinfo {author} {\bibfnamefont {M.~H.}\
  \bibnamefont {Freedman}},\ }\bibfield  {title} {\bibinfo {title} {Scalable
  designs for quasiparticle-poisoning-protected topological quantum computation
  with {M}ajorana zero modes},\ }\href
  {https://doi.org/10.1103/PHYSREVB.95.235305/FIGURES/17/MEDIUM} {\bibfield
  {journal} {\bibinfo  {journal} {Phys. Rev. B}\ }\textbf {\bibinfo {volume}
  {95}},\ \bibinfo {pages} {235305} (\bibinfo {year} {2017})}\BibitemShut
  {NoStop}%
\bibitem [{\citenamefont {Smith}\ \emph {et~al.}(2020)\citenamefont {Smith},
  \citenamefont {Cassidy}, \citenamefont {Reilly}, \citenamefont {Bartlett},\
  and\ \citenamefont {Grimsmo}}]{Smith2020}%
  \BibitemOpen
  \bibfield  {author} {\bibinfo {author} {\bibfnamefont {T.~B.}\ \bibnamefont
  {Smith}}, \bibinfo {author} {\bibfnamefont {M.~C.}\ \bibnamefont {Cassidy}},
  \bibinfo {author} {\bibfnamefont {D.~J.}\ \bibnamefont {Reilly}}, \bibinfo
  {author} {\bibfnamefont {S.~D.}\ \bibnamefont {Bartlett}},\ and\ \bibinfo
  {author} {\bibfnamefont {A.~L.}\ \bibnamefont {Grimsmo}},\ }\bibfield
  {title} {\bibinfo {title} {Dispersive readout of {M}ajorana qubits},\ }\href
  {https://doi.org/10.1103/PRXQuantum.1.020313} {\bibfield  {journal} {\bibinfo
   {journal} {PRX Quantum}\ }\textbf {\bibinfo {volume} {1}},\ \bibinfo {pages}
  {020313} (\bibinfo {year} {2020})}\BibitemShut {NoStop}%
\bibitem [{\citenamefont {de~Lange}\ \emph {et~al.}(2015)\citenamefont
  {de~Lange}, \citenamefont {van Heck}, \citenamefont {Bruno}, \citenamefont
  {van Woerkom}, \citenamefont {Geresdi}, \citenamefont {Plissard},
  \citenamefont {Bakkers}, \citenamefont {Akhmerov},\ and\ \citenamefont
  {DiCarlo}}]{deLange2015}%
  \BibitemOpen
  \bibfield  {author} {\bibinfo {author} {\bibfnamefont {G.}~\bibnamefont
  {de~Lange}}, \bibinfo {author} {\bibfnamefont {B.}~\bibnamefont {van Heck}},
  \bibinfo {author} {\bibfnamefont {A.}~\bibnamefont {Bruno}}, \bibinfo
  {author} {\bibfnamefont {D.~J.}\ \bibnamefont {van Woerkom}}, \bibinfo
  {author} {\bibfnamefont {A.}~\bibnamefont {Geresdi}}, \bibinfo {author}
  {\bibfnamefont {S.~R.}\ \bibnamefont {Plissard}}, \bibinfo {author}
  {\bibfnamefont {E.~P. A.~M.}\ \bibnamefont {Bakkers}}, \bibinfo {author}
  {\bibfnamefont {A.~R.}\ \bibnamefont {Akhmerov}},\ and\ \bibinfo {author}
  {\bibfnamefont {L.}~\bibnamefont {DiCarlo}},\ }\bibfield  {title} {\bibinfo
  {title} {Realization of microwave quantum circuits using hybrid
  superconducting-semiconducting nanowire {J}osephson elements},\ }\href
  {https://doi.org/10.1103/PhysRevLett.115.127002} {\bibfield  {journal}
  {\bibinfo  {journal} {Phys. Rev. Lett.}\ }\textbf {\bibinfo {volume} {115}},\
  \bibinfo {pages} {127002} (\bibinfo {year} {2015})}\BibitemShut {NoStop}%
\bibitem [{\citenamefont {Larsen}\ \emph {et~al.}(2015)\citenamefont {Larsen},
  \citenamefont {Petersson}, \citenamefont {Kuemmeth}, \citenamefont
  {Jespersen}, \citenamefont {Krogstrup}, \citenamefont {Nyg\aa{}rd},\ and\
  \citenamefont {Marcus}}]{Larsen2015}%
  \BibitemOpen
  \bibfield  {author} {\bibinfo {author} {\bibfnamefont {T.~W.}\ \bibnamefont
  {Larsen}}, \bibinfo {author} {\bibfnamefont {K.~D.}\ \bibnamefont
  {Petersson}}, \bibinfo {author} {\bibfnamefont {F.}~\bibnamefont {Kuemmeth}},
  \bibinfo {author} {\bibfnamefont {T.~S.}\ \bibnamefont {Jespersen}}, \bibinfo
  {author} {\bibfnamefont {P.}~\bibnamefont {Krogstrup}}, \bibinfo {author}
  {\bibfnamefont {J.}~\bibnamefont {Nyg\aa{}rd}},\ and\ \bibinfo {author}
  {\bibfnamefont {C.~M.}\ \bibnamefont {Marcus}},\ }\bibfield  {title}
  {\bibinfo {title} {Semiconductor-nanowire-based superconducting qubit},\
  }\href {https://doi.org/10.1103/PhysRevLett.115.127001} {\bibfield  {journal}
  {\bibinfo  {journal} {Phys. Rev. Lett.}\ }\textbf {\bibinfo {volume} {115}},\
  \bibinfo {pages} {127001} (\bibinfo {year} {2015})}\BibitemShut {NoStop}%
\bibitem [{\citenamefont {Tosi}\ \emph {et~al.}(2019)\citenamefont {Tosi},
  \citenamefont {Metzger}, \citenamefont {Goffman}, \citenamefont {Urbina},
  \citenamefont {Pothier}, \citenamefont {Park}, \citenamefont {Yeyati},
  \citenamefont {Nyg{\aa}rd},\ and\ \citenamefont {Krogstrup}}]{Tosi2018}%
  \BibitemOpen
  \bibfield  {author} {\bibinfo {author} {\bibfnamefont {L.}~\bibnamefont
  {Tosi}}, \bibinfo {author} {\bibfnamefont {C.}~\bibnamefont {Metzger}},
  \bibinfo {author} {\bibfnamefont {M.}~\bibnamefont {Goffman}}, \bibinfo
  {author} {\bibfnamefont {C.}~\bibnamefont {Urbina}}, \bibinfo {author}
  {\bibfnamefont {H.}~\bibnamefont {Pothier}}, \bibinfo {author} {\bibfnamefont
  {S.}~\bibnamefont {Park}}, \bibinfo {author} {\bibfnamefont {A.~L.}\
  \bibnamefont {Yeyati}}, \bibinfo {author} {\bibfnamefont {J.}~\bibnamefont
  {Nyg{\aa}rd}},\ and\ \bibinfo {author} {\bibfnamefont {P.}~\bibnamefont
  {Krogstrup}},\ }\bibfield  {title} {\bibinfo {title} {{Spin-Orbit splitting
  of {A}ndreev states revealed by microwave spectroscopy}},\ }\href
  {https://doi.org/10.1103/physrevx.9.011010} {\bibfield  {journal} {\bibinfo
  {journal} {Phys. Rev. X}\ }\textbf {\bibinfo {volume} {9}},\ \bibinfo {pages}
  {011010} (\bibinfo {year} {2019})}\BibitemShut {NoStop}%
\bibitem [{\citenamefont {Hays}\ \emph {et~al.}(2018)\citenamefont {Hays},
  \citenamefont {de~Lange}, \citenamefont {Serniak}, \citenamefont {van
  Woerkom}, \citenamefont {Bouman}, \citenamefont {Krogstrup}, \citenamefont
  {Nyg\aa{}rd}, \citenamefont {Geresdi},\ and\ \citenamefont
  {Devoret}}]{Hays2018}%
  \BibitemOpen
  \bibfield  {author} {\bibinfo {author} {\bibfnamefont {M.}~\bibnamefont
  {Hays}}, \bibinfo {author} {\bibfnamefont {G.}~\bibnamefont {de~Lange}},
  \bibinfo {author} {\bibfnamefont {K.}~\bibnamefont {Serniak}}, \bibinfo
  {author} {\bibfnamefont {D.~J.}\ \bibnamefont {van Woerkom}}, \bibinfo
  {author} {\bibfnamefont {D.}~\bibnamefont {Bouman}}, \bibinfo {author}
  {\bibfnamefont {P.}~\bibnamefont {Krogstrup}}, \bibinfo {author}
  {\bibfnamefont {J.}~\bibnamefont {Nyg\aa{}rd}}, \bibinfo {author}
  {\bibfnamefont {A.}~\bibnamefont {Geresdi}},\ and\ \bibinfo {author}
  {\bibfnamefont {M.~H.}\ \bibnamefont {Devoret}},\ }\bibfield  {title}
  {\bibinfo {title} {Direct microwave measurement of {A}ndreev-bound-state
  dynamics in a semiconductor-nanowire {J}osephson junction},\ }\href
  {https://doi.org/10.1103/PhysRevLett.121.047001} {\bibfield  {journal}
  {\bibinfo  {journal} {Phys. Rev. Lett.}\ }\textbf {\bibinfo {volume} {121}},\
  \bibinfo {pages} {047001} (\bibinfo {year} {2018})}\BibitemShut {NoStop}%
\bibitem [{\citenamefont {Bargerbos}\ \emph {et~al.}(2020)\citenamefont
  {Bargerbos}, \citenamefont {Uilhoorn}, \citenamefont {Yang}, \citenamefont
  {Krogstrup}, \citenamefont {Kouwenhoven}, \citenamefont {de~Lange},
  \citenamefont {van Heck},\ and\ \citenamefont {Kou}}]{Bargerbos2020}%
  \BibitemOpen
  \bibfield  {author} {\bibinfo {author} {\bibfnamefont {A.}~\bibnamefont
  {Bargerbos}}, \bibinfo {author} {\bibfnamefont {W.}~\bibnamefont {Uilhoorn}},
  \bibinfo {author} {\bibfnamefont {C.-K.}\ \bibnamefont {Yang}}, \bibinfo
  {author} {\bibfnamefont {P.}~\bibnamefont {Krogstrup}}, \bibinfo {author}
  {\bibfnamefont {L.~P.}\ \bibnamefont {Kouwenhoven}}, \bibinfo {author}
  {\bibfnamefont {G.}~\bibnamefont {de~Lange}}, \bibinfo {author}
  {\bibfnamefont {B.}~\bibnamefont {van Heck}},\ and\ \bibinfo {author}
  {\bibfnamefont {A.}~\bibnamefont {Kou}},\ }\bibfield  {title} {\bibinfo
  {title} {Observation of vanishing charge dispersion of a nearly open
  superconducting island},\ }\href
  {https://doi.org/10.1103/PhysRevLett.124.246802} {\bibfield  {journal}
  {\bibinfo  {journal} {Phys. Rev. Lett.}\ }\textbf {\bibinfo {volume} {124}},\
  \bibinfo {pages} {246802} (\bibinfo {year} {2020})}\BibitemShut {NoStop}%
\bibitem [{\citenamefont {Kringh\o{}j}\ \emph {et~al.}(2020)\citenamefont
  {Kringh\o{}j}, \citenamefont {van Heck}, \citenamefont {Larsen},
  \citenamefont {Erlandsson}, \citenamefont {Sabonis}, \citenamefont
  {Krogstrup}, \citenamefont {Casparis}, \citenamefont {Petersson},\ and\
  \citenamefont {Marcus}}]{Kringhoj2020}%
  \BibitemOpen
  \bibfield  {author} {\bibinfo {author} {\bibfnamefont {A.}~\bibnamefont
  {Kringh\o{}j}}, \bibinfo {author} {\bibfnamefont {B.}~\bibnamefont {van
  Heck}}, \bibinfo {author} {\bibfnamefont {T.~W.}\ \bibnamefont {Larsen}},
  \bibinfo {author} {\bibfnamefont {O.}~\bibnamefont {Erlandsson}}, \bibinfo
  {author} {\bibfnamefont {D.}~\bibnamefont {Sabonis}}, \bibinfo {author}
  {\bibfnamefont {P.}~\bibnamefont {Krogstrup}}, \bibinfo {author}
  {\bibfnamefont {L.}~\bibnamefont {Casparis}}, \bibinfo {author}
  {\bibfnamefont {K.~D.}\ \bibnamefont {Petersson}},\ and\ \bibinfo {author}
  {\bibfnamefont {C.~M.}\ \bibnamefont {Marcus}},\ }\bibfield  {title}
  {\bibinfo {title} {Suppressed charge dispersion via resonant tunneling in a
  single-channel transmon},\ }\href
  {https://doi.org/10.1103/PhysRevLett.124.246803} {\bibfield  {journal}
  {\bibinfo  {journal} {Phys. Rev. Lett.}\ }\textbf {\bibinfo {volume} {124}},\
  \bibinfo {pages} {246803} (\bibinfo {year} {2020})}\BibitemShut {NoStop}%
\bibitem [{\citenamefont {Uilhoorn}\ \emph {et~al.}(2021)\citenamefont
  {Uilhoorn}, \citenamefont {Kroll}, \citenamefont {Bargerbos}, \citenamefont
  {Nabi}, \citenamefont {Yang}, \citenamefont {Krogstrup}, \citenamefont
  {Kouwenhoven}, \citenamefont {Kou},\ and\ \citenamefont
  {de~Lange}}]{Uilhoorn2021}%
  \BibitemOpen
  \bibfield  {author} {\bibinfo {author} {\bibfnamefont {W.}~\bibnamefont
  {Uilhoorn}}, \bibinfo {author} {\bibfnamefont {J.~G.}\ \bibnamefont {Kroll}},
  \bibinfo {author} {\bibfnamefont {A.}~\bibnamefont {Bargerbos}}, \bibinfo
  {author} {\bibfnamefont {S.~D.}\ \bibnamefont {Nabi}}, \bibinfo {author}
  {\bibfnamefont {C.-K.}\ \bibnamefont {Yang}}, \bibinfo {author}
  {\bibfnamefont {P.}~\bibnamefont {Krogstrup}}, \bibinfo {author}
  {\bibfnamefont {L.~P.}\ \bibnamefont {Kouwenhoven}}, \bibinfo {author}
  {\bibfnamefont {A.}~\bibnamefont {Kou}},\ and\ \bibinfo {author}
  {\bibfnamefont {G.}~\bibnamefont {de~Lange}},\ }\bibfield  {title} {\bibinfo
  {title} {Quasiparticle trapping by orbital effect in a hybrid
  superconducting-semiconducting circuit},\ }\href@noop {} {\bibfield
  {journal} {\bibinfo  {journal} {arXiv e-prints}\ } (\bibinfo {year}
  {2021})},\ \Eprint {https://arxiv.org/abs/2105.11038} {arXiv:2105.11038}
  \BibitemShut {NoStop}%
\bibitem [{\citenamefont {Cañadas}\ \emph {et~al.}(2021)\citenamefont
  {Cañadas}, \citenamefont {Metzger}, \citenamefont {Park}, \citenamefont
  {Tosi}, \citenamefont {Krogstrup}, \citenamefont {Nygård}, \citenamefont
  {Goffman}, \citenamefont {Urbina}, \citenamefont {Pothier},\ and\
  \citenamefont {Yeyati}}]{canadas2021}%
  \BibitemOpen
  \bibfield  {author} {\bibinfo {author} {\bibfnamefont {F.~J.~M.}\
  \bibnamefont {Cañadas}}, \bibinfo {author} {\bibfnamefont {C.}~\bibnamefont
  {Metzger}}, \bibinfo {author} {\bibfnamefont {S.}~\bibnamefont {Park}},
  \bibinfo {author} {\bibfnamefont {L.}~\bibnamefont {Tosi}}, \bibinfo {author}
  {\bibfnamefont {P.}~\bibnamefont {Krogstrup}}, \bibinfo {author}
  {\bibfnamefont {J.}~\bibnamefont {Nygård}}, \bibinfo {author} {\bibfnamefont
  {M.~F.}\ \bibnamefont {Goffman}}, \bibinfo {author} {\bibfnamefont
  {C.}~\bibnamefont {Urbina}}, \bibinfo {author} {\bibfnamefont
  {H.}~\bibnamefont {Pothier}},\ and\ \bibinfo {author} {\bibfnamefont {A.~L.}\
  \bibnamefont {Yeyati}},\ }\href@noop {} {\bibinfo {title} {Signatures of
  interactions in the {Andreev} spectrum of nanowire {Josephson} junctions}}
  (\bibinfo {year} {2021}),\ \Eprint {https://arxiv.org/abs/2112.05625}
  {arXiv:2112.05625} \BibitemShut {NoStop}%
\bibitem [{\citenamefont {Kurilovich}\ \emph {et~al.}(2021)\citenamefont
  {Kurilovich}, \citenamefont {Kurilovich}, \citenamefont {Fatemi},
  \citenamefont {Devoret},\ and\ \citenamefont {Glazman}}]{Kurilovich2021}%
  \BibitemOpen
  \bibfield  {author} {\bibinfo {author} {\bibfnamefont {P.~D.}\ \bibnamefont
  {Kurilovich}}, \bibinfo {author} {\bibfnamefont {V.~D.}\ \bibnamefont
  {Kurilovich}}, \bibinfo {author} {\bibfnamefont {V.}~\bibnamefont {Fatemi}},
  \bibinfo {author} {\bibfnamefont {M.~H.}\ \bibnamefont {Devoret}},\ and\
  \bibinfo {author} {\bibfnamefont {L.~I.}\ \bibnamefont {Glazman}},\
  }\bibfield  {title} {\bibinfo {title} {Microwave response of an {Andreev}
  bound state},\ }\href {https://doi.org/10.1103/PhysRevB.104.174517}
  {\bibfield  {journal} {\bibinfo  {journal} {Phys. Rev. B}\ }\textbf {\bibinfo
  {volume} {104}},\ \bibinfo {pages} {174517} (\bibinfo {year}
  {2021})}\BibitemShut {NoStop}%
\bibitem [{\citenamefont {Hermansen}\ \emph {et~al.}(2022)\citenamefont
  {Hermansen}, \citenamefont {Yeyati},\ and\ \citenamefont
  {Paaske}}]{Hermansen2022}%
  \BibitemOpen
  \bibfield  {author} {\bibinfo {author} {\bibfnamefont {C.}~\bibnamefont
  {Hermansen}}, \bibinfo {author} {\bibfnamefont {A.~L.}\ \bibnamefont
  {Yeyati}},\ and\ \bibinfo {author} {\bibfnamefont {J.}~\bibnamefont
  {Paaske}},\ }\bibfield  {title} {\bibinfo {title} {Inductive microwave
  response of {Yu-Shiba-Rusinov} states},\ }\href
  {https://doi.org/10.1103/PhysRevB.105.054503} {\bibfield  {journal} {\bibinfo
   {journal} {Phys. Rev. B}\ }\textbf {\bibinfo {volume} {105}},\ \bibinfo
  {pages} {054503} (\bibinfo {year} {2022})}\BibitemShut {NoStop}%
\bibitem [{\citenamefont {Fatemi}\ \emph {et~al.}(2021)\citenamefont {Fatemi},
  \citenamefont {Kurilovich}, \citenamefont {Hays}, \citenamefont {Bouman},
  \citenamefont {Connolly}, \citenamefont {Diamond}, \citenamefont {Frattini},
  \citenamefont {Kurilovich}, \citenamefont {Krogstrup}, \citenamefont
  {Nygard}, \citenamefont {Geresdi}, \citenamefont {Glazman},\ and\
  \citenamefont {Devoret}}]{Fatemi2021}%
  \BibitemOpen
  \bibfield  {author} {\bibinfo {author} {\bibfnamefont {V.}~\bibnamefont
  {Fatemi}}, \bibinfo {author} {\bibfnamefont {P.~D.}\ \bibnamefont
  {Kurilovich}}, \bibinfo {author} {\bibfnamefont {M.}~\bibnamefont {Hays}},
  \bibinfo {author} {\bibfnamefont {D.}~\bibnamefont {Bouman}}, \bibinfo
  {author} {\bibfnamefont {T.}~\bibnamefont {Connolly}}, \bibinfo {author}
  {\bibfnamefont {S.}~\bibnamefont {Diamond}}, \bibinfo {author} {\bibfnamefont
  {N.~E.}\ \bibnamefont {Frattini}}, \bibinfo {author} {\bibfnamefont {V.~D.}\
  \bibnamefont {Kurilovich}}, \bibinfo {author} {\bibfnamefont
  {P.}~\bibnamefont {Krogstrup}}, \bibinfo {author} {\bibfnamefont
  {J.}~\bibnamefont {Nygard}}, \bibinfo {author} {\bibfnamefont
  {A.}~\bibnamefont {Geresdi}}, \bibinfo {author} {\bibfnamefont {L.~I.}\
  \bibnamefont {Glazman}},\ and\ \bibinfo {author} {\bibfnamefont {M.~H.}\
  \bibnamefont {Devoret}},\ }\bibfield  {title} {\bibinfo {title} {Microwave
  susceptibility observation of interacting many-body {A}ndreev states},\
  }\href@noop {} {\bibfield  {journal} {\bibinfo  {journal} {arXiv e-prints}\ }
  (\bibinfo {year} {2021})},\ \Eprint {https://arxiv.org/abs/2112.05624}
  {arXiv:2112.05624} \BibitemShut {NoStop}%
\bibitem [{\citenamefont {Spivak}\ and\ \citenamefont
  {Kivelson}(1991)}]{spivak1991}%
  \BibitemOpen
  \bibfield  {author} {\bibinfo {author} {\bibfnamefont {B.~I.}\ \bibnamefont
  {Spivak}}\ and\ \bibinfo {author} {\bibfnamefont {S.~A.}\ \bibnamefont
  {Kivelson}},\ }\bibfield  {title} {\bibinfo {title} {Negative local
  superfluid densities: The difference between dirty superconductors and dirty
  {Bose} liquids},\ }\href {https://doi.org/10.1103/PhysRevB.43.3740}
  {\bibfield  {journal} {\bibinfo  {journal} {Phys. Rev. B}\ }\textbf {\bibinfo
  {volume} {43}},\ \bibinfo {pages} {3740} (\bibinfo {year}
  {1991})}\BibitemShut {NoStop}%
\bibitem [{\citenamefont {Blais}\ \emph {et~al.}(2004)\citenamefont {Blais},
  \citenamefont {Huang}, \citenamefont {Wallraff}, \citenamefont {Girvin},\
  and\ \citenamefont {Schoelkopf}}]{Blais2004}%
  \BibitemOpen
  \bibfield  {author} {\bibinfo {author} {\bibfnamefont {A.}~\bibnamefont
  {Blais}}, \bibinfo {author} {\bibfnamefont {R.-S.}\ \bibnamefont {Huang}},
  \bibinfo {author} {\bibfnamefont {A.}~\bibnamefont {Wallraff}}, \bibinfo
  {author} {\bibfnamefont {S.~M.}\ \bibnamefont {Girvin}},\ and\ \bibinfo
  {author} {\bibfnamefont {R.~J.}\ \bibnamefont {Schoelkopf}},\ }\bibfield
  {title} {\bibinfo {title} {Cavity quantum electrodynamics for superconducting
  electrical circuits: An architecture for quantum computation},\ }\href
  {https://doi.org/10.1103/PhysRevA.69.062320} {\bibfield  {journal} {\bibinfo
  {journal} {Phys. Rev. A}\ }\textbf {\bibinfo {volume} {69}},\ \bibinfo
  {pages} {062320} (\bibinfo {year} {2004})}\BibitemShut {NoStop}%
\bibitem [{\citenamefont {Wilson}(1975)}]{wilson1975}%
  \BibitemOpen
  \bibfield  {author} {\bibinfo {author} {\bibfnamefont {K.~G.}\ \bibnamefont
  {Wilson}},\ }\bibfield  {title} {\bibinfo {title} {The renormalization group:
  {Critical} phenomena and the {{K}ondo} problem},\ }\href
  {https://doi.org/10.1103/revmodphys.47.773} {\bibfield  {journal} {\bibinfo
  {journal} {Rev. Mod. Phys.}\ }\textbf {\bibinfo {volume} {47}},\ \bibinfo
  {pages} {773} (\bibinfo {year} {1975})}\BibitemShut {NoStop}%
\bibitem [{\citenamefont {Satori}\ \emph {et~al.}(1992)\citenamefont {Satori},
  \citenamefont {Shiba}, \citenamefont {Sakai},\ and\ \citenamefont
  {Shimizu}}]{satori1992}%
  \BibitemOpen
  \bibfield  {author} {\bibinfo {author} {\bibfnamefont {K.}~\bibnamefont
  {Satori}}, \bibinfo {author} {\bibfnamefont {H.}~\bibnamefont {Shiba}},
  \bibinfo {author} {\bibfnamefont {O.}~\bibnamefont {Sakai}},\ and\ \bibinfo
  {author} {\bibfnamefont {Y.}~\bibnamefont {Shimizu}},\ }\bibfield  {title}
  {\bibinfo {title} {Numerical renormalization group study of magnetic
  impurities in superconductors},\ }\href
  {https://doi.org/10.1143/jpsj.61.3239} {\bibfield  {journal} {\bibinfo
  {journal} {J. Phys. Soc. Japan}\ }\textbf {\bibinfo {volume} {61}},\ \bibinfo
  {pages} {3239} (\bibinfo {year} {1992})}\BibitemShut {NoStop}%
\bibitem [{\citenamefont {Bulla}\ \emph {et~al.}(2008)\citenamefont {Bulla},
  \citenamefont {Costi},\ and\ \citenamefont {Pruschke}}]{bulla2008}%
  \BibitemOpen
  \bibfield  {author} {\bibinfo {author} {\bibfnamefont {R.}~\bibnamefont
  {Bulla}}, \bibinfo {author} {\bibfnamefont {T.~A.}\ \bibnamefont {Costi}},\
  and\ \bibinfo {author} {\bibfnamefont {T.}~\bibnamefont {Pruschke}},\
  }\bibfield  {title} {\bibinfo {title} {The numerical renormalization group
  method for quantum impurity systems},\ }\href
  {https://doi.org/10.1103/revmodphys.80.395} {\bibfield  {journal} {\bibinfo
  {journal} {Rev. Mod. Phys.}\ }\textbf {\bibinfo {volume} {80}},\ \bibinfo
  {pages} {395} (\bibinfo {year} {2008})}\BibitemShut {NoStop}%
\bibitem [{\citenamefont {Krogstrup}\ \emph {et~al.}(2015)\citenamefont
  {Krogstrup}, \citenamefont {Ziino}, \citenamefont {Chang}, \citenamefont
  {Albrecht}, \citenamefont {Madsen}, \citenamefont {Johnson}, \citenamefont
  {Nyg{\aa}rd}, \citenamefont {Marcus},\ and\ \citenamefont
  {Jespersen}}]{Krogstrup2015}%
  \BibitemOpen
  \bibfield  {author} {\bibinfo {author} {\bibfnamefont {P.}~\bibnamefont
  {Krogstrup}}, \bibinfo {author} {\bibfnamefont {N.~L.~B.}\ \bibnamefont
  {Ziino}}, \bibinfo {author} {\bibfnamefont {W.}~\bibnamefont {Chang}},
  \bibinfo {author} {\bibfnamefont {S.~M.}\ \bibnamefont {Albrecht}}, \bibinfo
  {author} {\bibfnamefont {M.~H.}\ \bibnamefont {Madsen}}, \bibinfo {author}
  {\bibfnamefont {E.}~\bibnamefont {Johnson}}, \bibinfo {author} {\bibfnamefont
  {J.}~\bibnamefont {Nyg{\aa}rd}}, \bibinfo {author} {\bibfnamefont
  {C.}~\bibnamefont {Marcus}},\ and\ \bibinfo {author} {\bibfnamefont {T.~S.}\
  \bibnamefont {Jespersen}},\ }\bibfield  {title} {\bibinfo {title} {Epitaxy of
  semiconductor-superconductor nanowires},\ }\href
  {https://doi.org/10.1038/nmat4176} {\bibfield  {journal} {\bibinfo  {journal}
  {Nat. Mater.}\ }\textbf {\bibinfo {volume} {14}},\ \bibinfo {pages} {400}
  (\bibinfo {year} {2015})}\BibitemShut {NoStop}%
\bibitem [{\citenamefont {Koch}\ \emph {et~al.}(2007)\citenamefont {Koch},
  \citenamefont {Yu}, \citenamefont {Gambetta}, \citenamefont {Houck},
  \citenamefont {Schuster}, \citenamefont {Majer}, \citenamefont {Blais},
  \citenamefont {Devoret}, \citenamefont {Girvin},\ and\ \citenamefont
  {Schoelkopf}}]{Koch2007}%
  \BibitemOpen
  \bibfield  {author} {\bibinfo {author} {\bibfnamefont {J.}~\bibnamefont
  {Koch}}, \bibinfo {author} {\bibfnamefont {T.~M.}\ \bibnamefont {Yu}},
  \bibinfo {author} {\bibfnamefont {J.}~\bibnamefont {Gambetta}}, \bibinfo
  {author} {\bibfnamefont {A.~A.}\ \bibnamefont {Houck}}, \bibinfo {author}
  {\bibfnamefont {D.~I.}\ \bibnamefont {Schuster}}, \bibinfo {author}
  {\bibfnamefont {J.}~\bibnamefont {Majer}}, \bibinfo {author} {\bibfnamefont
  {A.}~\bibnamefont {Blais}}, \bibinfo {author} {\bibfnamefont {M.~H.}\
  \bibnamefont {Devoret}}, \bibinfo {author} {\bibfnamefont {S.~M.}\
  \bibnamefont {Girvin}},\ and\ \bibinfo {author} {\bibfnamefont {R.~J.}\
  \bibnamefont {Schoelkopf}},\ }\bibfield  {title} {\bibinfo {title}
  {Charge-insensitive qubit design derived from the cooper pair box},\ }\href
  {https://doi.org/10.1103/physreva.76.042319} {\bibfield  {journal} {\bibinfo
  {journal} {Phys. Rev. A}\ }\textbf {\bibinfo {volume} {76}},\ \bibinfo
  {pages} {042319} (\bibinfo {year} {2007})}\BibitemShut {NoStop}%
\bibitem [{Note2()}]{Note2}%
  \BibitemOpen
  \bibinfo {note} {See Supplemental Material at [URL], which contains further
  details about theoretical modeling, device fabrication, experimental setup,
  device tune-up, analysis for in-field data, processing of time-domain data,
  as well as power, tunnel gate and temperature dependence of the parity
  lifetimes. It includes Refs. \cite {kadlecova2017,rok_zitko_2022_5874832,
  Schroer2011, Kringhoj2018, Antipov2018, Winkler2019, Kringhoj2020b, Jong2021,
  Spanton2017, Hart2019, Splitthoff2022, Welch1967, Wesdorp2021b}}\BibitemShut
  {NoStop}%
\bibitem [{\citenamefont {Della~Rocca}\ \emph {et~al.}(2007)\citenamefont
  {Della~Rocca}, \citenamefont {Chauvin}, \citenamefont {Huard}, \citenamefont
  {Pothier}, \citenamefont {Esteve},\ and\ \citenamefont {Urbina}}]{Della2007}%
  \BibitemOpen
  \bibfield  {author} {\bibinfo {author} {\bibfnamefont {M.~L.}\ \bibnamefont
  {Della~Rocca}}, \bibinfo {author} {\bibfnamefont {M.}~\bibnamefont
  {Chauvin}}, \bibinfo {author} {\bibfnamefont {B.}~\bibnamefont {Huard}},
  \bibinfo {author} {\bibfnamefont {H.}~\bibnamefont {Pothier}}, \bibinfo
  {author} {\bibfnamefont {D.}~\bibnamefont {Esteve}},\ and\ \bibinfo {author}
  {\bibfnamefont {C.}~\bibnamefont {Urbina}},\ }\bibfield  {title} {\bibinfo
  {title} {Measurement of the current-phase relation of superconducting atomic
  contacts},\ }\href {https://doi.org/10.1103/PhysRevLett.99.127005} {\bibfield
   {journal} {\bibinfo  {journal} {Phys. Rev. Lett.}\ }\textbf {\bibinfo
  {volume} {99}},\ \bibinfo {pages} {127005} (\bibinfo {year}
  {2007})}\BibitemShut {NoStop}%
\bibitem [{\citenamefont {Blais}\ \emph {et~al.}(2021)\citenamefont {Blais},
  \citenamefont {Grimsmo}, \citenamefont {Girvin},\ and\ \citenamefont
  {Wallraff}}]{Blais2021}%
  \BibitemOpen
  \bibfield  {author} {\bibinfo {author} {\bibfnamefont {A.}~\bibnamefont
  {Blais}}, \bibinfo {author} {\bibfnamefont {A.~L.}\ \bibnamefont {Grimsmo}},
  \bibinfo {author} {\bibfnamefont {S.~M.}\ \bibnamefont {Girvin}},\ and\
  \bibinfo {author} {\bibfnamefont {A.}~\bibnamefont {Wallraff}},\ }\bibfield
  {title} {\bibinfo {title} {Circuit quantum electrodynamics},\ }\href
  {https://doi.org/10.1103/REVMODPHYS.93.025005/FIGURES/34/MEDIUM} {\bibfield
  {journal} {\bibinfo  {journal} {Rev. Mod. Phys.}\ }\textbf {\bibinfo {volume}
  {93}},\ \bibinfo {pages} {025005} (\bibinfo {year} {2021})}\BibitemShut
  {NoStop}%
\bibitem [{\citenamefont {Luthi}\ \emph {et~al.}(2018)\citenamefont {Luthi},
  \citenamefont {Stavenga}, \citenamefont {Enzing}, \citenamefont {Bruno},
  \citenamefont {Dickel}, \citenamefont {Langford}, \citenamefont {Rol},
  \citenamefont {Jespersen}, \citenamefont {Nyg\aa{}rd}, \citenamefont
  {Krogstrup},\ and\ \citenamefont {DiCarlo}}]{Luthi2018}%
  \BibitemOpen
  \bibfield  {author} {\bibinfo {author} {\bibfnamefont {F.}~\bibnamefont
  {Luthi}}, \bibinfo {author} {\bibfnamefont {T.}~\bibnamefont {Stavenga}},
  \bibinfo {author} {\bibfnamefont {O.~W.}\ \bibnamefont {Enzing}}, \bibinfo
  {author} {\bibfnamefont {A.}~\bibnamefont {Bruno}}, \bibinfo {author}
  {\bibfnamefont {C.}~\bibnamefont {Dickel}}, \bibinfo {author} {\bibfnamefont
  {N.~K.}\ \bibnamefont {Langford}}, \bibinfo {author} {\bibfnamefont {M.~A.}\
  \bibnamefont {Rol}}, \bibinfo {author} {\bibfnamefont {T.~S.}\ \bibnamefont
  {Jespersen}}, \bibinfo {author} {\bibfnamefont {J.}~\bibnamefont
  {Nyg\aa{}rd}}, \bibinfo {author} {\bibfnamefont {P.}~\bibnamefont
  {Krogstrup}},\ and\ \bibinfo {author} {\bibfnamefont {L.}~\bibnamefont
  {DiCarlo}},\ }\bibfield  {title} {\bibinfo {title} {Evolution of nanowire
  transmon qubits and their coherence in a magnetic field},\ }\href
  {https://doi.org/10.1103/PhysRevLett.120.100502} {\bibfield  {journal}
  {\bibinfo  {journal} {Phys. Rev. Lett.}\ }\textbf {\bibinfo {volume} {120}},\
  \bibinfo {pages} {100502} (\bibinfo {year} {2018})}\BibitemShut {NoStop}%
\bibitem [{\citenamefont {Kroll}\ \emph {et~al.}(2019)\citenamefont {Kroll},
  \citenamefont {Borsoi}, \citenamefont {van~der Enden}, \citenamefont
  {Uilhoorn}, \citenamefont {de~Jong}, \citenamefont {Quintero-P\'erez},
  \citenamefont {van Woerkom}, \citenamefont {Bruno}, \citenamefont {Plissard},
  \citenamefont {Car}, \citenamefont {Bakkers}, \citenamefont {Cassidy},\ and\
  \citenamefont {Kouwenhoven}}]{Kroll2019}%
  \BibitemOpen
  \bibfield  {author} {\bibinfo {author} {\bibfnamefont {J.~G.}\ \bibnamefont
  {Kroll}}, \bibinfo {author} {\bibfnamefont {F.}~\bibnamefont {Borsoi}},
  \bibinfo {author} {\bibfnamefont {K.~L.}\ \bibnamefont {van~der Enden}},
  \bibinfo {author} {\bibfnamefont {W.}~\bibnamefont {Uilhoorn}}, \bibinfo
  {author} {\bibfnamefont {D.}~\bibnamefont {de~Jong}}, \bibinfo {author}
  {\bibfnamefont {M.}~\bibnamefont {Quintero-P\'erez}}, \bibinfo {author}
  {\bibfnamefont {D.~J.}\ \bibnamefont {van Woerkom}}, \bibinfo {author}
  {\bibfnamefont {A.}~\bibnamefont {Bruno}}, \bibinfo {author} {\bibfnamefont
  {S.~R.}\ \bibnamefont {Plissard}}, \bibinfo {author} {\bibfnamefont
  {D.}~\bibnamefont {Car}}, \bibinfo {author} {\bibfnamefont {E.~P. A.~M.}\
  \bibnamefont {Bakkers}}, \bibinfo {author} {\bibfnamefont {M.~C.}\
  \bibnamefont {Cassidy}},\ and\ \bibinfo {author} {\bibfnamefont {L.~P.}\
  \bibnamefont {Kouwenhoven}},\ }\bibfield  {title} {\bibinfo {title}
  {Magnetic-field-resilient superconducting coplanar-waveguide resonators for
  hybrid circuit quantum electrodynamics experiments},\ }\href
  {https://doi.org/10.1103/PhysRevApplied.11.064053} {\bibfield  {journal}
  {\bibinfo  {journal} {Phys. Rev. Applied}\ }\textbf {\bibinfo {volume}
  {11}},\ \bibinfo {pages} {064053} (\bibinfo {year} {2019})}\BibitemShut
  {NoStop}%
\bibitem [{\citenamefont {Pita-Vidal}\ \emph {et~al.}(2020)\citenamefont
  {Pita-Vidal}, \citenamefont {Bargerbos}, \citenamefont {Yang}, \citenamefont
  {van Woerkom}, \citenamefont {Pfaff}, \citenamefont {Haider}, \citenamefont
  {Krogstrup}, \citenamefont {Kouwenhoven}, \citenamefont {de~Lange},\ and\
  \citenamefont {Kou}}]{PitaVidal2020}%
  \BibitemOpen
  \bibfield  {author} {\bibinfo {author} {\bibfnamefont {M.}~\bibnamefont
  {Pita-Vidal}}, \bibinfo {author} {\bibfnamefont {A.}~\bibnamefont
  {Bargerbos}}, \bibinfo {author} {\bibfnamefont {C.-K.}\ \bibnamefont {Yang}},
  \bibinfo {author} {\bibfnamefont {D.~J.}\ \bibnamefont {van Woerkom}},
  \bibinfo {author} {\bibfnamefont {W.}~\bibnamefont {Pfaff}}, \bibinfo
  {author} {\bibfnamefont {N.}~\bibnamefont {Haider}}, \bibinfo {author}
  {\bibfnamefont {P.}~\bibnamefont {Krogstrup}}, \bibinfo {author}
  {\bibfnamefont {L.~P.}\ \bibnamefont {Kouwenhoven}}, \bibinfo {author}
  {\bibfnamefont {G.}~\bibnamefont {de~Lange}},\ and\ \bibinfo {author}
  {\bibfnamefont {A.}~\bibnamefont {Kou}},\ }\bibfield  {title} {\bibinfo
  {title} {Gate-tunable field-compatible fluxonium},\ }\href
  {https://doi.org/10.1103/PhysRevApplied.14.064038} {\bibfield  {journal}
  {\bibinfo  {journal} {Phys. Rev. Applied}\ }\textbf {\bibinfo {volume}
  {14}},\ \bibinfo {pages} {064038} (\bibinfo {year} {2020})}\BibitemShut
  {NoStop}%
\bibitem [{\citenamefont {Samkharadze}\ \emph {et~al.}(2016)\citenamefont
  {Samkharadze}, \citenamefont {Bruno}, \citenamefont {Scarlino}, \citenamefont
  {Zheng}, \citenamefont {DiVincenzo}, \citenamefont {DiCarlo},\ and\
  \citenamefont {Vandersypen}}]{Samkharadze2016}%
  \BibitemOpen
  \bibfield  {author} {\bibinfo {author} {\bibfnamefont {N.}~\bibnamefont
  {Samkharadze}}, \bibinfo {author} {\bibfnamefont {A.}~\bibnamefont {Bruno}},
  \bibinfo {author} {\bibfnamefont {P.}~\bibnamefont {Scarlino}}, \bibinfo
  {author} {\bibfnamefont {G.}~\bibnamefont {Zheng}}, \bibinfo {author}
  {\bibfnamefont {D.~P.}\ \bibnamefont {DiVincenzo}}, \bibinfo {author}
  {\bibfnamefont {L.}~\bibnamefont {DiCarlo}},\ and\ \bibinfo {author}
  {\bibfnamefont {L.~M.~K.}\ \bibnamefont {Vandersypen}},\ }\bibfield  {title}
  {\bibinfo {title} {High-kinetic-inductance superconducting nanowire
  resonators for circuit {Q}{E}{D} in a magnetic field},\ }\href
  {https://doi.org/10.1103/PhysRevApplied.5.044004} {\bibfield  {journal}
  {\bibinfo  {journal} {Phys. Rev. Applied}\ }\textbf {\bibinfo {volume} {5}},\
  \bibinfo {pages} {044004} (\bibinfo {year} {2016})}\BibitemShut {NoStop}%
\bibitem [{\citenamefont {\v{Z}itko}\ and\ \citenamefont
  {Pruschke}(2009)}]{zitko2009}%
  \BibitemOpen
  \bibfield  {author} {\bibinfo {author} {\bibfnamefont {R.}~\bibnamefont
  {\v{Z}itko}}\ and\ \bibinfo {author} {\bibfnamefont {T.}~\bibnamefont
  {Pruschke}},\ }\bibfield  {title} {\bibinfo {title} {Energy resolution and
  discretization artefacts in the numerical renormalization group},\
  }\href@noop {} {\bibfield  {journal} {\bibinfo  {journal} {Phys. Rev. B}\
  }\textbf {\bibinfo {volume} {79}},\ \bibinfo {pages} {085106} (\bibinfo
  {year} {2009})}\BibitemShut {NoStop}%
\bibitem [{\citenamefont {Zitko}(2021)}]{zitko_rok_2021_4841076}%
  \BibitemOpen
  \bibfield  {author} {\bibinfo {author} {\bibfnamefont {R.}~\bibnamefont
  {Zitko}},\ }\href {https://doi.org/10.5281/zenodo.4841076} {\bibinfo {title}
  {{NRG} {Ljubljana}}} (\bibinfo {year} {2021})\BibitemShut {NoStop}%
\bibitem [{\citenamefont {Keselman}\ \emph {et~al.}(2019)\citenamefont
  {Keselman}, \citenamefont {Murthy}, \citenamefont {van Heck},\ and\
  \citenamefont {Bauer}}]{Keselman2019}%
  \BibitemOpen
  \bibfield  {author} {\bibinfo {author} {\bibfnamefont {A.}~\bibnamefont
  {Keselman}}, \bibinfo {author} {\bibfnamefont {C.}~\bibnamefont {Murthy}},
  \bibinfo {author} {\bibfnamefont {B.}~\bibnamefont {van Heck}},\ and\
  \bibinfo {author} {\bibfnamefont {B.}~\bibnamefont {Bauer}},\ }\bibfield
  {title} {\bibinfo {title} {{Spectral response of {J}osephson junctions with
  low-energy quasiparticles}},\ }\href
  {https://doi.org/10.21468/SciPostPhys.7.4.050} {\bibfield  {journal}
  {\bibinfo  {journal} {SciPost Phys.}\ }\textbf {\bibinfo {volume} {7}},\
  \bibinfo {pages} {50} (\bibinfo {year} {2019})}\BibitemShut {NoStop}%
\bibitem [{\citenamefont {{\'{A}}vila}\ \emph
  {et~al.}(2020{\natexlab{a}})\citenamefont {{\'{A}}vila}, \citenamefont
  {Prada}, \citenamefont {San-Jose},\ and\ \citenamefont {Aguado}}]{Avila2020}%
  \BibitemOpen
  \bibfield  {author} {\bibinfo {author} {\bibfnamefont {J.}~\bibnamefont
  {{\'{A}}vila}}, \bibinfo {author} {\bibfnamefont {E.}~\bibnamefont {Prada}},
  \bibinfo {author} {\bibfnamefont {P.}~\bibnamefont {San-Jose}},\ and\
  \bibinfo {author} {\bibfnamefont {R.}~\bibnamefont {Aguado}},\ }\bibfield
  {title} {\bibinfo {title} {{Superconducting islands with topological
  {J}osephson junctions based on semiconductor nanowires}},\ }\href
  {https://doi.org/10.1103/physrevb.102.094518} {\bibfield  {journal} {\bibinfo
   {journal} {Phys. Rev. B}\ }\textbf {\bibinfo {volume} {102}},\ \bibinfo
  {pages} {094518} (\bibinfo {year} {2020}{\natexlab{a}})}\BibitemShut
  {NoStop}%
\bibitem [{\citenamefont {{\'{A}}vila}\ \emph
  {et~al.}(2020{\natexlab{b}})\citenamefont {{\'{A}}vila}, \citenamefont
  {Prada}, \citenamefont {San-Jose},\ and\ \citenamefont
  {Aguado}}]{Avila2020a}%
  \BibitemOpen
  \bibfield  {author} {\bibinfo {author} {\bibfnamefont {J.}~\bibnamefont
  {{\'{A}}vila}}, \bibinfo {author} {\bibfnamefont {E.}~\bibnamefont {Prada}},
  \bibinfo {author} {\bibfnamefont {P.}~\bibnamefont {San-Jose}},\ and\
  \bibinfo {author} {\bibfnamefont {R.}~\bibnamefont {Aguado}},\ }\bibfield
  {title} {\bibinfo {title} {{{M}ajorana oscillations and parity crossings in
  semiconductor nanowire-based transmon qubits}},\ }\href
  {https://doi.org/10.1103/physrevresearch.2.033493} {\bibfield  {journal}
  {\bibinfo  {journal} {Phys. Rev. Research}\ }\textbf {\bibinfo {volume}
  {2}},\ \bibinfo {pages} {033493} (\bibinfo {year}
  {2020}{\natexlab{b}})}\BibitemShut {NoStop}%
\bibitem [{Note1()}]{Note1}%
  \BibitemOpen
  \bibinfo {note} {Our assumption that regimes with $0$-junction and $\pi
  $-junction behaviour correspond to the quantum dot junction being in a
  singlet or doublet state, respectively, is only valid in the single-level
  regime, where the level spacing of the quantum dot is significantly larger
  than $\Delta $ and $U$. In the multi-level regime, where excited states of
  the quantum dot are involved, the presence or absence of the $\pi $ offset
  also depends on the character of the orbital wavefunctions in addition to the
  fermion parity \cite {VanDam2006a}.}\BibitemShut {Stop}%
\bibitem [{\citenamefont {Kir\ifmmode~\check{s}\else \v{s}\fi{}anskas}\ \emph
  {et~al.}(2015)\citenamefont {Kir\ifmmode~\check{s}\else \v{s}\fi{}anskas},
  \citenamefont {Goldstein}, \citenamefont {Flensberg}, \citenamefont
  {Glazman},\ and\ \citenamefont {Paaske}}]{kirsanskas2015}%
  \BibitemOpen
  \bibfield  {author} {\bibinfo {author} {\bibfnamefont {G.}~\bibnamefont
  {Kir\ifmmode~\check{s}\else \v{s}\fi{}anskas}}, \bibinfo {author}
  {\bibfnamefont {M.}~\bibnamefont {Goldstein}}, \bibinfo {author}
  {\bibfnamefont {K.}~\bibnamefont {Flensberg}}, \bibinfo {author}
  {\bibfnamefont {L.~I.}\ \bibnamefont {Glazman}},\ and\ \bibinfo {author}
  {\bibfnamefont {J.}~\bibnamefont {Paaske}},\ }\bibfield  {title} {\bibinfo
  {title} {{Yu-Shiba-Rusinov} states in phase-biased superconductor--quantum
  dot--superconductor junctions},\ }\href
  {https://doi.org/10.1103/PhysRevB.92.235422} {\bibfield  {journal} {\bibinfo
  {journal} {Phys. Rev. B}\ }\textbf {\bibinfo {volume} {92}},\ \bibinfo
  {pages} {235422} (\bibinfo {year} {2015})}\BibitemShut {NoStop}%
\bibitem [{\citenamefont {Kadlecov{\'{a}}}\ \emph {et~al.}(2017)\citenamefont
  {Kadlecov{\'{a}}}, \citenamefont {{\v{Z}}onda},\ and\ \citenamefont
  {Novotn{\'{y}}}}]{kadlecova2017}%
  \BibitemOpen
  \bibfield  {author} {\bibinfo {author} {\bibfnamefont {A.}~\bibnamefont
  {Kadlecov{\'{a}}}}, \bibinfo {author} {\bibfnamefont {M.}~\bibnamefont
  {{\v{Z}}onda}},\ and\ \bibinfo {author} {\bibfnamefont {T.}~\bibnamefont
  {Novotn{\'{y}}}},\ }\bibfield  {title} {\bibinfo {title} {Quantum dot
  attached to superconducting leads: Relation between symmetric and asymmetric
  coupling},\ }\href {https://doi.org/10.1103/physrevb.95.195114} {\bibfield
  {journal} {\bibinfo  {journal} {Phys. Rev. B}\ }\textbf {\bibinfo {volume}
  {95}},\ \bibinfo {pages} {195114} (\bibinfo {year} {2017})}\BibitemShut
  {NoStop}%
\bibitem [{\citenamefont {Zalom}\ \emph {et~al.}(2021)\citenamefont {Zalom},
  \citenamefont {Pokorn\'y},\ and\ \citenamefont {Novotn\'y}}]{zalom2021}%
  \BibitemOpen
  \bibfield  {author} {\bibinfo {author} {\bibfnamefont {P.}~\bibnamefont
  {Zalom}}, \bibinfo {author} {\bibfnamefont {V.}~\bibnamefont {Pokorn\'y}},\
  and\ \bibinfo {author} {\bibfnamefont {T.~c.~v.}\ \bibnamefont {Novotn\'y}},\
  }\bibfield  {title} {\bibinfo {title} {Spectral and transport properties of a
  half-filled {Anderson} impurity coupled to phase-biased superconducting and
  metallic leads},\ }\href {https://doi.org/10.1103/PhysRevB.103.035419}
  {\bibfield  {journal} {\bibinfo  {journal} {Phys. Rev. B}\ }\textbf {\bibinfo
  {volume} {103}},\ \bibinfo {pages} {035419} (\bibinfo {year}
  {2021})}\BibitemShut {NoStop}%
\bibitem [{\citenamefont {Escribano}\ \emph {et~al.}(2022)\citenamefont
  {Escribano}, \citenamefont {Levy~Yeyati}, \citenamefont {Aguado},
  \citenamefont {Prada},\ and\ \citenamefont {San-Jose}}]{escribano2022}%
  \BibitemOpen
  \bibfield  {author} {\bibinfo {author} {\bibfnamefont {S.~D.}\ \bibnamefont
  {Escribano}}, \bibinfo {author} {\bibfnamefont {A.}~\bibnamefont
  {Levy~Yeyati}}, \bibinfo {author} {\bibfnamefont {R.}~\bibnamefont {Aguado}},
  \bibinfo {author} {\bibfnamefont {E.}~\bibnamefont {Prada}},\ and\ \bibinfo
  {author} {\bibfnamefont {P.}~\bibnamefont {San-Jose}},\ }\bibfield  {title}
  {\bibinfo {title} {Fluxoid-induced pairing suppression and near-zero modes in
  quantum dots coupled to full-shell nanowires},\ }\href
  {https://doi.org/10.1103/PhysRevB.105.045418} {\bibfield  {journal} {\bibinfo
   {journal} {Phys. Rev. B}\ }\textbf {\bibinfo {volume} {105}},\ \bibinfo
  {pages} {045418} (\bibinfo {year} {2022})}\BibitemShut {NoStop}%
\bibitem [{\citenamefont {\ifmmode~\check{Z}\else \v{Z}\fi{}itko}\ and\
  \citenamefont {Fabrizio}(2017)}]{zitko2017}%
  \BibitemOpen
  \bibfield  {author} {\bibinfo {author} {\bibfnamefont {R.}~\bibnamefont
  {\ifmmode~\check{Z}\else \v{Z}\fi{}itko}}\ and\ \bibinfo {author}
  {\bibfnamefont {M.}~\bibnamefont {Fabrizio}},\ }\bibfield  {title} {\bibinfo
  {title} {Non-{{F}ermi}-liquid behavior in quantum impurity models with
  superconducting channels},\ }\href
  {https://doi.org/10.1103/PhysRevB.95.085121} {\bibfield  {journal} {\bibinfo
  {journal} {Phys. Rev. B}\ }\textbf {\bibinfo {volume} {95}},\ \bibinfo
  {pages} {085121} (\bibinfo {year} {2017})}\BibitemShut {NoStop}%
\bibitem [{\citenamefont {Serniak}\ \emph {et~al.}(2019)\citenamefont
  {Serniak}, \citenamefont {Diamond}, \citenamefont {Hays}, \citenamefont
  {Fatemi}, \citenamefont {Shankar}, \citenamefont {Frunzio}, \citenamefont
  {Schoelkopf},\ and\ \citenamefont {Devoret}}]{Serniak2019}%
  \BibitemOpen
  \bibfield  {author} {\bibinfo {author} {\bibfnamefont {K.}~\bibnamefont
  {Serniak}}, \bibinfo {author} {\bibfnamefont {S.}~\bibnamefont {Diamond}},
  \bibinfo {author} {\bibfnamefont {M.}~\bibnamefont {Hays}}, \bibinfo {author}
  {\bibfnamefont {V.}~\bibnamefont {Fatemi}}, \bibinfo {author} {\bibfnamefont
  {S.}~\bibnamefont {Shankar}}, \bibinfo {author} {\bibfnamefont
  {L.}~\bibnamefont {Frunzio}}, \bibinfo {author} {\bibfnamefont {R.~J.}\
  \bibnamefont {Schoelkopf}},\ and\ \bibinfo {author} {\bibfnamefont {M.~H.}\
  \bibnamefont {Devoret}},\ }\bibfield  {title} {\bibinfo {title} {Direct
  dispersive monitoring of charge parity in offset-charge-sensitive
  transmons},\ }\href {https://doi.org/10.1103/PhysRevApplied.12.014052}
  {\bibfield  {journal} {\bibinfo  {journal} {Phys. Rev. Applied}\ }\textbf
  {\bibinfo {volume} {12}},\ \bibinfo {pages} {014052} (\bibinfo {year}
  {2019})}\BibitemShut {NoStop}%
\bibitem [{\citenamefont {Wesdorp}\ \emph {et~al.}(2021)\citenamefont
  {Wesdorp}, \citenamefont {Gr{\"u}nhaupt}, \citenamefont {Vaartjes},
  \citenamefont {Pita-Vidal}, \citenamefont {Bargerbos}, \citenamefont
  {Splitthoff}, \citenamefont {Krogstrup}, \citenamefont {van Heck},\ and\
  \citenamefont {de~Lange}}]{Wesdorp2021}%
  \BibitemOpen
  \bibfield  {author} {\bibinfo {author} {\bibfnamefont {J.~J.}\ \bibnamefont
  {Wesdorp}}, \bibinfo {author} {\bibfnamefont {L.}~\bibnamefont
  {Gr{\"u}nhaupt}}, \bibinfo {author} {\bibfnamefont {A.}~\bibnamefont
  {Vaartjes}}, \bibinfo {author} {\bibfnamefont {M.}~\bibnamefont
  {Pita-Vidal}}, \bibinfo {author} {\bibfnamefont {A.}~\bibnamefont
  {Bargerbos}}, \bibinfo {author} {\bibfnamefont {L.~J.}\ \bibnamefont
  {Splitthoff}}, \bibinfo {author} {\bibfnamefont {P.}~\bibnamefont
  {Krogstrup}}, \bibinfo {author} {\bibfnamefont {B.}~\bibnamefont {van
  Heck}},\ and\ \bibinfo {author} {\bibfnamefont {G.}~\bibnamefont
  {de~Lange}},\ }\bibfield  {title} {\bibinfo {title} {Dynamical polarization
  of the fermion parity in a nanowire {J}osephson junction},\ }\href@noop {}
  {\bibfield  {journal} {\bibinfo  {journal} {arXiv e-prints}\ } (\bibinfo
  {year} {2021})},\ \Eprint {https://arxiv.org/abs/2112.01936}
  {arXiv:2112.01936} \BibitemShut {NoStop}%
\bibitem [{\citenamefont {Heinsoo}\ \emph {et~al.}(2018)\citenamefont
  {Heinsoo}, \citenamefont {Andersen}, \citenamefont {Remm}, \citenamefont
  {Krinner}, \citenamefont {Walter}, \citenamefont {Salath\'e}, \citenamefont
  {Gasparinetti}, \citenamefont {Besse}, \citenamefont
  {Poto\ifmmode~\check{c}\else \v{c}\fi{}nik}, \citenamefont {Wallraff},\ and\
  \citenamefont {Eichler}}]{Heinsoo2018}%
  \BibitemOpen
  \bibfield  {author} {\bibinfo {author} {\bibfnamefont {J.}~\bibnamefont
  {Heinsoo}}, \bibinfo {author} {\bibfnamefont {C.~K.}\ \bibnamefont
  {Andersen}}, \bibinfo {author} {\bibfnamefont {A.}~\bibnamefont {Remm}},
  \bibinfo {author} {\bibfnamefont {S.}~\bibnamefont {Krinner}}, \bibinfo
  {author} {\bibfnamefont {T.}~\bibnamefont {Walter}}, \bibinfo {author}
  {\bibfnamefont {Y.}~\bibnamefont {Salath\'e}}, \bibinfo {author}
  {\bibfnamefont {S.}~\bibnamefont {Gasparinetti}}, \bibinfo {author}
  {\bibfnamefont {J.-C.}\ \bibnamefont {Besse}}, \bibinfo {author}
  {\bibfnamefont {A.}~\bibnamefont {Poto\ifmmode~\check{c}\else
  \v{c}\fi{}nik}}, \bibinfo {author} {\bibfnamefont {A.}~\bibnamefont
  {Wallraff}},\ and\ \bibinfo {author} {\bibfnamefont {C.}~\bibnamefont
  {Eichler}},\ }\bibfield  {title} {\bibinfo {title} {Rapid high-fidelity
  multiplexed readout of superconducting qubits},\ }\href
  {https://doi.org/10.1103/PhysRevApplied.10.034040} {\bibfield  {journal}
  {\bibinfo  {journal} {Phys. Rev. Applied}\ }\textbf {\bibinfo {volume}
  {10}},\ \bibinfo {pages} {034040} (\bibinfo {year} {2018})}\BibitemShut
  {NoStop}%
\bibitem [{\citenamefont {Pop}\ \emph {et~al.}(2014)\citenamefont {Pop},
  \citenamefont {Geerlings}, \citenamefont {Catelani}, \citenamefont
  {Schoelkopf}, \citenamefont {Glazman},\ and\ \citenamefont
  {Devoret}}]{Pop2014}%
  \BibitemOpen
  \bibfield  {author} {\bibinfo {author} {\bibfnamefont {I.~M.}\ \bibnamefont
  {Pop}}, \bibinfo {author} {\bibfnamefont {K.}~\bibnamefont {Geerlings}},
  \bibinfo {author} {\bibfnamefont {G.}~\bibnamefont {Catelani}}, \bibinfo
  {author} {\bibfnamefont {R.~J.}\ \bibnamefont {Schoelkopf}}, \bibinfo
  {author} {\bibfnamefont {L.~I.}\ \bibnamefont {Glazman}},\ and\ \bibinfo
  {author} {\bibfnamefont {M.~H.}\ \bibnamefont {Devoret}},\ }\bibfield
  {title} {\bibinfo {title} {Coherent suppression of electromagnetic
  dissipation due to superconducting quasiparticles},\ }\href
  {https://doi.org/10.1038/nature13017} {\bibfield  {journal} {\bibinfo
  {journal} {Nature}\ }\textbf {\bibinfo {volume} {508}},\ \bibinfo {pages}
  {369} (\bibinfo {year} {2014})}\BibitemShut {NoStop}%
\bibitem [{\citenamefont {Glazman}\ and\ \citenamefont
  {Catelani}(2021)}]{Glazman2021}%
  \BibitemOpen
  \bibfield  {author} {\bibinfo {author} {\bibfnamefont {L.~I.}\ \bibnamefont
  {Glazman}}\ and\ \bibinfo {author} {\bibfnamefont {G.}~\bibnamefont
  {Catelani}},\ }\bibfield  {title} {\bibinfo {title} {{Bogoliubov
  quasiparticles in superconducting qubits}},\ }\href
  {https://doi.org/10.21468/SciPostPhysLectNotes.31} {\bibfield  {journal}
  {\bibinfo  {journal} {SciPost Phys. Lect. Notes}\ ,\ \bibinfo {pages} {31}}
  (\bibinfo {year} {2021})}\BibitemShut {NoStop}%
\bibitem [{\citenamefont {Serniak}\ \emph {et~al.}(2018)\citenamefont
  {Serniak}, \citenamefont {Hays}, \citenamefont {de~Lange}, \citenamefont
  {Diamond}, \citenamefont {Shankar}, \citenamefont {Burkhart}, \citenamefont
  {Frunzio}, \citenamefont {Houzet},\ and\ \citenamefont
  {Devoret}}]{Serniak2018}%
  \BibitemOpen
  \bibfield  {author} {\bibinfo {author} {\bibfnamefont {K.}~\bibnamefont
  {Serniak}}, \bibinfo {author} {\bibfnamefont {M.}~\bibnamefont {Hays}},
  \bibinfo {author} {\bibfnamefont {G.}~\bibnamefont {de~Lange}}, \bibinfo
  {author} {\bibfnamefont {S.}~\bibnamefont {Diamond}}, \bibinfo {author}
  {\bibfnamefont {S.}~\bibnamefont {Shankar}}, \bibinfo {author} {\bibfnamefont
  {L.~D.}\ \bibnamefont {Burkhart}}, \bibinfo {author} {\bibfnamefont
  {L.}~\bibnamefont {Frunzio}}, \bibinfo {author} {\bibfnamefont
  {M.}~\bibnamefont {Houzet}},\ and\ \bibinfo {author} {\bibfnamefont {M.~H.}\
  \bibnamefont {Devoret}},\ }\bibfield  {title} {\bibinfo {title} {Hot
  nonequilibrium quasiparticles in transmon qubits},\ }\href
  {https://doi.org/10.1103/PhysRevLett.121.157701} {\bibfield  {journal}
  {\bibinfo  {journal} {Phys. Rev. Lett.}\ }\textbf {\bibinfo {volume} {121}},\
  \bibinfo {pages} {157701} (\bibinfo {year} {2018})}\BibitemShut {NoStop}%
\bibitem [{\citenamefont {Larsen}\ \emph {et~al.}(2020)\citenamefont {Larsen},
  \citenamefont {Gershenson}, \citenamefont {Casparis}, \citenamefont
  {Kringh\o{}j}, \citenamefont {Pearson}, \citenamefont {McNeil}, \citenamefont
  {Kuemmeth}, \citenamefont {Krogstrup}, \citenamefont {Petersson},\ and\
  \citenamefont {Marcus}}]{Larsen2020}%
  \BibitemOpen
  \bibfield  {author} {\bibinfo {author} {\bibfnamefont {T.~W.}\ \bibnamefont
  {Larsen}}, \bibinfo {author} {\bibfnamefont {M.~E.}\ \bibnamefont
  {Gershenson}}, \bibinfo {author} {\bibfnamefont {L.}~\bibnamefont
  {Casparis}}, \bibinfo {author} {\bibfnamefont {A.}~\bibnamefont
  {Kringh\o{}j}}, \bibinfo {author} {\bibfnamefont {N.~J.}\ \bibnamefont
  {Pearson}}, \bibinfo {author} {\bibfnamefont {R.~P.~G.}\ \bibnamefont
  {McNeil}}, \bibinfo {author} {\bibfnamefont {F.}~\bibnamefont {Kuemmeth}},
  \bibinfo {author} {\bibfnamefont {P.}~\bibnamefont {Krogstrup}}, \bibinfo
  {author} {\bibfnamefont {K.~D.}\ \bibnamefont {Petersson}},\ and\ \bibinfo
  {author} {\bibfnamefont {C.~M.}\ \bibnamefont {Marcus}},\ }\bibfield  {title}
  {\bibinfo {title} {Parity-protected superconductor-semiconductor qubit},\
  }\href {https://doi.org/10.1103/PhysRevLett.125.056801} {\bibfield  {journal}
  {\bibinfo  {journal} {Phys. Rev. Lett.}\ }\textbf {\bibinfo {volume} {125}},\
  \bibinfo {pages} {056801} (\bibinfo {year} {2020})}\BibitemShut {NoStop}%
\bibitem [{\citenamefont {Sau}\ and\ \citenamefont {Sarma}(2012)}]{Sau2012}%
  \BibitemOpen
  \bibfield  {author} {\bibinfo {author} {\bibfnamefont {J.~D.}\ \bibnamefont
  {Sau}}\ and\ \bibinfo {author} {\bibfnamefont {S.~D.}\ \bibnamefont
  {Sarma}},\ }\bibfield  {title} {\bibinfo {title} {Realizing a robust
  practical {M}ajorana chain in a quantum-dot-superconductor linear array},\
  }\href {https://doi.org/10.1038/ncomms1966} {\bibfield  {journal} {\bibinfo
  {journal} {Nature Communications}\ }\textbf {\bibinfo {volume} {3}},\
  \bibinfo {pages} {1} (\bibinfo {year} {2012})}\BibitemShut {NoStop}%
\bibitem [{\citenamefont {Fulga}\ \emph {et~al.}(2013)\citenamefont {Fulga},
  \citenamefont {Haim}, \citenamefont {Akhmerov},\ and\ \citenamefont
  {Oreg}}]{Fulga2013}%
  \BibitemOpen
  \bibfield  {author} {\bibinfo {author} {\bibfnamefont {I.~C.}\ \bibnamefont
  {Fulga}}, \bibinfo {author} {\bibfnamefont {A.}~\bibnamefont {Haim}},
  \bibinfo {author} {\bibfnamefont {A.~R.}\ \bibnamefont {Akhmerov}},\ and\
  \bibinfo {author} {\bibfnamefont {Y.}~\bibnamefont {Oreg}},\ }\bibfield
  {title} {\bibinfo {title} {Adaptive tuning of {M}ajorana fermions in a
  quantum dot chain},\ }\href {https://doi.org/10.1088/1367-2630/15/4/045020}
  {\bibfield  {journal} {\bibinfo  {journal} {New J. Phys.}\ }\textbf {\bibinfo
  {volume} {15}},\ \bibinfo {pages} {045020} (\bibinfo {year}
  {2013})}\BibitemShut {NoStop}%
\bibitem [{\citenamefont {Stenger}\ \emph {et~al.}(2018)\citenamefont
  {Stenger}, \citenamefont {Woods}, \citenamefont {Frolov},\ and\ \citenamefont
  {Stanescu}}]{Stenger2018}%
  \BibitemOpen
  \bibfield  {author} {\bibinfo {author} {\bibfnamefont {J.~P.~T.}\
  \bibnamefont {Stenger}}, \bibinfo {author} {\bibfnamefont {B.~D.}\
  \bibnamefont {Woods}}, \bibinfo {author} {\bibfnamefont {S.~M.}\ \bibnamefont
  {Frolov}},\ and\ \bibinfo {author} {\bibfnamefont {T.~D.}\ \bibnamefont
  {Stanescu}},\ }\bibfield  {title} {\bibinfo {title} {Control and detection of
  {M}ajorana bound states in quantum dot arrays},\ }\href
  {https://doi.org/10.1103/PhysRevB.98.085407} {\bibfield  {journal} {\bibinfo
  {journal} {Phys. Rev. B}\ }\textbf {\bibinfo {volume} {98}},\ \bibinfo
  {pages} {085407} (\bibinfo {year} {2018})}\BibitemShut {NoStop}%
\bibitem [{\citenamefont {Dayton}\ \emph {et~al.}(2018)\citenamefont {Dayton},
  \citenamefont {Sage}, \citenamefont {Gingrich}, \citenamefont {Loving},
  \citenamefont {Ambrose}, \citenamefont {Siwak}, \citenamefont {Keebaugh},
  \citenamefont {Kirby}, \citenamefont {Miller}, \citenamefont {Herr},
  \citenamefont {Herr},\ and\ \citenamefont {Naaman}}]{Dayton2018}%
  \BibitemOpen
  \bibfield  {author} {\bibinfo {author} {\bibfnamefont {I.~M.}\ \bibnamefont
  {Dayton}}, \bibinfo {author} {\bibfnamefont {T.}~\bibnamefont {Sage}},
  \bibinfo {author} {\bibfnamefont {E.~C.}\ \bibnamefont {Gingrich}}, \bibinfo
  {author} {\bibfnamefont {M.~G.}\ \bibnamefont {Loving}}, \bibinfo {author}
  {\bibfnamefont {T.~F.}\ \bibnamefont {Ambrose}}, \bibinfo {author}
  {\bibfnamefont {N.~P.}\ \bibnamefont {Siwak}}, \bibinfo {author}
  {\bibfnamefont {S.}~\bibnamefont {Keebaugh}}, \bibinfo {author}
  {\bibfnamefont {C.}~\bibnamefont {Kirby}}, \bibinfo {author} {\bibfnamefont
  {D.~L.}\ \bibnamefont {Miller}}, \bibinfo {author} {\bibfnamefont {A.~Y.}\
  \bibnamefont {Herr}}, \bibinfo {author} {\bibfnamefont {Q.~P.}\ \bibnamefont
  {Herr}},\ and\ \bibinfo {author} {\bibfnamefont {O.}~\bibnamefont {Naaman}},\
  }\bibfield  {title} {\bibinfo {title} {Experimental demonstration of a
  {J}osephson magnetic memory cell with a programmable $\pi$-junction},\ }\href
  {https://doi.org/10.1109/LMAG.2018.2801820} {\bibfield  {journal} {\bibinfo
  {journal} {IEEE Magnetics Letters}\ }\textbf {\bibinfo {volume} {9}},\
  \bibinfo {pages} {1} (\bibinfo {year} {2018})}\BibitemShut {NoStop}%
\bibitem [{\citenamefont {Ginossar}\ and\ \citenamefont
  {Grosfeld}(2014)}]{Ginossar2014}%
  \BibitemOpen
  \bibfield  {author} {\bibinfo {author} {\bibfnamefont {E.}~\bibnamefont
  {Ginossar}}\ and\ \bibinfo {author} {\bibfnamefont {E.}~\bibnamefont
  {Grosfeld}},\ }\bibfield  {title} {\bibinfo {title} {Microwave transitions as
  a signature of coherent parity mixing effects in the {M}ajorana-transmon
  qubit},\ }\href {https://doi.org/10.1038/ncomms5772} {\bibfield  {journal}
  {\bibinfo  {journal} {Nature Communications}\ }\textbf {\bibinfo {volume}
  {5}},\ \bibinfo {pages} {1} (\bibinfo {year} {2014})}\BibitemShut {NoStop}%
\bibitem [{\citenamefont {Pikulin}\ \emph {et~al.}(2021)\citenamefont
  {Pikulin}, \citenamefont {van Heck}, \citenamefont {Karzig}, \citenamefont
  {Martinez}, \citenamefont {Nijholt}, \citenamefont {Laeven}, \citenamefont
  {Winkler}, \citenamefont {Watson}, \citenamefont {Heedt}, \citenamefont
  {Temurhan}, \citenamefont {Svidenko}, \citenamefont {Lutchyn}, \citenamefont
  {Thomas}, \citenamefont {de~Lange}, \citenamefont {Casparis},\ and\
  \citenamefont {Nayak}}]{Pikulin2021}%
  \BibitemOpen
  \bibfield  {author} {\bibinfo {author} {\bibfnamefont {D.~I.}\ \bibnamefont
  {Pikulin}}, \bibinfo {author} {\bibfnamefont {B.}~\bibnamefont {van Heck}},
  \bibinfo {author} {\bibfnamefont {T.}~\bibnamefont {Karzig}}, \bibinfo
  {author} {\bibfnamefont {E.~A.}\ \bibnamefont {Martinez}}, \bibinfo {author}
  {\bibfnamefont {B.}~\bibnamefont {Nijholt}}, \bibinfo {author} {\bibfnamefont
  {T.}~\bibnamefont {Laeven}}, \bibinfo {author} {\bibfnamefont {G.~W.}\
  \bibnamefont {Winkler}}, \bibinfo {author} {\bibfnamefont {J.~D.}\
  \bibnamefont {Watson}}, \bibinfo {author} {\bibfnamefont {S.}~\bibnamefont
  {Heedt}}, \bibinfo {author} {\bibfnamefont {M.}~\bibnamefont {Temurhan}},
  \bibinfo {author} {\bibfnamefont {V.}~\bibnamefont {Svidenko}}, \bibinfo
  {author} {\bibfnamefont {R.~M.}\ \bibnamefont {Lutchyn}}, \bibinfo {author}
  {\bibfnamefont {M.}~\bibnamefont {Thomas}}, \bibinfo {author} {\bibfnamefont
  {G.}~\bibnamefont {de~Lange}}, \bibinfo {author} {\bibfnamefont
  {L.}~\bibnamefont {Casparis}},\ and\ \bibinfo {author} {\bibfnamefont
  {C.}~\bibnamefont {Nayak}},\ }\bibfield  {title} {\bibinfo {title} {Protocol
  to identify a topological superconducting phase in a three-terminal device},\
  }\href@noop {} {\bibfield  {journal} {\bibinfo  {journal} {arXiv e-prints}\ }
  (\bibinfo {year} {2021})},\ \Eprint {https://arxiv.org/abs/2103.12217}
  {arXiv:2103.12217} \BibitemShut {NoStop}%
\bibitem [{\citenamefont {Ahn}\ \emph {et~al.}(2021)\citenamefont {Ahn},
  \citenamefont {Pan}, \citenamefont {Woods}, \citenamefont {Stanescu},\ and\
  \citenamefont {Das~Sarma}}]{Stanescu2021}%
  \BibitemOpen
  \bibfield  {author} {\bibinfo {author} {\bibfnamefont {S.}~\bibnamefont
  {Ahn}}, \bibinfo {author} {\bibfnamefont {H.}~\bibnamefont {Pan}}, \bibinfo
  {author} {\bibfnamefont {B.}~\bibnamefont {Woods}}, \bibinfo {author}
  {\bibfnamefont {T.~D.}\ \bibnamefont {Stanescu}},\ and\ \bibinfo {author}
  {\bibfnamefont {S.}~\bibnamefont {Das~Sarma}},\ }\bibfield  {title} {\bibinfo
  {title} {Estimating disorder and its adverse effects in semiconductor
  {M}ajorana nanowires},\ }\href
  {https://doi.org/10.1103/PhysRevMaterials.5.124602} {\bibfield  {journal}
  {\bibinfo  {journal} {Phys. Rev. Materials}\ }\textbf {\bibinfo {volume}
  {5}},\ \bibinfo {pages} {124602} (\bibinfo {year} {2021})}\BibitemShut
  {NoStop}%
\bibitem [{\citenamefont {Schulenborg}\ and\ \citenamefont
  {Flensberg}(2020)}]{Schulenborg2020}%
  \BibitemOpen
  \bibfield  {author} {\bibinfo {author} {\bibfnamefont {J.}~\bibnamefont
  {Schulenborg}}\ and\ \bibinfo {author} {\bibfnamefont {K.}~\bibnamefont
  {Flensberg}},\ }\bibfield  {title} {\bibinfo {title} {{Absence of
  supercurrent sign reversal in a topological junction with a quantum dot}},\
  }\href {https://doi.org/10.1103/PhysRevB.101.014512} {\bibfield  {journal}
  {\bibinfo  {journal} {Phys. Rev. B}\ }\textbf {\bibinfo {volume} {101}},\
  \bibinfo {pages} {014512} (\bibinfo {year} {2020})}\BibitemShut {NoStop}%
\bibitem [{\citenamefont {Prada}\ \emph {et~al.}(2020)\citenamefont {Prada},
  \citenamefont {San-Jose}, \citenamefont {de~Moor}, \citenamefont {Geresdi},
  \citenamefont {Lee}, \citenamefont {Klinovaja}, \citenamefont {Loss},
  \citenamefont {Nyg{\aa}rd}, \citenamefont {Aguado},\ and\ \citenamefont
  {Kouwenhoven}}]{prada2020}%
  \BibitemOpen
  \bibfield  {author} {\bibinfo {author} {\bibfnamefont {E.}~\bibnamefont
  {Prada}}, \bibinfo {author} {\bibfnamefont {P.}~\bibnamefont {San-Jose}},
  \bibinfo {author} {\bibfnamefont {M.}~\bibnamefont {de~Moor}}, \bibinfo
  {author} {\bibfnamefont {A.}~\bibnamefont {Geresdi}}, \bibinfo {author}
  {\bibfnamefont {E.}~\bibnamefont {Lee}}, \bibinfo {author} {\bibfnamefont
  {J.}~\bibnamefont {Klinovaja}}, \bibinfo {author} {\bibfnamefont
  {D.}~\bibnamefont {Loss}}, \bibinfo {author} {\bibfnamefont {J.}~\bibnamefont
  {Nyg{\aa}rd}}, \bibinfo {author} {\bibfnamefont {R.}~\bibnamefont {Aguado}},\
  and\ \bibinfo {author} {\bibfnamefont {L.}~\bibnamefont {Kouwenhoven}},\
  }\bibfield  {title} {\bibinfo {title} {From {Andreev} to {Majorana} bound
  states in hybrid superconductor--semiconductor nanowires},\ }\href
  {https://doi.org/10.1038/s42254-020-0228-y} {\bibfield  {journal} {\bibinfo
  {journal} {Nature Reviews Physics}\ }\textbf {\bibinfo {volume} {2}},\
  \bibinfo {pages} {575} (\bibinfo {year} {2020})}\BibitemShut {NoStop}%
\bibitem [{\citenamefont {Zitko}(2022)}]{rok_zitko_2022_5874832}%
  \BibitemOpen
  \bibfield  {author} {\bibinfo {author} {\bibfnamefont {R.}~\bibnamefont
  {Zitko}},\ }\bibfield  {title} {\bibinfo {title} {{{J}osephson potentials for
  single impurity {A}nderson impurity in a junction between two
  superconductors}},\ }\href {https://doi.org/10.5281/zenodo.5874832}
  {10.5281/zenodo.5874832} (\bibinfo {year} {2022})\BibitemShut {NoStop}%
\bibitem [{\citenamefont {Schroer}\ \emph {et~al.}(2011)\citenamefont
  {Schroer}, \citenamefont {Petersson}, \citenamefont {Jung},\ and\
  \citenamefont {Petta}}]{Schroer2011}%
  \BibitemOpen
  \bibfield  {author} {\bibinfo {author} {\bibfnamefont {M.~D.}\ \bibnamefont
  {Schroer}}, \bibinfo {author} {\bibfnamefont {K.~D.}\ \bibnamefont
  {Petersson}}, \bibinfo {author} {\bibfnamefont {M.}~\bibnamefont {Jung}},\
  and\ \bibinfo {author} {\bibfnamefont {J.~R.}\ \bibnamefont {Petta}},\
  }\bibfield  {title} {\bibinfo {title} {Field tuning the $g$ factor in {InAs}
  nanowire double quantum dots},\ }\href
  {https://doi.org/10.1103/physrevlett.107.176811} {\bibfield  {journal}
  {\bibinfo  {journal} {Phys. Rev. Lett.}\ }\textbf {\bibinfo {volume} {107}},\
  \bibinfo {pages} {176811} (\bibinfo {year} {2011})}\BibitemShut {NoStop}%
\bibitem [{\citenamefont {Kringh{\o}j}\ \emph {et~al.}(2018)\citenamefont
  {Kringh{\o}j}, \citenamefont {Casparis}, \citenamefont {Hell}, \citenamefont
  {Larsen}, \citenamefont {Kuemmeth}, \citenamefont {Leijnse}, \citenamefont
  {Flensberg}, \citenamefont {Krogstrup}, \citenamefont {Nyg{\aa}rd},
  \citenamefont {Petersson},\ and\ \citenamefont {Marcus}}]{Kringhoj2018}%
  \BibitemOpen
  \bibfield  {author} {\bibinfo {author} {\bibfnamefont {A.}~\bibnamefont
  {Kringh{\o}j}}, \bibinfo {author} {\bibfnamefont {L.}~\bibnamefont
  {Casparis}}, \bibinfo {author} {\bibfnamefont {M.}~\bibnamefont {Hell}},
  \bibinfo {author} {\bibfnamefont {T.~W.}\ \bibnamefont {Larsen}}, \bibinfo
  {author} {\bibfnamefont {F.}~\bibnamefont {Kuemmeth}}, \bibinfo {author}
  {\bibfnamefont {M.}~\bibnamefont {Leijnse}}, \bibinfo {author} {\bibfnamefont
  {K.}~\bibnamefont {Flensberg}}, \bibinfo {author} {\bibfnamefont
  {P.}~\bibnamefont {Krogstrup}}, \bibinfo {author} {\bibfnamefont
  {J.}~\bibnamefont {Nyg{\aa}rd}}, \bibinfo {author} {\bibfnamefont {K.~D.}\
  \bibnamefont {Petersson}},\ and\ \bibinfo {author} {\bibfnamefont {C.~M.}\
  \bibnamefont {Marcus}},\ }\bibfield  {title} {\bibinfo {title} {Anharmonicity
  of a superconducting qubit with a few-mode {J}osephson junction},\ }\href
  {https://doi.org/10.1103/physrevb.97.060508} {\bibfield  {journal} {\bibinfo
  {journal} {Phys. Rev. B}\ }\textbf {\bibinfo {volume} {97}},\ \bibinfo
  {pages} {060508} (\bibinfo {year} {2018})}\BibitemShut {NoStop}%
\bibitem [{\citenamefont {Antipov}\ \emph {et~al.}(2018)\citenamefont
  {Antipov}, \citenamefont {Bargerbos}, \citenamefont {Winkler}, \citenamefont
  {Bauer}, \citenamefont {Rossi},\ and\ \citenamefont {Lutchyn}}]{Antipov2018}%
  \BibitemOpen
  \bibfield  {author} {\bibinfo {author} {\bibfnamefont {A.~E.}\ \bibnamefont
  {Antipov}}, \bibinfo {author} {\bibfnamefont {A.}~\bibnamefont {Bargerbos}},
  \bibinfo {author} {\bibfnamefont {G.~W.}\ \bibnamefont {Winkler}}, \bibinfo
  {author} {\bibfnamefont {B.}~\bibnamefont {Bauer}}, \bibinfo {author}
  {\bibfnamefont {E.}~\bibnamefont {Rossi}},\ and\ \bibinfo {author}
  {\bibfnamefont {R.~M.}\ \bibnamefont {Lutchyn}},\ }\bibfield  {title}
  {\bibinfo {title} {Effects of gate-induced electric fields on semiconductor
  {M}ajorana nanowires},\ }\href {https://doi.org/10.1103/physrevx.8.031041}
  {\bibfield  {journal} {\bibinfo  {journal} {Phys. Rev. X}\ }\textbf {\bibinfo
  {volume} {8}},\ \bibinfo {pages} {031041} (\bibinfo {year}
  {2018})}\BibitemShut {NoStop}%
\bibitem [{\citenamefont {Winkler}\ \emph {et~al.}(2019)\citenamefont
  {Winkler}, \citenamefont {Antipov}, \citenamefont {van Heck}, \citenamefont
  {Soluyanov}, \citenamefont {Glazman}, \citenamefont {Wimmer},\ and\
  \citenamefont {Lutchyn}}]{Winkler2019}%
  \BibitemOpen
  \bibfield  {author} {\bibinfo {author} {\bibfnamefont {G.~W.}\ \bibnamefont
  {Winkler}}, \bibinfo {author} {\bibfnamefont {A.~E.}\ \bibnamefont
  {Antipov}}, \bibinfo {author} {\bibfnamefont {B.}~\bibnamefont {van Heck}},
  \bibinfo {author} {\bibfnamefont {A.~A.}\ \bibnamefont {Soluyanov}}, \bibinfo
  {author} {\bibfnamefont {L.~I.}\ \bibnamefont {Glazman}}, \bibinfo {author}
  {\bibfnamefont {M.}~\bibnamefont {Wimmer}},\ and\ \bibinfo {author}
  {\bibfnamefont {R.~M.}\ \bibnamefont {Lutchyn}},\ }\bibfield  {title}
  {\bibinfo {title} {Unified numerical approach to topological
  semiconductor-superconductor heterostructures},\ }\href
  {https://doi.org/10.1103/physrevb.99.245408} {\bibfield  {journal} {\bibinfo
  {journal} {Phys. Rev. B}\ }\textbf {\bibinfo {volume} {99}},\ \bibinfo
  {pages} {245408} (\bibinfo {year} {2019})}\BibitemShut {NoStop}%
\bibitem [{\citenamefont {Kringh{\o}j}\ \emph {et~al.}(2020)\citenamefont
  {Kringh{\o}j}, \citenamefont {Larsen}, \citenamefont {van Heck},
  \citenamefont {Sabonis}, \citenamefont {Erlandsson}, \citenamefont
  {Petkovic}, \citenamefont {Pikulin}, \citenamefont {Krogstrup}, \citenamefont
  {Petersson},\ and\ \citenamefont {Marcus}}]{Kringhoj2020b}%
  \BibitemOpen
  \bibfield  {author} {\bibinfo {author} {\bibfnamefont {A.}~\bibnamefont
  {Kringh{\o}j}}, \bibinfo {author} {\bibfnamefont {T.}~\bibnamefont {Larsen}},
  \bibinfo {author} {\bibfnamefont {B.}~\bibnamefont {van Heck}}, \bibinfo
  {author} {\bibfnamefont {D.}~\bibnamefont {Sabonis}}, \bibinfo {author}
  {\bibfnamefont {O.}~\bibnamefont {Erlandsson}}, \bibinfo {author}
  {\bibfnamefont {I.}~\bibnamefont {Petkovic}}, \bibinfo {author}
  {\bibfnamefont {D.}~\bibnamefont {Pikulin}}, \bibinfo {author} {\bibfnamefont
  {P.}~\bibnamefont {Krogstrup}}, \bibinfo {author} {\bibfnamefont
  {K.}~\bibnamefont {Petersson}},\ and\ \bibinfo {author} {\bibfnamefont
  {C.}~\bibnamefont {Marcus}},\ }\bibfield  {title} {\bibinfo {title}
  {Controlled dc monitoring of a superconducting qubit},\ }\href
  {https://doi.org/10.1103/physrevlett.124.056801} {\bibfield  {journal}
  {\bibinfo  {journal} {Phys. Rev. Lett.}\ }\textbf {\bibinfo {volume} {124}},\
  \bibinfo {pages} {056801} (\bibinfo {year} {2020})}\BibitemShut {NoStop}%
\bibitem [{\citenamefont {de~Jong}\ \emph {et~al.}(2021)\citenamefont
  {de~Jong}, \citenamefont {Prosko}, \citenamefont {Waardenburg}, \citenamefont
  {Han}, \citenamefont {Malinowski}, \citenamefont {Krogstrup}, \citenamefont
  {Kouwenhoven}, \citenamefont {Koski},\ and\ \citenamefont
  {Pfaff}}]{Jong2021}%
  \BibitemOpen
  \bibfield  {author} {\bibinfo {author} {\bibfnamefont {D.}~\bibnamefont
  {de~Jong}}, \bibinfo {author} {\bibfnamefont {C.~G.}\ \bibnamefont {Prosko}},
  \bibinfo {author} {\bibfnamefont {D.~M.~A.}\ \bibnamefont {Waardenburg}},
  \bibinfo {author} {\bibfnamefont {L.}~\bibnamefont {Han}}, \bibinfo {author}
  {\bibfnamefont {F.~K.}\ \bibnamefont {Malinowski}}, \bibinfo {author}
  {\bibfnamefont {P.}~\bibnamefont {Krogstrup}}, \bibinfo {author}
  {\bibfnamefont {L.~P.}\ \bibnamefont {Kouwenhoven}}, \bibinfo {author}
  {\bibfnamefont {J.~V.}\ \bibnamefont {Koski}},\ and\ \bibinfo {author}
  {\bibfnamefont {W.}~\bibnamefont {Pfaff}},\ }\bibfield  {title} {\bibinfo
  {title} {Rapid microwave-only characterization and readout of quantum dots
  using multiplexed gigahertz-frequency resonators},\ }\href
  {https://doi.org/10.1103/physrevapplied.16.014007} {\bibfield  {journal}
  {\bibinfo  {journal} {Phys. Rev. Applied}\ }\textbf {\bibinfo {volume}
  {16}},\ \bibinfo {pages} {014007} (\bibinfo {year} {2021})}\BibitemShut
  {NoStop}%
\bibitem [{\citenamefont {Spanton}\ \emph {et~al.}(2017)\citenamefont
  {Spanton}, \citenamefont {Deng}, \citenamefont {Vaitiek{\.{e}}nas},
  \citenamefont {Krogstrup}, \citenamefont {Nyg{\aa}rd}, \citenamefont
  {Marcus},\ and\ \citenamefont {Moler}}]{Spanton2017}%
  \BibitemOpen
  \bibfield  {author} {\bibinfo {author} {\bibfnamefont {E.~M.}\ \bibnamefont
  {Spanton}}, \bibinfo {author} {\bibfnamefont {M.}~\bibnamefont {Deng}},
  \bibinfo {author} {\bibfnamefont {S.}~\bibnamefont {Vaitiek{\.{e}}nas}},
  \bibinfo {author} {\bibfnamefont {P.}~\bibnamefont {Krogstrup}}, \bibinfo
  {author} {\bibfnamefont {J.}~\bibnamefont {Nyg{\aa}rd}}, \bibinfo {author}
  {\bibfnamefont {C.~M.}\ \bibnamefont {Marcus}},\ and\ \bibinfo {author}
  {\bibfnamefont {K.~A.}\ \bibnamefont {Moler}},\ }\bibfield  {title} {\bibinfo
  {title} {Current{\textendash}phase relations of few-mode {InAs} nanowire
  {J}osephson junctions},\ }\href {https://doi.org/10.1038/nphys4224}
  {\bibfield  {journal} {\bibinfo  {journal} {Nature Physics}\ }\textbf
  {\bibinfo {volume} {13}},\ \bibinfo {pages} {1177} (\bibinfo {year}
  {2017})}\BibitemShut {NoStop}%
\bibitem [{\citenamefont {Hart}\ \emph {et~al.}(2019)\citenamefont {Hart},
  \citenamefont {Cui}, \citenamefont {M{\'{e}}nard}, \citenamefont {Deng},
  \citenamefont {Antipov}, \citenamefont {Lutchyn}, \citenamefont {Krogstrup},
  \citenamefont {Marcus},\ and\ \citenamefont {Moler}}]{Hart2019}%
  \BibitemOpen
  \bibfield  {author} {\bibinfo {author} {\bibfnamefont {S.}~\bibnamefont
  {Hart}}, \bibinfo {author} {\bibfnamefont {Z.}~\bibnamefont {Cui}}, \bibinfo
  {author} {\bibfnamefont {G.}~\bibnamefont {M{\'{e}}nard}}, \bibinfo {author}
  {\bibfnamefont {M.}~\bibnamefont {Deng}}, \bibinfo {author} {\bibfnamefont
  {A.~E.}\ \bibnamefont {Antipov}}, \bibinfo {author} {\bibfnamefont {R.~M.}\
  \bibnamefont {Lutchyn}}, \bibinfo {author} {\bibfnamefont {P.}~\bibnamefont
  {Krogstrup}}, \bibinfo {author} {\bibfnamefont {C.~M.}\ \bibnamefont
  {Marcus}},\ and\ \bibinfo {author} {\bibfnamefont {K.~A.}\ \bibnamefont
  {Moler}},\ }\bibfield  {title} {\bibinfo {title} {Current-phase relations of
  {InAs} nanowire {J}osephson junctions: From interacting to multimode
  regimes},\ }\href {https://doi.org/10.1103/physrevb.100.064523} {\bibfield
  {journal} {\bibinfo  {journal} {Phys. Rev. B}\ }\textbf {\bibinfo {volume}
  {100}},\ \bibinfo {pages} {064523} (\bibinfo {year} {2019})}\BibitemShut
  {NoStop}%
\bibitem [{\citenamefont {Splitthoff}\ \emph {et~al.}(2022)\citenamefont
  {Splitthoff}, \citenamefont {Bargerbos}, \citenamefont {Grunhaupt},
  \citenamefont {Pita-Vidal}, \citenamefont {Wesdorp}, \citenamefont {Liu},
  \citenamefont {Kou}, \citenamefont {Andersen},\ and\ \citenamefont {van
  Heck}}]{Splitthoff2022}%
  \BibitemOpen
  \bibfield  {author} {\bibinfo {author} {\bibfnamefont {L.~J.}\ \bibnamefont
  {Splitthoff}}, \bibinfo {author} {\bibfnamefont {A.}~\bibnamefont
  {Bargerbos}}, \bibinfo {author} {\bibfnamefont {L.}~\bibnamefont
  {Grunhaupt}}, \bibinfo {author} {\bibfnamefont {M.}~\bibnamefont
  {Pita-Vidal}}, \bibinfo {author} {\bibfnamefont {J.}~\bibnamefont {Wesdorp}},
  \bibinfo {author} {\bibfnamefont {Y.}~\bibnamefont {Liu}}, \bibinfo {author}
  {\bibfnamefont {A.}~\bibnamefont {Kou}}, \bibinfo {author} {\bibfnamefont
  {C.~K.}\ \bibnamefont {Andersen}},\ and\ \bibinfo {author} {\bibfnamefont
  {B.}~\bibnamefont {van Heck}},\ }\bibfield  {title} {\bibinfo {title}
  {Gate-tunable kinetic inductance in proximitized nanowires},\ }\href@noop {}
  {\bibfield  {journal} {\bibinfo  {journal} {arXiv e-prints}\ } (\bibinfo
  {year} {2022})},\ \Eprint {https://arxiv.org/abs/2202.08729}
  {arXiv:2202.08729} \BibitemShut {NoStop}%
\bibitem [{\citenamefont {Welch}(1967)}]{Welch1967}%
  \BibitemOpen
  \bibfield  {author} {\bibinfo {author} {\bibfnamefont {P.}~\bibnamefont
  {Welch}},\ }\bibfield  {title} {\bibinfo {title} {The use of fast fourier
  transform for the estimation of power spectra: A method based on time
  averaging over short, modified periodograms},\ }\href
  {https://doi.org/10.1109/tau.1967.1161901} {\bibfield  {journal} {\bibinfo
  {journal} {{IEEE} Transactions on Audio and Electroacoustics}\ }\textbf
  {\bibinfo {volume} {15}},\ \bibinfo {pages} {70} (\bibinfo {year}
  {1967})}\BibitemShut {NoStop}%
\bibitem [{\citenamefont {Wesdorp}\ \emph {et~al.}(2022)\citenamefont
  {Wesdorp}, \citenamefont {Gr{\"u}nhaupt}, \citenamefont {Vaartjes},
  \citenamefont {Pita-Vidal}, \citenamefont {Bargerbos}, \citenamefont
  {Splitthoff}, \citenamefont {van Heck},\ and\ \citenamefont
  {de~Lange}}]{Wesdorp2021b}%
  \BibitemOpen
  \bibfield  {author} {\bibinfo {author} {\bibfnamefont {J.}~\bibnamefont
  {Wesdorp}}, \bibinfo {author} {\bibfnamefont {L.}~\bibnamefont
  {Gr{\"u}nhaupt}}, \bibinfo {author} {\bibfnamefont {A.}~\bibnamefont
  {Vaartjes}}, \bibinfo {author} {\bibfnamefont {M.}~\bibnamefont
  {Pita-Vidal}}, \bibinfo {author} {\bibfnamefont {A.}~\bibnamefont
  {Bargerbos}}, \bibinfo {author} {\bibfnamefont {L.~J.}\ \bibnamefont
  {Splitthoff}}, \bibinfo {author} {\bibfnamefont {B.}~\bibnamefont {van
  Heck}},\ and\ \bibinfo {author} {\bibfnamefont {G.}~\bibnamefont
  {de~Lange}},\ }\bibfield  {title} {\bibinfo {title} {{A}ndreev spectroscopy
  of an {InAs} junction in a magnetic field},\ }\href@noop {} {\bibfield
  {journal} {\bibinfo  {journal} {In preparation}\ } (\bibinfo {year}
  {2022})}\BibitemShut {NoStop}%
\end{thebibliography}%


\begin{thebibliography}{26}%
\makeatletter
\providecommand \@ifxundefined [1]{%
 \@ifx{#1\undefined}
}%
\providecommand \@ifnum [1]{%
 \ifnum #1\expandafter \@firstoftwo
 \else \expandafter \@secondoftwo
 \fi
}%
\providecommand \@ifx [1]{%
 \ifx #1\expandafter \@firstoftwo
 \else \expandafter \@secondoftwo
 \fi
}%
\providecommand \natexlab [1]{#1}%
\providecommand \enquote  [1]{``#1''}%
\providecommand \bibnamefont  [1]{#1}%
\providecommand \bibfnamefont [1]{#1}%
\providecommand \citenamefont [1]{#1}%
\providecommand \href@noop [0]{\@secondoftwo}%
\providecommand \href [0]{\begingroup \@sanitize@url \@href}%
\providecommand \@href[1]{\@@startlink{#1}\@@href}%
\providecommand \@@href[1]{\endgroup#1\@@endlink}%
\providecommand \@sanitize@url [0]{\catcode `\\12\catcode `\$12\catcode
  `\&12\catcode `\#12\catcode `\^12\catcode `\_12\catcode `\%12\relax}%
\providecommand \@@startlink[1]{}%
\providecommand \@@endlink[0]{}%
\providecommand \url  [0]{\begingroup\@sanitize@url \@url }%
\providecommand \@url [1]{\endgroup\@href {#1}{\urlprefix }}%
\providecommand \urlprefix  [0]{URL }%
\providecommand \Eprint [0]{\href }%
\providecommand \doibase [0]{https://doi.org/}%
\providecommand \selectlanguage [0]{\@gobble}%
\providecommand \bibinfo  [0]{\@secondoftwo}%
\providecommand \bibfield  [0]{\@secondoftwo}%
\providecommand \translation [1]{[#1]}%
\providecommand \BibitemOpen [0]{}%
\providecommand \bibitemStop [0]{}%
\providecommand \bibitemNoStop [0]{.\EOS\space}%
\providecommand \EOS [0]{\spacefactor3000\relax}%
\providecommand \BibitemShut  [1]{\csname bibitem#1\endcsname}%
\let\auto@bib@innerbib\@empty
\bibitem [{\citenamefont {Wilson}(1975)}]{wilson1975}%
  \BibitemOpen
  \bibfield  {author} {\bibinfo {author} {\bibfnamefont {K.~G.}\ \bibnamefont
  {Wilson}},\ }\bibfield  {title} {\bibinfo {title} {The renormalization group:
  {Critical} phenomena and the {{K}ondo} problem},\ }\href
  {https://doi.org/10.1103/revmodphys.47.773} {\bibfield  {journal} {\bibinfo
  {journal} {Rev. Mod. Phys.}\ }\textbf {\bibinfo {volume} {47}},\ \bibinfo
  {pages} {773} (\bibinfo {year} {1975})}\BibitemShut {NoStop}%
\bibitem [{\citenamefont {Krishna-murthy}\ \emph {et~al.}(1980)\citenamefont
  {Krishna-murthy}, \citenamefont {Wilkins},\ and\ \citenamefont
  {Wilson}}]{krishna1980a}%
  \BibitemOpen
  \bibfield  {author} {\bibinfo {author} {\bibfnamefont {H.}~\bibnamefont
  {Krishna-murthy}}, \bibinfo {author} {\bibfnamefont {J.}~\bibnamefont
  {Wilkins}},\ and\ \bibinfo {author} {\bibfnamefont {K.}~\bibnamefont
  {Wilson}},\ }\bibfield  {title} {\bibinfo {title} {Renormalization-group
  approach to the {A}nderson model of dilute magnetic alloys. i. static
  properties for the symmetric case},\ }\href
  {https://doi.org/10.1103/physrevb.21.1003} {\bibfield  {journal} {\bibinfo
  {journal} {Phys. Rev. B}\ }\textbf {\bibinfo {volume} {21}},\ \bibinfo
  {pages} {1003} (\bibinfo {year} {1980})}\BibitemShut {NoStop}%
\bibitem [{\citenamefont {Satori}\ \emph {et~al.}(1992)\citenamefont {Satori},
  \citenamefont {Shiba}, \citenamefont {Sakai},\ and\ \citenamefont
  {Shimizu}}]{satori1992}%
  \BibitemOpen
  \bibfield  {author} {\bibinfo {author} {\bibfnamefont {K.}~\bibnamefont
  {Satori}}, \bibinfo {author} {\bibfnamefont {H.}~\bibnamefont {Shiba}},
  \bibinfo {author} {\bibfnamefont {O.}~\bibnamefont {Sakai}},\ and\ \bibinfo
  {author} {\bibfnamefont {Y.}~\bibnamefont {Shimizu}},\ }\bibfield  {title}
  {\bibinfo {title} {Numerical renormalization group study of magnetic
  impurities in superconductors},\ }\href
  {https://doi.org/10.1143/jpsj.61.3239} {\bibfield  {journal} {\bibinfo
  {journal} {J. Phys. Soc. Japan}\ }\textbf {\bibinfo {volume} {61}},\ \bibinfo
  {pages} {3239} (\bibinfo {year} {1992})}\BibitemShut {NoStop}%
\bibitem [{\citenamefont {Yoshioka}\ and\ \citenamefont
  {Ohashi}(2000)}]{yoshioka2000}%
  \BibitemOpen
  \bibfield  {author} {\bibinfo {author} {\bibfnamefont {T.}~\bibnamefont
  {Yoshioka}}\ and\ \bibinfo {author} {\bibfnamefont {Y.}~\bibnamefont
  {Ohashi}},\ }\bibfield  {title} {\bibinfo {title} {Numerical renormalization
  group studies on single impurity {{A}nderson} model in superconductivity: a
  unified treatment of magnetic, nonmagnetic impurities, and resonance
  scattering},\ }\href {https://doi.org/10.1143/jpsj.69.1812} {\bibfield
  {journal} {\bibinfo  {journal} {J. Phys. Soc. Japan}\ }\textbf {\bibinfo
  {volume} {69}},\ \bibinfo {pages} {1812} (\bibinfo {year}
  {2000})}\BibitemShut {NoStop}%
\bibitem [{\citenamefont {Bulla}\ \emph {et~al.}(2008)\citenamefont {Bulla},
  \citenamefont {Costi},\ and\ \citenamefont {Pruschke}}]{bulla2008}%
  \BibitemOpen
  \bibfield  {author} {\bibinfo {author} {\bibfnamefont {R.}~\bibnamefont
  {Bulla}}, \bibinfo {author} {\bibfnamefont {T.~A.}\ \bibnamefont {Costi}},\
  and\ \bibinfo {author} {\bibfnamefont {T.}~\bibnamefont {Pruschke}},\
  }\bibfield  {title} {\bibinfo {title} {The numerical renormalization group
  method for quantum impurity systems},\ }\href
  {https://doi.org/10.1103/revmodphys.80.395} {\bibfield  {journal} {\bibinfo
  {journal} {Rev. Mod. Phys.}\ }\textbf {\bibinfo {volume} {80}},\ \bibinfo
  {pages} {395} (\bibinfo {year} {2008})}\BibitemShut {NoStop}%
\bibitem [{\citenamefont {\v{Z}itko}\ and\ \citenamefont
  {Pruschke}(2009)}]{zitko2009}%
  \BibitemOpen
  \bibfield  {author} {\bibinfo {author} {\bibfnamefont {R.}~\bibnamefont
  {\v{Z}itko}}\ and\ \bibinfo {author} {\bibfnamefont {T.}~\bibnamefont
  {Pruschke}},\ }\bibfield  {title} {\bibinfo {title} {Energy resolution and
  discretization artefacts in the numerical renormalization group},\
  }\href@noop {} {\bibfield  {journal} {\bibinfo  {journal} {Phys. Rev. B}\
  }\textbf {\bibinfo {volume} {79}},\ \bibinfo {pages} {085106} (\bibinfo
  {year} {2009})}\BibitemShut {NoStop}%
\bibitem [{\citenamefont {Zitko}(2021)}]{zitko_rok_2021_4841076}%
  \BibitemOpen
  \bibfield  {author} {\bibinfo {author} {\bibfnamefont {R.}~\bibnamefont
  {Zitko}},\ }\href {https://doi.org/10.5281/zenodo.4841076} {\bibinfo {title}
  {{NRG} {Ljubljana}}} (\bibinfo {year} {2021})\BibitemShut {NoStop}%
\bibitem [{\citenamefont {Kadlecov{\'{a}}}\ \emph {et~al.}(2017)\citenamefont
  {Kadlecov{\'{a}}}, \citenamefont {{\v{Z}}onda},\ and\ \citenamefont
  {Novotn{\'{y}}}}]{kadlecova2017}%
  \BibitemOpen
  \bibfield  {author} {\bibinfo {author} {\bibfnamefont {A.}~\bibnamefont
  {Kadlecov{\'{a}}}}, \bibinfo {author} {\bibfnamefont {M.}~\bibnamefont
  {{\v{Z}}onda}},\ and\ \bibinfo {author} {\bibfnamefont {T.}~\bibnamefont
  {Novotn{\'{y}}}},\ }\bibfield  {title} {\bibinfo {title} {Quantum dot
  attached to superconducting leads: Relation between symmetric and asymmetric
  coupling},\ }\href {https://doi.org/10.1103/physrevb.95.195114} {\bibfield
  {journal} {\bibinfo  {journal} {Phys. Rev. B}\ }\textbf {\bibinfo {volume}
  {95}},\ \bibinfo {pages} {195114} (\bibinfo {year} {2017})}\BibitemShut
  {NoStop}%
\bibitem [{\citenamefont {Zitko}(2022)}]{rok_zitko_2022_5874832}%
  \BibitemOpen
  \bibfield  {author} {\bibinfo {author} {\bibfnamefont {R.}~\bibnamefont
  {Zitko}},\ }\bibfield  {title} {\bibinfo {title} {{{J}osephson potentials for
  single impurity {A}nderson impurity in a junction between two
  superconductors}},\ }\href {https://doi.org/10.5281/zenodo.5874832}
  {10.5281/zenodo.5874832} (\bibinfo {year} {2022})\BibitemShut {NoStop}%
\bibitem [{\citenamefont {Koch}\ \emph {et~al.}(2007)\citenamefont {Koch},
  \citenamefont {Yu}, \citenamefont {Gambetta}, \citenamefont {Houck},
  \citenamefont {Schuster}, \citenamefont {Majer}, \citenamefont {Blais},
  \citenamefont {Devoret}, \citenamefont {Girvin},\ and\ \citenamefont
  {Schoelkopf}}]{Koch2007}%
  \BibitemOpen
  \bibfield  {author} {\bibinfo {author} {\bibfnamefont {J.}~\bibnamefont
  {Koch}}, \bibinfo {author} {\bibfnamefont {T.~M.}\ \bibnamefont {Yu}},
  \bibinfo {author} {\bibfnamefont {J.}~\bibnamefont {Gambetta}}, \bibinfo
  {author} {\bibfnamefont {A.~A.}\ \bibnamefont {Houck}}, \bibinfo {author}
  {\bibfnamefont {D.~I.}\ \bibnamefont {Schuster}}, \bibinfo {author}
  {\bibfnamefont {J.}~\bibnamefont {Majer}}, \bibinfo {author} {\bibfnamefont
  {A.}~\bibnamefont {Blais}}, \bibinfo {author} {\bibfnamefont {M.~H.}\
  \bibnamefont {Devoret}}, \bibinfo {author} {\bibfnamefont {S.~M.}\
  \bibnamefont {Girvin}},\ and\ \bibinfo {author} {\bibfnamefont {R.~J.}\
  \bibnamefont {Schoelkopf}},\ }\bibfield  {title} {\bibinfo {title}
  {Charge-insensitive qubit design derived from the cooper pair box},\ }\href
  {https://doi.org/10.1103/physreva.76.042319} {\bibfield  {journal} {\bibinfo
  {journal} {Phys. Rev. A}\ }\textbf {\bibinfo {volume} {76}},\ \bibinfo
  {pages} {042319} (\bibinfo {year} {2007})}\BibitemShut {NoStop}%
\bibitem [{\citenamefont {Schroer}\ \emph {et~al.}(2011)\citenamefont
  {Schroer}, \citenamefont {Petersson}, \citenamefont {Jung},\ and\
  \citenamefont {Petta}}]{Schroer2011}%
  \BibitemOpen
  \bibfield  {author} {\bibinfo {author} {\bibfnamefont {M.~D.}\ \bibnamefont
  {Schroer}}, \bibinfo {author} {\bibfnamefont {K.~D.}\ \bibnamefont
  {Petersson}}, \bibinfo {author} {\bibfnamefont {M.}~\bibnamefont {Jung}},\
  and\ \bibinfo {author} {\bibfnamefont {J.~R.}\ \bibnamefont {Petta}},\
  }\bibfield  {title} {\bibinfo {title} {Field tuning the $g$ factor in {InAs}
  nanowire double quantum dots},\ }\href
  {https://doi.org/10.1103/physrevlett.107.176811} {\bibfield  {journal}
  {\bibinfo  {journal} {Phys. Rev. Lett.}\ }\textbf {\bibinfo {volume} {107}},\
  \bibinfo {pages} {176811} (\bibinfo {year} {2011})}\BibitemShut {NoStop}%
\bibitem [{\citenamefont {Kringh{\o}j}\ \emph {et~al.}(2018)\citenamefont
  {Kringh{\o}j}, \citenamefont {Casparis}, \citenamefont {Hell}, \citenamefont
  {Larsen}, \citenamefont {Kuemmeth}, \citenamefont {Leijnse}, \citenamefont
  {Flensberg}, \citenamefont {Krogstrup}, \citenamefont {Nyg{\aa}rd},
  \citenamefont {Petersson},\ and\ \citenamefont {Marcus}}]{Kringhoj2018}%
  \BibitemOpen
  \bibfield  {author} {\bibinfo {author} {\bibfnamefont {A.}~\bibnamefont
  {Kringh{\o}j}}, \bibinfo {author} {\bibfnamefont {L.}~\bibnamefont
  {Casparis}}, \bibinfo {author} {\bibfnamefont {M.}~\bibnamefont {Hell}},
  \bibinfo {author} {\bibfnamefont {T.~W.}\ \bibnamefont {Larsen}}, \bibinfo
  {author} {\bibfnamefont {F.}~\bibnamefont {Kuemmeth}}, \bibinfo {author}
  {\bibfnamefont {M.}~\bibnamefont {Leijnse}}, \bibinfo {author} {\bibfnamefont
  {K.}~\bibnamefont {Flensberg}}, \bibinfo {author} {\bibfnamefont
  {P.}~\bibnamefont {Krogstrup}}, \bibinfo {author} {\bibfnamefont
  {J.}~\bibnamefont {Nyg{\aa}rd}}, \bibinfo {author} {\bibfnamefont {K.~D.}\
  \bibnamefont {Petersson}},\ and\ \bibinfo {author} {\bibfnamefont {C.~M.}\
  \bibnamefont {Marcus}},\ }\bibfield  {title} {\bibinfo {title} {Anharmonicity
  of a superconducting qubit with a few-mode {J}osephson junction},\ }\href
  {https://doi.org/10.1103/physrevb.97.060508} {\bibfield  {journal} {\bibinfo
  {journal} {Phys. Rev. B}\ }\textbf {\bibinfo {volume} {97}},\ \bibinfo
  {pages} {060508} (\bibinfo {year} {2018})}\BibitemShut {NoStop}%
\bibitem [{\citenamefont {Antipov}\ \emph {et~al.}(2018)\citenamefont
  {Antipov}, \citenamefont {Bargerbos}, \citenamefont {Winkler}, \citenamefont
  {Bauer}, \citenamefont {Rossi},\ and\ \citenamefont {Lutchyn}}]{Antipov2018}%
  \BibitemOpen
  \bibfield  {author} {\bibinfo {author} {\bibfnamefont {A.~E.}\ \bibnamefont
  {Antipov}}, \bibinfo {author} {\bibfnamefont {A.}~\bibnamefont {Bargerbos}},
  \bibinfo {author} {\bibfnamefont {G.~W.}\ \bibnamefont {Winkler}}, \bibinfo
  {author} {\bibfnamefont {B.}~\bibnamefont {Bauer}}, \bibinfo {author}
  {\bibfnamefont {E.}~\bibnamefont {Rossi}},\ and\ \bibinfo {author}
  {\bibfnamefont {R.~M.}\ \bibnamefont {Lutchyn}},\ }\bibfield  {title}
  {\bibinfo {title} {Effects of gate-induced electric fields on semiconductor
  {M}ajorana nanowires},\ }\href {https://doi.org/10.1103/physrevx.8.031041}
  {\bibfield  {journal} {\bibinfo  {journal} {Phys. Rev. X}\ }\textbf {\bibinfo
  {volume} {8}},\ \bibinfo {pages} {031041} (\bibinfo {year}
  {2018})}\BibitemShut {NoStop}%
\bibitem [{\citenamefont {Winkler}\ \emph {et~al.}(2019)\citenamefont
  {Winkler}, \citenamefont {Antipov}, \citenamefont {van Heck}, \citenamefont
  {Soluyanov}, \citenamefont {Glazman}, \citenamefont {Wimmer},\ and\
  \citenamefont {Lutchyn}}]{Winkler2019}%
  \BibitemOpen
  \bibfield  {author} {\bibinfo {author} {\bibfnamefont {G.~W.}\ \bibnamefont
  {Winkler}}, \bibinfo {author} {\bibfnamefont {A.~E.}\ \bibnamefont
  {Antipov}}, \bibinfo {author} {\bibfnamefont {B.}~\bibnamefont {van Heck}},
  \bibinfo {author} {\bibfnamefont {A.~A.}\ \bibnamefont {Soluyanov}}, \bibinfo
  {author} {\bibfnamefont {L.~I.}\ \bibnamefont {Glazman}}, \bibinfo {author}
  {\bibfnamefont {M.}~\bibnamefont {Wimmer}},\ and\ \bibinfo {author}
  {\bibfnamefont {R.~M.}\ \bibnamefont {Lutchyn}},\ }\bibfield  {title}
  {\bibinfo {title} {Unified numerical approach to topological
  semiconductor-superconductor heterostructures},\ }\href
  {https://doi.org/10.1103/physrevb.99.245408} {\bibfield  {journal} {\bibinfo
  {journal} {Phys. Rev. B}\ }\textbf {\bibinfo {volume} {99}},\ \bibinfo
  {pages} {245408} (\bibinfo {year} {2019})}\BibitemShut {NoStop}%
\bibitem [{\citenamefont {Kringh{\o}j}\ \emph {et~al.}(2020)\citenamefont
  {Kringh{\o}j}, \citenamefont {Larsen}, \citenamefont {van Heck},
  \citenamefont {Sabonis}, \citenamefont {Erlandsson}, \citenamefont
  {Petkovic}, \citenamefont {Pikulin}, \citenamefont {Krogstrup}, \citenamefont
  {Petersson},\ and\ \citenamefont {Marcus}}]{Kringhoj2020b}%
  \BibitemOpen
  \bibfield  {author} {\bibinfo {author} {\bibfnamefont {A.}~\bibnamefont
  {Kringh{\o}j}}, \bibinfo {author} {\bibfnamefont {T.}~\bibnamefont {Larsen}},
  \bibinfo {author} {\bibfnamefont {B.}~\bibnamefont {van Heck}}, \bibinfo
  {author} {\bibfnamefont {D.}~\bibnamefont {Sabonis}}, \bibinfo {author}
  {\bibfnamefont {O.}~\bibnamefont {Erlandsson}}, \bibinfo {author}
  {\bibfnamefont {I.}~\bibnamefont {Petkovic}}, \bibinfo {author}
  {\bibfnamefont {D.}~\bibnamefont {Pikulin}}, \bibinfo {author} {\bibfnamefont
  {P.}~\bibnamefont {Krogstrup}}, \bibinfo {author} {\bibfnamefont
  {K.}~\bibnamefont {Petersson}},\ and\ \bibinfo {author} {\bibfnamefont
  {C.}~\bibnamefont {Marcus}},\ }\bibfield  {title} {\bibinfo {title}
  {Controlled dc monitoring of a superconducting qubit},\ }\href
  {https://doi.org/10.1103/physrevlett.124.056801} {\bibfield  {journal}
  {\bibinfo  {journal} {Phys. Rev. Lett.}\ }\textbf {\bibinfo {volume} {124}},\
  \bibinfo {pages} {056801} (\bibinfo {year} {2020})}\BibitemShut {NoStop}%
\bibitem [{\citenamefont {de~Jong}\ \emph {et~al.}(2021)\citenamefont
  {de~Jong}, \citenamefont {Prosko}, \citenamefont {Waardenburg}, \citenamefont
  {Han}, \citenamefont {Malinowski}, \citenamefont {Krogstrup}, \citenamefont
  {Kouwenhoven}, \citenamefont {Koski},\ and\ \citenamefont
  {Pfaff}}]{Jong2021}%
  \BibitemOpen
  \bibfield  {author} {\bibinfo {author} {\bibfnamefont {D.}~\bibnamefont
  {de~Jong}}, \bibinfo {author} {\bibfnamefont {C.~G.}\ \bibnamefont {Prosko}},
  \bibinfo {author} {\bibfnamefont {D.~M.~A.}\ \bibnamefont {Waardenburg}},
  \bibinfo {author} {\bibfnamefont {L.}~\bibnamefont {Han}}, \bibinfo {author}
  {\bibfnamefont {F.~K.}\ \bibnamefont {Malinowski}}, \bibinfo {author}
  {\bibfnamefont {P.}~\bibnamefont {Krogstrup}}, \bibinfo {author}
  {\bibfnamefont {L.~P.}\ \bibnamefont {Kouwenhoven}}, \bibinfo {author}
  {\bibfnamefont {J.~V.}\ \bibnamefont {Koski}},\ and\ \bibinfo {author}
  {\bibfnamefont {W.}~\bibnamefont {Pfaff}},\ }\bibfield  {title} {\bibinfo
  {title} {Rapid microwave-only characterization and readout of quantum dots
  using multiplexed gigahertz-frequency resonators},\ }\href
  {https://doi.org/10.1103/physrevapplied.16.014007} {\bibfield  {journal}
  {\bibinfo  {journal} {Phys. Rev. Applied}\ }\textbf {\bibinfo {volume}
  {16}},\ \bibinfo {pages} {014007} (\bibinfo {year} {2021})}\BibitemShut
  {NoStop}%
\bibitem [{\citenamefont {Glazman}\ and\ \citenamefont
  {Catelani}(2021)}]{Glazman2021}%
  \BibitemOpen
  \bibfield  {author} {\bibinfo {author} {\bibfnamefont {L.~I.}\ \bibnamefont
  {Glazman}}\ and\ \bibinfo {author} {\bibfnamefont {G.}~\bibnamefont
  {Catelani}},\ }\bibfield  {title} {\bibinfo {title} {{Bogoliubov
  quasiparticles in superconducting qubits}},\ }\href
  {https://doi.org/10.21468/SciPostPhysLectNotes.31} {\bibfield  {journal}
  {\bibinfo  {journal} {SciPost Phys. Lect. Notes}\ ,\ \bibinfo {pages} {31}}
  (\bibinfo {year} {2021})}\BibitemShut {NoStop}%
\bibitem [{\citenamefont {Krogstrup}\ \emph {et~al.}(2015)\citenamefont
  {Krogstrup}, \citenamefont {Ziino}, \citenamefont {Chang}, \citenamefont
  {Albrecht}, \citenamefont {Madsen}, \citenamefont {Johnson}, \citenamefont
  {Nyg{\aa}rd}, \citenamefont {Marcus},\ and\ \citenamefont
  {Jespersen}}]{Krogstrup2015}%
  \BibitemOpen
  \bibfield  {author} {\bibinfo {author} {\bibfnamefont {P.}~\bibnamefont
  {Krogstrup}}, \bibinfo {author} {\bibfnamefont {N.~L.~B.}\ \bibnamefont
  {Ziino}}, \bibinfo {author} {\bibfnamefont {W.}~\bibnamefont {Chang}},
  \bibinfo {author} {\bibfnamefont {S.~M.}\ \bibnamefont {Albrecht}}, \bibinfo
  {author} {\bibfnamefont {M.~H.}\ \bibnamefont {Madsen}}, \bibinfo {author}
  {\bibfnamefont {E.}~\bibnamefont {Johnson}}, \bibinfo {author} {\bibfnamefont
  {J.}~\bibnamefont {Nyg{\aa}rd}}, \bibinfo {author} {\bibfnamefont
  {C.}~\bibnamefont {Marcus}},\ and\ \bibinfo {author} {\bibfnamefont {T.~S.}\
  \bibnamefont {Jespersen}},\ }\bibfield  {title} {\bibinfo {title} {Epitaxy of
  semiconductor-superconductor nanowires},\ }\href
  {https://doi.org/10.1038/nmat4176} {\bibfield  {journal} {\bibinfo  {journal}
  {Nat. Mater.}\ }\textbf {\bibinfo {volume} {14}},\ \bibinfo {pages} {400}
  (\bibinfo {year} {2015})}\BibitemShut {NoStop}%
\bibitem [{\citenamefont {Wesdorp}\ \emph {et~al.}(2022)\citenamefont
  {Wesdorp}, \citenamefont {Gr{\"u}nhaupt}, \citenamefont {Vaartjes},
  \citenamefont {Pita-Vidal}, \citenamefont {Bargerbos}, \citenamefont
  {Splitthoff}, \citenamefont {van Heck},\ and\ \citenamefont
  {de~Lange}}]{Wesdorp2021b}%
  \BibitemOpen
  \bibfield  {author} {\bibinfo {author} {\bibfnamefont {J.}~\bibnamefont
  {Wesdorp}}, \bibinfo {author} {\bibfnamefont {L.}~\bibnamefont
  {Gr{\"u}nhaupt}}, \bibinfo {author} {\bibfnamefont {A.}~\bibnamefont
  {Vaartjes}}, \bibinfo {author} {\bibfnamefont {M.}~\bibnamefont
  {Pita-Vidal}}, \bibinfo {author} {\bibfnamefont {A.}~\bibnamefont
  {Bargerbos}}, \bibinfo {author} {\bibfnamefont {L.~J.}\ \bibnamefont
  {Splitthoff}}, \bibinfo {author} {\bibfnamefont {B.}~\bibnamefont {van
  Heck}},\ and\ \bibinfo {author} {\bibfnamefont {G.}~\bibnamefont
  {de~Lange}},\ }\bibfield  {title} {\bibinfo {title} {{A}ndreev spectroscopy
  of an {InAs} junction in a magnetic field},\ }\href@noop {} {\bibfield
  {journal} {\bibinfo  {journal} {In preparation}\ } (\bibinfo {year}
  {2022})}\BibitemShut {NoStop}%
\bibitem [{\citenamefont {Luthi}\ \emph {et~al.}(2018)\citenamefont {Luthi},
  \citenamefont {Stavenga}, \citenamefont {Enzing}, \citenamefont {Bruno},
  \citenamefont {Dickel}, \citenamefont {Langford}, \citenamefont {Rol},
  \citenamefont {Jespersen}, \citenamefont {Nyg\aa{}rd}, \citenamefont
  {Krogstrup},\ and\ \citenamefont {DiCarlo}}]{Luthi2018}%
  \BibitemOpen
  \bibfield  {author} {\bibinfo {author} {\bibfnamefont {F.}~\bibnamefont
  {Luthi}}, \bibinfo {author} {\bibfnamefont {T.}~\bibnamefont {Stavenga}},
  \bibinfo {author} {\bibfnamefont {O.~W.}\ \bibnamefont {Enzing}}, \bibinfo
  {author} {\bibfnamefont {A.}~\bibnamefont {Bruno}}, \bibinfo {author}
  {\bibfnamefont {C.}~\bibnamefont {Dickel}}, \bibinfo {author} {\bibfnamefont
  {N.~K.}\ \bibnamefont {Langford}}, \bibinfo {author} {\bibfnamefont {M.~A.}\
  \bibnamefont {Rol}}, \bibinfo {author} {\bibfnamefont {T.~S.}\ \bibnamefont
  {Jespersen}}, \bibinfo {author} {\bibfnamefont {J.}~\bibnamefont
  {Nyg\aa{}rd}}, \bibinfo {author} {\bibfnamefont {P.}~\bibnamefont
  {Krogstrup}},\ and\ \bibinfo {author} {\bibfnamefont {L.}~\bibnamefont
  {DiCarlo}},\ }\bibfield  {title} {\bibinfo {title} {Evolution of nanowire
  transmon qubits and their coherence in a magnetic field},\ }\href
  {https://doi.org/10.1103/PhysRevLett.120.100502} {\bibfield  {journal}
  {\bibinfo  {journal} {Phys. Rev. Lett.}\ }\textbf {\bibinfo {volume} {120}},\
  \bibinfo {pages} {100502} (\bibinfo {year} {2018})}\BibitemShut {NoStop}%
\bibitem [{\citenamefont {Spanton}\ \emph {et~al.}(2017)\citenamefont
  {Spanton}, \citenamefont {Deng}, \citenamefont {Vaitiek{\.{e}}nas},
  \citenamefont {Krogstrup}, \citenamefont {Nyg{\aa}rd}, \citenamefont
  {Marcus},\ and\ \citenamefont {Moler}}]{Spanton2017}%
  \BibitemOpen
  \bibfield  {author} {\bibinfo {author} {\bibfnamefont {E.~M.}\ \bibnamefont
  {Spanton}}, \bibinfo {author} {\bibfnamefont {M.}~\bibnamefont {Deng}},
  \bibinfo {author} {\bibfnamefont {S.}~\bibnamefont {Vaitiek{\.{e}}nas}},
  \bibinfo {author} {\bibfnamefont {P.}~\bibnamefont {Krogstrup}}, \bibinfo
  {author} {\bibfnamefont {J.}~\bibnamefont {Nyg{\aa}rd}}, \bibinfo {author}
  {\bibfnamefont {C.~M.}\ \bibnamefont {Marcus}},\ and\ \bibinfo {author}
  {\bibfnamefont {K.~A.}\ \bibnamefont {Moler}},\ }\bibfield  {title} {\bibinfo
  {title} {Current{\textendash}phase relations of few-mode {InAs} nanowire
  {J}osephson junctions},\ }\href {https://doi.org/10.1038/nphys4224}
  {\bibfield  {journal} {\bibinfo  {journal} {Nature Physics}\ }\textbf
  {\bibinfo {volume} {13}},\ \bibinfo {pages} {1177} (\bibinfo {year}
  {2017})}\BibitemShut {NoStop}%
\bibitem [{\citenamefont {Hart}\ \emph {et~al.}(2019)\citenamefont {Hart},
  \citenamefont {Cui}, \citenamefont {M{\'{e}}nard}, \citenamefont {Deng},
  \citenamefont {Antipov}, \citenamefont {Lutchyn}, \citenamefont {Krogstrup},
  \citenamefont {Marcus},\ and\ \citenamefont {Moler}}]{Hart2019}%
  \BibitemOpen
  \bibfield  {author} {\bibinfo {author} {\bibfnamefont {S.}~\bibnamefont
  {Hart}}, \bibinfo {author} {\bibfnamefont {Z.}~\bibnamefont {Cui}}, \bibinfo
  {author} {\bibfnamefont {G.}~\bibnamefont {M{\'{e}}nard}}, \bibinfo {author}
  {\bibfnamefont {M.}~\bibnamefont {Deng}}, \bibinfo {author} {\bibfnamefont
  {A.~E.}\ \bibnamefont {Antipov}}, \bibinfo {author} {\bibfnamefont {R.~M.}\
  \bibnamefont {Lutchyn}}, \bibinfo {author} {\bibfnamefont {P.}~\bibnamefont
  {Krogstrup}}, \bibinfo {author} {\bibfnamefont {C.~M.}\ \bibnamefont
  {Marcus}},\ and\ \bibinfo {author} {\bibfnamefont {K.~A.}\ \bibnamefont
  {Moler}},\ }\bibfield  {title} {\bibinfo {title} {Current-phase relations of
  {InAs} nanowire {J}osephson junctions: From interacting to multimode
  regimes},\ }\href {https://doi.org/10.1103/physrevb.100.064523} {\bibfield
  {journal} {\bibinfo  {journal} {Phys. Rev. B}\ }\textbf {\bibinfo {volume}
  {100}},\ \bibinfo {pages} {064523} (\bibinfo {year} {2019})}\BibitemShut
  {NoStop}%
\bibitem [{\citenamefont {Splitthoff}\ \emph {et~al.}(2022)\citenamefont
  {Splitthoff}, \citenamefont {Bargerbos}, \citenamefont {Grunhaupt},
  \citenamefont {Pita-Vidal}, \citenamefont {Wesdorp}, \citenamefont {Liu},
  \citenamefont {Kou}, \citenamefont {Andersen},\ and\ \citenamefont {van
  Heck}}]{Splitthoff2022}%
  \BibitemOpen
  \bibfield  {author} {\bibinfo {author} {\bibfnamefont {L.~J.}\ \bibnamefont
  {Splitthoff}}, \bibinfo {author} {\bibfnamefont {A.}~\bibnamefont
  {Bargerbos}}, \bibinfo {author} {\bibfnamefont {L.}~\bibnamefont
  {Grunhaupt}}, \bibinfo {author} {\bibfnamefont {M.}~\bibnamefont
  {Pita-Vidal}}, \bibinfo {author} {\bibfnamefont {J.}~\bibnamefont {Wesdorp}},
  \bibinfo {author} {\bibfnamefont {Y.}~\bibnamefont {Liu}}, \bibinfo {author}
  {\bibfnamefont {A.}~\bibnamefont {Kou}}, \bibinfo {author} {\bibfnamefont
  {C.~K.}\ \bibnamefont {Andersen}},\ and\ \bibinfo {author} {\bibfnamefont
  {B.}~\bibnamefont {van Heck}},\ }\bibfield  {title} {\bibinfo {title}
  {Gate-tunable kinetic inductance in proximitized nanowires},\ }\href@noop {}
  {\bibfield  {journal} {\bibinfo  {journal} {arXiv e-prints}\ } (\bibinfo
  {year} {2022})},\ \Eprint {https://arxiv.org/abs/2202.08729}
  {arXiv:2202.08729} \BibitemShut {NoStop}%
\bibitem [{\citenamefont {Welch}(1967)}]{Welch1967}%
  \BibitemOpen
  \bibfield  {author} {\bibinfo {author} {\bibfnamefont {P.}~\bibnamefont
  {Welch}},\ }\bibfield  {title} {\bibinfo {title} {The use of fast fourier
  transform for the estimation of power spectra: A method based on time
  averaging over short, modified periodograms},\ }\href
  {https://doi.org/10.1109/tau.1967.1161901} {\bibfield  {journal} {\bibinfo
  {journal} {{IEEE} Transactions on Audio and Electroacoustics}\ }\textbf
  {\bibinfo {volume} {15}},\ \bibinfo {pages} {70} (\bibinfo {year}
  {1967})}\BibitemShut {NoStop}%
\bibitem [{\citenamefont {Pop}\ \emph {et~al.}(2014)\citenamefont {Pop},
  \citenamefont {Geerlings}, \citenamefont {Catelani}, \citenamefont
  {Schoelkopf}, \citenamefont {Glazman},\ and\ \citenamefont
  {Devoret}}]{Pop2014}%
  \BibitemOpen
  \bibfield  {author} {\bibinfo {author} {\bibfnamefont {I.~M.}\ \bibnamefont
  {Pop}}, \bibinfo {author} {\bibfnamefont {K.}~\bibnamefont {Geerlings}},
  \bibinfo {author} {\bibfnamefont {G.}~\bibnamefont {Catelani}}, \bibinfo
  {author} {\bibfnamefont {R.~J.}\ \bibnamefont {Schoelkopf}}, \bibinfo
  {author} {\bibfnamefont {L.~I.}\ \bibnamefont {Glazman}},\ and\ \bibinfo
  {author} {\bibfnamefont {M.~H.}\ \bibnamefont {Devoret}},\ }\bibfield
  {title} {\bibinfo {title} {Coherent suppression of electromagnetic
  dissipation due to superconducting quasiparticles},\ }\href
  {https://doi.org/10.1038/nature13017} {\bibfield  {journal} {\bibinfo
  {journal} {Nature}\ }\textbf {\bibinfo {volume} {508}},\ \bibinfo {pages}
  {369} (\bibinfo {year} {2014})}\BibitemShut {NoStop}%
\bibitem [{\citenamefont {Wesdorp}\ \emph {et~al.}(2021)\citenamefont
  {Wesdorp}, \citenamefont {Gr{\"u}nhaupt}, \citenamefont {Vaartjes},
  \citenamefont {Pita-Vidal}, \citenamefont {Bargerbos}, \citenamefont
  {Splitthoff}, \citenamefont {Krogstrup}, \citenamefont {van Heck},\ and\
  \citenamefont {de~Lange}}]{Wesdorp2021}%
  \BibitemOpen
  \bibfield  {author} {\bibinfo {author} {\bibfnamefont {J.~J.}\ \bibnamefont
  {Wesdorp}}, \bibinfo {author} {\bibfnamefont {L.}~\bibnamefont
  {Gr{\"u}nhaupt}}, \bibinfo {author} {\bibfnamefont {A.}~\bibnamefont
  {Vaartjes}}, \bibinfo {author} {\bibfnamefont {M.}~\bibnamefont
  {Pita-Vidal}}, \bibinfo {author} {\bibfnamefont {A.}~\bibnamefont
  {Bargerbos}}, \bibinfo {author} {\bibfnamefont {L.~J.}\ \bibnamefont
  {Splitthoff}}, \bibinfo {author} {\bibfnamefont {P.}~\bibnamefont
  {Krogstrup}}, \bibinfo {author} {\bibfnamefont {B.}~\bibnamefont {van
  Heck}},\ and\ \bibinfo {author} {\bibfnamefont {G.}~\bibnamefont
  {de~Lange}},\ }\bibfield  {title} {\bibinfo {title} {Dynamical polarization
  of the fermion parity in a nanowire {J}osephson junction},\ }\href@noop {}
  {\bibfield  {journal} {\bibinfo  {journal} {arXiv e-prints}\ } (\bibinfo
  {year} {2021})},\ \Eprint {https://arxiv.org/abs/2112.01936}
  {arXiv:2112.01936} \BibitemShut {NoStop}%
\end{thebibliography}%

\end{document}


\beginsupplement

\title{Supplementary information for ``Singlet-doublet transitions of a quantum dot Josephson junction revealed in a transmon circuit''}
\author{Arno Bargerbos}
\thanks{These two authors contributed equally.}
\affiliation{QuTech and Kavli Institute of Nanoscience, Delft University of Technology, 2600 GA Delft, The Netherlands}

\author{Marta Pita-Vidal}
\thanks{These two authors contributed equally.}
\affiliation{QuTech and Kavli Institute of Nanoscience, Delft University of Technology, 2600 GA Delft, The Netherlands}

\author{Rok Žitko}
\affiliation{Jožef Stefan Institute, Jamova 39, SI-1000 Ljubljana, Slovenia}
\affiliation{Faculty of Mathematics and Physics, University of Ljubljana, Jadranska 19, SI-1000 Ljubljana, Slovenia}

\author{Jesús Ávila}
\affiliation{Instituto de Ciencia de Materiales de Madrid (ICMM),
Consejo Superior de Investigaciones Cientificas (CSIC), Sor Juana Ines de la Cruz 3, 28049 Madrid, Spain}

\author{Lukas J. Splitthoff}
\affiliation{QuTech and Kavli Institute of Nanoscience, Delft University of Technology, 2600 GA Delft, The Netherlands}

\author{Lukas Grünhaupt}
\affiliation{QuTech and Kavli Institute of Nanoscience, Delft University of Technology, 2600 GA Delft, The Netherlands}

\author{Jaap J. Wesdorp}
\affiliation{QuTech and Kavli Institute of Nanoscience, Delft University of Technology, 2600 GA Delft, The Netherlands}

\author{Christian K. Andersen}
\affiliation{QuTech and Kavli Institute of Nanoscience, Delft University of Technology, 2600 GA Delft, The Netherlands}

\author{Yu Liu}
\affiliation{Center for Quantum Devices, Niels Bohr Institute, University of Copenhagen, 2100 Copenhagen, Denmark}

\author{Leo P. Kouwenhoven}
\affiliation{QuTech and Kavli Institute of Nanoscience, Delft University of Technology, 2600 GA Delft, The Netherlands}

\author{Ramón Aguado}
\affiliation{Instituto de Ciencia de Materiales de Madrid (ICMM),
Consejo Superior de Investigaciones Cientificas (CSIC), Sor Juana Ines de la Cruz 3, 28049 Madrid, Spain}

\author{Angela Kou}
\affiliation{Department of Physics and Frederick Seitz Materials Research Laboratory,
University of Illinois Urbana-Champaign, Urbana, IL 61801, USA}

\author{Bernard van Heck}
\affiliation{Leiden Institute of Physics, Leiden University, Niels Bohrweg 2, 2333 CA Leiden, The Netherlands}

\date{\today}

\maketitle

\tableofcontents

\vspace{2 cm}

\newpage

\section{Numerical modeling}
As discussed in the main text, we model the quantum dot junction as a single Anderson impurity coupled to two superconducting leads. Using numerical renormalization group (NRG) methods we extract the energies of the singlet and doublet states for any combination of the model parameters, and subsequently incorporate them into a DC SQUID transmon Hamiltonian. This Hamiltonian is then used to match the experimental data and extract estimates of the model parameters. These procedures are detailed in this section.

\subsection{NRG calculation}

The NRG method is an iterative procedure for solving quantum impurity problems involving a localized few-level system coupled to a continuum of itinerant electrons (fermionic bath, normal-state or mean-field BCS superconductor). It consists of several steps: 1) discretization of the continuum parts of the Hamiltonian using a geometric-progression mesh with an accumulation point at the Fermi level (the so-called logarithmic discretization), 2) unitary transformation of the resulting discretized Hamiltonian from the star-geometry (impurity coupling to each representative mesh point) to a linear tight-binding chain representation (the so-called Wilson chain), 3) iterative diagonalization in which the Wilson chain sites are taken into account consecutively \cite{wilson1975,krishna1980a,satori1992,yoshioka2000,bulla2008}. The discretization is controlled by the discretization parameter $\Lambda>1$ which controls the coarseness of the grid. When the discretization is coarse, the results can be improved by twist averaging, which consists of performing the same calculation for several different discretization grids and averaging the results \cite{bulla2008,zitko2009}. The growth of the Hilbert space is controlled by the truncation parameters which control the number of states retained after each step of the iteration.

The calculations in this work have been performed with the NRG Ljubljana code \cite{zitko_rok_2021_4841076}. Since the main quantities of interest are the ground state energies in each spin sector, very high quality results can be obtained even
with coarse discretization ($\Lambda=8$) and keeping no more than 3000 states (spin multiplets)
in the truncation. We have verified that the twist averaging is not required. 
The BCS gap was chosen to be $\Delta=0.1D$, where $D$ is the half-bandwidth.
The calculations were performed for a problem with symmetric hybridisations,
$\Gamma_{\mathrm{L}}=\Gamma_{\mathrm{R}}$. This is sufficient, because the results for an arbitrary
coupling asymmetry can be obtained from the following mapping \cite{kadlecova2017}:
\begin{equation}
\phi_S(\phi,a) = 2\arccos\sqrt{1-\frac{4a}{(a+1)^2} \sin^2(\phi/2)},
\end{equation}
where $a=\Gamma_{\mathrm{L}}/\Gamma_{\mathrm{R}}$ is the asymmetry, $\phi$ is the BCS phase difference
in the asymmetric problem, and $\phi_S$ is the effective BCS phase difference in
the effective symmetric problem.

Such calculations were performed for a set of values of the interaction strength
$U$ (from very low values $U=0.1\Delta$ that correspond to ABS-like subgap states, up to $U=30\Delta$ that correspond to YSR-like subgap states). In every value of $U$, a grid of $\xi$ and $\Gamma$ parameters was set up, and a sweep of $\phi$ between 0 and $\pi$ (50 points) has been performed for each ($\xi$, $\Gamma$) pair. The ground state energies are obtained as the sum of all energy shifts \cite{krishna1980a} performed during the NRG evolution, which has been shown to produce extremely accurate results \cite{zitko2009}. Some calculations have also been performed in the presence of a small Zeeman splitting. The results have been collected, documented, and made available on a public repository \cite{rok_zitko_2022_5874832}. The full set of input files and scripts is provided for running the calculations for different parameters or for different Hamiltonians.

\begin{figure}[h!]
    \centering
    \includegraphics[scale=1.0]{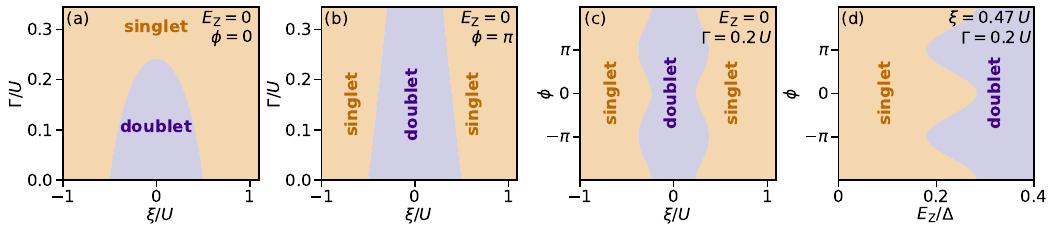}\caption{ {\bf Boundaries between singlet and doublet ground states extracted from NRG calculations.} (a) Boundary in the $\xi-\Gamma$ plane at $\phi=0$ for $\Gamma_{\mathrm{L}} = \Gamma_{\mathrm{R}}$. (b) Same as (a) for $\phi=\pi$. (c) Boundary in the $\xi-\phi$ plane at $\Gamma = \SI{0.2}{U}$. (d) Boundary in the $E_{\rm Z}-\phi$ plane for $\xi=\SI{0.47}{U}$. All panels are for $U/\Delta = 5$.}
    \label{fig:phasediagram1}
\end{figure}

Having developed the NRG calculation, we can gain insight into the expected boundaries between singlet and doublet occupation. In Fig.~\ref{fig:phasediagram1}(a), we show the phase diagram for the symmetric configuration $\Gamma_{\rm L}=\Gamma_{\rm R}$ at fixed $\phi=0$ and $U/\Delta=5$.
In the $(\xi, \Gamma)$ plane, the phase diagram takes a dome-like shape with the transition value of $\Gamma$ being the highest at the electron-hole symmetry point $\xi=0$.
At this point, the transition value of $\Gamma$ diverges if the phase difference between the reservoirs is changed to $\phi=\pi$, because in this case a destructive interference between tunneling events to the left or right occurs.
This causes the ``dome'' in the $(\xi, \Gamma)$ plane to turn into the ``chimney'' shown in  Fig.~\ref{fig:phasediagram1}(b).

As mentioned above, at $\Gamma=0$ the ground state is in the doublet sector for $\abs{\xi/U}<1/2$. Upon increasing $\Gamma$, the Kondo coupling favours the binding of a Bogoliubov quasiparticle in the superconductor to the impurity local moment (``Yu-Shiba-Rusinov" screening), ultimately determining the transition to a singlet ground state at a value $\Gamma_c$.
The value of $\Gamma_c$ depends on $\xi$, $\phi$, $U$ and $\Delta$, as well as on the asymmetry between $\Gamma_{\mathrm{L}}$ and $\Gamma_{\mathrm{R}}$.
This implies that the singlet-doublet transition can be observed varying any of these parameters individually.
Since in the experiment the values of $U$ and $\Delta$ are fixed, being determined by the materials and the geometry of the physical device, we focus here on variations in $\xi$, $\phi$, $\Gamma_{\mathrm{L}}$ and $\Gamma_{\mathrm{R}}$.

In Fig.~\ref{fig:phasediagram1}(c), we show the singlet-doublet transition boundary in the $\xi-\phi$ plane.
The interference effect is modulated continuously by the value of the phase difference $\phi$, resulting in periodic oscillations of the boundary.
The average position of the oscillating boundary is determined by $\Gamma$.
In Fig.~\ref{fig:phasediagram1}(d), we show the effect of a Zeeman energy $E_{\rm Z}$ in the case when the ground state is singlet at $B=0$.
As mentioned in the main text, a singlet-doublet transition is induced at finite $E_{\rm Z}$ due to the spin-splitting of energy levels in the doublet sector.

\subsection{Transmon diagonalization}
Having established how to calculate singlet and doublet potentials using the NRG method, we now turn to their inclusion in the Hamiltonian of the transmon circuit (main text Eq.~(1)). To numerically solve the Hamiltonian for an arbitrary potential term $V(\phi)$ we make use of the Fourier decomposition (note that the potential can include an external flux \flux):
\begin{equation}
V(\phi) = E_{\mathrm{J},0} + \sum_n{E_{\mathrm{J},n}^{\cos} \cos{\left(n \phi \right)}} + \sum_n{E_{\mathrm{J},n}^{\sin} \sin{\left(n \phi \right)}}
\end{equation}
with the components
\begin{subequations}
\begin{align} 
E_{\mathrm{J},0} &= \frac{1}{2\pi} \int_{-\pi}^{\pi} V(\phi) d\phi \\ 
E_{\mathrm{J},n}^{\cos} &= \frac{1}{\pi} \int_{-\pi}^{\pi} V(\phi) \cos{\left(n \phi \right)} d\phi \\ 
E_{\mathrm{J},n}^{\sin} &= \frac{1}{\pi} \int_{-\pi}^{\pi} V(\phi) \sin{\left(n \phi \right)} d\phi
\end{align}
\end{subequations}
where we assume the potential to be a real-valued $2\pi$-periodic function. We can then express the full Hamiltonian in the charge basis as 
\begin{equation}
H = 4 E_c \hat{N}^2 + E_{\mathrm{J},0} + \sum_n \tfrac{1}{2}\,E_{\mathrm{J},n} \hat{N}_+^n+ \mathrm{h.c.}
\label{eq:transmon_ham_charge}
\end{equation}
with $E_{\mathrm{J},n} = E_{\mathrm{J},n}^{\cos} - i E_{\mathrm{J},n}^{\sin}$, $\hat{N}$ the charge operator and $\hat{N}_+$ the charge raising operator defined via $\hat{N}_+\ket{N}=\ket{N+1}$.

Upon substituting the potential of main text Eq.~(2) into Eq.~\eqref{eq:transmon_ham_charge} and diagonalizing the Hamiltonian, we find the eigenvalues and obtain the energy levels of the combined reference junction and quantum dot junction system. Their difference then results in the transmon's transition frequencies. To numerically compute the eigenvalues we truncate the number of charge states and Fourier coefficients to $N=35$ for all calculations \cite{Koch2007}. We verify that this leads to good convergence for the eigenvalues. We further note that while the presence of the potential offset $E_{\mathrm{J},0}$ does not affect the transmon transition frequencies, its inclusion is crucial: it plays a large role in determining whether the ground state of the combined system corresponds to singlet or doublet occupation for a given set of quantum dot junction parameters.

\subsection{Parameter matching routine}
\label{sec:matching}
To match the numerical model to the experimental data we have to overcome several complications. First, the mapping between experimental control parameters and those present in the model is not always trivial. As discussed in the main text, $V_{\mathrm{p}}$ appears to not only tune $\xi$ but also $\Gamma_{\mathrm{L},\mathrm{R}}$. In turn $V_{\mathrm{t}}$ is constructed in such a way that (to first approximation) it does not tune $\xi$, but it does act on both tunnel rates simultaneously with different, unknown lever arms. For mapping the magnetic field axis to the Zeeman energy the challenge lies in determination of the effective g-factor of the quantum dot, known to be a strongly gate and angle-dependent quantity \cite{Schroer2011}. Only the flux axis allows for a simpler identification, in particular if one assumes that in the singlet configuration the combined DC SQUID Josephson potential takes its minimal (maximal) values at $0$ ($\pi$), which should hold for even modest SQUID asymmetry. 
A separate challenge comes from the large number of parameters of the model: $\Delta$, $U$, $\xi$, $\Gamma_{\mathrm{L}}$, $\Gamma_{\mathrm{R}}$, and \flux. With 6 potentially correlated parameters to match one has to carefully assess whether the fit is under-determined.

\begin{figure}
    \centering
    \includegraphics[scale=1.0]{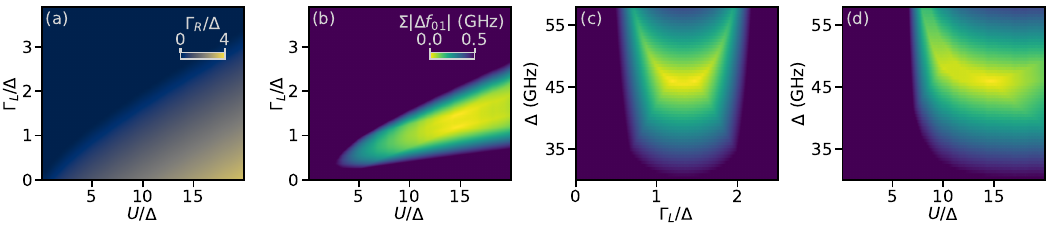}\caption{ {\bf Numerical matching of model parameters} (a) Calculation of the value of $\Gamma_{\mathrm{R}}$ that leads to a singlet-doublet transition with other model parameters held fixed. Here we fix $\phi_\textrm{ext}=0$ and $\xi=0$. A value of zero indicates that no such transition occurs. (b) Calculation Eq.~\eqref{eq:differences} in the $U-\Gamma_{\mathrm{L}}$ plane evaluated at $\Delta=\SI{46}{GHz}$. (c) Same as (b) in the $\Delta-\Gamma_{\mathrm{L}}$ plane for $U/\Delta=12.2$. (d) Same as (b) in the $U-\Delta$ plane for $\Gamma_{\mathrm{L}}/\Delta=1.19$.}
    \label{fig:bestfit}
\end{figure}

\begin{figure}
    \centering
    \includegraphics[scale=1.0]{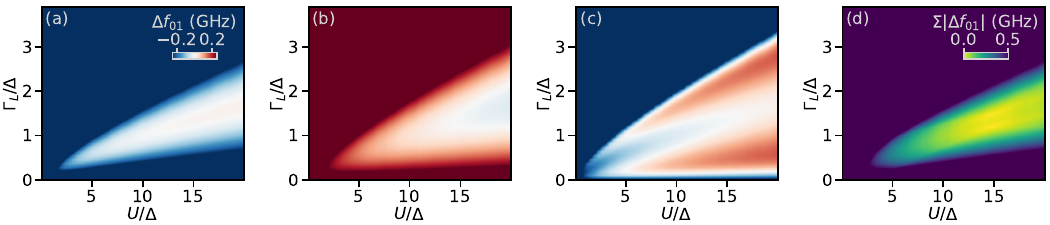}\caption{ {\bf Numerical matching in $\Gamma_{\mathrm{L}}-U$ plane} (a) Difference in the calculated and measured singlet qubit frequency at \flux~=~0 evaluated at $\Delta=\SI{46}{GHz}$. (b) Same as (a) for the doublet qubit frequency at \flux~=~0. (c) Same as (a) for the doublet qubit frequency at \flux~=$\pi$. (d) The absolute sum of the differences in panels (a-c).}
    \label{fig:bestfit_individual}
\end{figure}

 Given these considerations, we identify a specific gate point in the experimental data that could result in a well-constrained situation: the top of the dome shape of Fig.~5(a) in the main text. Here we have access to three measured quantities at a known flux \flux:  the singlet and doublet qubit frequencies $f_{01}^\mathrm{s}(0)$ and $f_{01}^\mathrm{d}(0)$ measured at the boundary of the transition, and also the doublet qubit frequency $f_{01}^\mathrm{d}(\pi)$. We furthermore know that here $E_\mathrm{s}\approx E_\mathrm{d}$ for \flux~=~0, since the data lies on the boundary of a singlet-doublet transition versus tunnel gate. Finally, based on the symmetry of the dome shape we identify that this \Vp~should correspond to $\xi\approx 0$. We can therefore eliminate two of the model parameters ($\xi$ and \flux) and are left to determine $\Delta$, $U$, $\Gamma_{\mathrm{L}}$ and $\Gamma_{\mathrm{R}}$.

In a first step we tackle the condition of a singlet-doublet transition occurring versus tunnel gate. For each value of $\Delta$, $U$ and $\Gamma_{\mathrm{L}}$ we numerically diagonalize the Hamiltonian of Eq.~\eqref{eq:transmon_ham_charge} to determine the lowest energy level of the total circuit for both the singlet and doublet states and find the value of $\Gamma_{\mathrm{R}}$ for which these energies are equal. For this we use a reference junction potential $V_{\rm J}=E_{\rm J}(1-\cos\phi)$ with $E_{\rm J} = 12.8$~GHz and $E_{\rm{c}}/h=\SI{210}{MHz}$ as determined in Sec.~\ref{sec:basic_char}. 
Shown in Fig.~\ref{fig:bestfit}(a), this results in a $U$-dependent range of $\Gamma_{\mathrm{L}}$ for which there is indeed a value of $\Gamma_{\mathrm{R}}$ that leads to a singlet-doublet transition. Outside of this range $\Gamma_{\mathrm{L}}$ is so large that the ground state is always a singlet. 

Having determined these possible values of $\Gamma_{\mathrm{R}}$ we calculate the three relevant transmon frequencies $f_{01}^\mathrm{s}(0)$, $f_{01}^\mathrm{d}(0)$, and $f_{01}^\mathrm{d}(\pi)$. These are then compared to the measured values, and an optimal solution is sought that minimizes the sum of the absolute difference between calculation and measurement of all three quantities 
\begin{equation}
\Sigma\vert\Delta f_{01} \vert = \vert f_{01}^\mathrm{s,\mathrm{exp.}}(0) - f_{01}^\mathrm{s,\mathrm{calc.}}(0) \vert + \vert f_{01}^\mathrm{d,\mathrm{exp.}}(0) - f_{01}^\mathrm{d,\mathrm{calc.}}(0) \vert + \vert f_{01}^\mathrm{d,\mathrm{exp.}}(\pi) - f_{01}^\mathrm{d,\mathrm{calc.}}(\pi) \vert.    
\label{eq:differences}
\end{equation} In Figs.~\ref{fig:bestfit}(b-d) we plot a sample of this three-dimensional optimization, while Fig.~\ref{fig:bestfit_individual} shows how each panel is constructed from the individual singlet and doublet qubit frequencies. Other than the trivial symmetry between $\Gamma_{\mathrm{L},\mathrm{R}}$, it appears that there is indeed a single region of parameters matching our data. At its global minimum we find $\Delta/h = \SI{46}{GHz}$ ($\SI{190}{\upmu eV}$), $U = 12.2 \Delta$, $\Gamma_{\mathrm{L}} = 1.19 \Delta$ and $\Gamma_{\mathrm{R}} = 1.47 \Delta$, which results in a precise match to the measured qubit frequencies. 

Having determined $\Delta$, $U$, $\Gamma_{\mathrm{L}}$, and $\Gamma_{\mathrm{R}}$ at this single point in gate space, we attempt to match the model to the \Vt~axis of the data.  
To do so we fix $\Delta$ and $U$ to the determined values and for each value of \Vt~find the best set of $\Gamma_{\mathrm{L},\mathrm{R}}$ to match the data. To determine these two parameters we have two measured quantities: up to the transition we have  $f_{01}^\mathrm{d}(0)$ and $f_{01}^\mathrm{d}(\pi)$, and after the transition we have $f_{01}^\mathrm{s}(0)$ and $f_{01}^\mathrm{d}(\pi)$. This procedure results in good correspondence to the experimental results, as shown in main text Fig.~5(c,d). We note that by construction this captures all the granularity and measurement uncertainty of the experimental data, even though the underlying quantities might have been more smooth. A subsequent procedure that attempts to match \Vp~to $\xi$ did not turn out to be unique, as \Vp~appears to also act on $\Gamma_{\mathrm{L},\mathrm{R}}$. We therefore leave this mapping undetermined. 

The uncertainty in the extracted quantities is affected by several factors. The first is the measurement accuracy; we measure the qubit frequency with MHz-scale  precision. Based on numerical evaluation of the model, this precision in qubit frequency should limit the extracted parameter accuracy to several GHz. A more substantial uncertainty comes from the determination of the transmon island charging energy $E_\mathrm{c}$, which is typically determined from the transmon transition anharmonicity $\alpha = f_{12}-f_{01}$. While the anharmonicity can be measured to high precision, a complication arises from the usage of a nanowire based Josephson junction as the reference junction. Up to now we have assumed its potential to take the form $V(\delta) = E_{\rm J}(1-\cos\delta)$; that of a conventional superconductor-insulator-superconductor (SIS) tunnel junction governed by many weakly transparent channels. In this case we find that $E_{\rm{c}}/h=\SI{210}{MHz}$, resulting in the parameter estimates given above. However, previous work has found that nanowire-based Josephson junctions are better described by several or even a single transport channel, such that $V(\delta) = -\sum_n \Delta \sqrt{1-T_n \sin^2{\delta/2}}$. This change in potential shape can lead to a strong reduction in the anharmonicity, and thus an underestimation of $E_{\rm{c}}$ when using the SIS potential \cite{Kringhoj2018}. We therefore also match our reference junction dependence to a single transport channel, which is the most extreme case for a reduction in the anharmonicity, finding good agreement with a single transport channel of $T=0.58$. This in turn leads to an extracted $E_{\rm{c}}/h=\SI{306}{MHz}$ (see Sec.~\ref{sec:basic_char}), resulting in a different set of extracted quantum dot parameters. In particular, we now find $\Delta=\SI{30.5}{GHz}$ and $U=17.3 \Delta$. This value of the induced gap in the InAs-Al nanowire is on the low end of what is typically found in  DC transport experiments, which might hint at a reduced proximity effect in the ungated leads \cite{Antipov2018,Winkler2019}.

Capacitance simulations of the full circuit do not provide an unambiguous answer for which of the two limits is more appropriate, as the circuit was designed to target $E_{\rm{c}}/h=\SI{250}{MHz}$ which falls in the center of the estimated range. As it stands we therefore do not have to uniquely determine the experimentally realized $E_{\rm{c}}$ and thereby resolve the uncertainty in the extracted quantities. However, future works could make use of additional circuit QED compatible quantum dot probes such as direct DC access \cite{Kringhoj2020b} or dispersive gate-sensing techniques \cite{Jong2021} to independently characterize several model parameters and further constrain the matching. 

\subsection{Calculated 2D maps}
\begin{figure}[h!]
    \centering
    \includegraphics[scale=1.0]{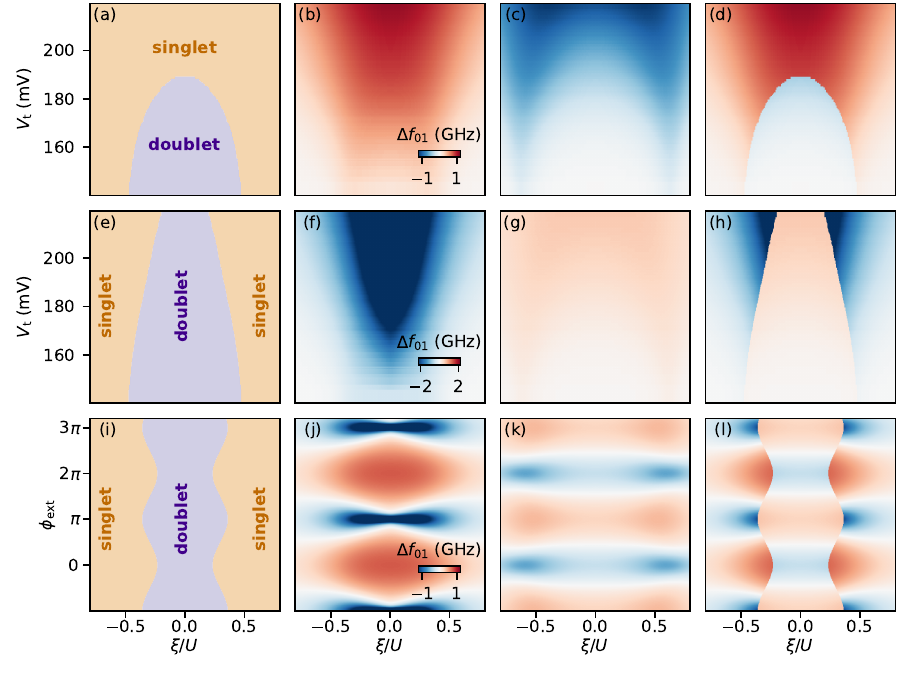}\caption{ {\bf Numerically calculated transmon frequency maps} (a,e,i) Boundaries between singlet and doublet ground states extracted from NRG calculations for \flux~=~0, \flux~=~$\pi$, and $\Gamma_{\mathrm{R}}=1.23\Gamma_{\mathrm{L}}$ respectively. Panels (b-d), (f-h), (j-l) show how the singlet qubit frequency, the doublet qubit frequency, and the combined result conditioned on the ground state of panels (a,e,i) respectively depend on the parameters. Each row shares the same color map. This leads to saturation of the color map in the panels corresponding to the unconditioned singlet and doublet qubit frequencies, but facilitates comparison to the experimental results. For all panels $U/\Delta = 12.2$ and $\Delta = \SI{46}{GHz}$.}
    \label{fig:calculated_maps}
\end{figure}

Having established how to match the model parameters to the data, we now turn to the reconstruction of the full 2D dependencies measured in the experiment (Fig.~\ref{fig:calculated_maps}). For the plunger versus tunnel gate dependence, we calculate both the singlet and doublet qubit frequencies for all values of $\Gamma_{\mathrm{L},\mathrm{R}}$ encoded by \Vt~ for a range of $\xi$ at both~\flux~=~0 and $\pi$. We subsequently mask the data according to the ground state of the combined transmon Hamiltonian, and obtain a result that closely approximates the measured data (main text Fig.~5). Using the same set of quantum dot junction parameters, we also perform a similar procedure for the 2D map of plunger gate and external flux, resulting good correspondence with main text Fig.~4. 

\subsection{State population}
We now turn to the singlet and doublet lifetimes determined in device B. For this device we could not identify a measurement point where a unique set of parameters matched the measured data, and can therefore not make a quantitative comparison to the numerics. Instead, we attempt to gain some intuition about the obtained results based on the parameters of device A. 

\begin{figure}[h!]
    \centering
    \includegraphics[scale=1.0]{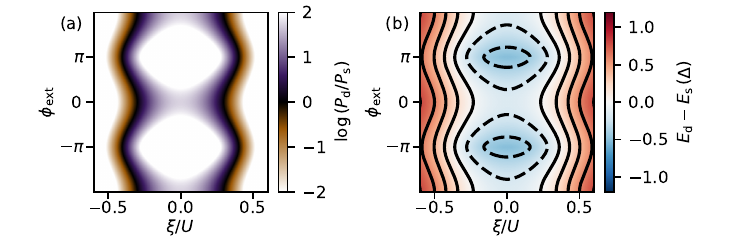}\caption{ {\bf State population versus $\xi$ and \flux} (a) Ratio of the expected state population as calculated from Eq.~\eqref{eq:maxwell} for a temperature of $\SI{400}{mK}$. The colormap is saturated to facilitate comparison to main text Fig.~7. (b) Difference between doublet and singlet energy. Each contour indicates a boundary of equal energy difference. Parameters are the same as that of Fig.~\ref{fig:calculated_maps}.}
    \label{fig:QPPtheory}
\end{figure}

In main text Fig.~7 we extract $\log_{10}\left(T_\mathrm{d}/T_{\mathrm{s}}\right)$, the ratio of the lifetimes of singlet and doublet occupation. If the system was in thermal equilibrium with a bath of temperature $T$, one would naively expect that the relative lifetimes should follow the state populations $P_{\mathrm{s},\mathrm{d}}$ as described by a Maxwell-Boltzmann distribution:
\begin{equation}
P_{i} = \frac{1}{Z} g_i \exp\left(-E_i/k_{\rm B} T\right)
\label{eq:maxwell}
\end{equation}
where $g_i$ is the degeneracy of the state, $E_{\mathrm{s},\mathrm{d}}$ are the singlet and doublet energies, and $k_{\rm B}$ is the Boltzmann constant. We take $Z = 2 \exp\left(-E_{\rm d}/k_{\rm B} T\right) + \exp\left(-E_{\rm s}/k_{\rm B} T\right)$, where we neglect potential other many-body states which should be unoccupied at the experimentally relevant temperatures. In Fig.~\ref{fig:QPPtheory}(a) we then plot $\log_{10} \left(P_\mathrm{d}/P_{\mathrm{s}}\right)$, choosing a bath temperature of $\SI{400}{mK}$. Qualitatively this follows the same trend as observed experimentally, with a sharp boundary at the phase transition and a saturated population imbalance away from that. We stress once-more that this is not a quantitative comparison. However, the need for a temperature far in excess of the refrigerator's base temperature of $\SI{20}{mK}$ could hint at a non-thermal origin such as non-equilibrium quasiparticles \cite{Glazman2021}.

In the main text we also speculate that non-thermal effects lie at the origin of the experimentally observed contours of fixed lifetime ratio's. We corroborate this in Fig.~\ref{fig:QPPtheory}(b), where we plot the energy difference between singlet and doublet occupation of the quantum dot junction. This quantity exhibits distinct contours of equal energy difference that qualitatively match those found in the experiment. If the environment has spectral components resonant with these specific energies, one could expect these to modify the dynamics.

\clearpage

\section{Device and experimental setup}
\subsection{Nanofabrication details}

The device fabrication occurs in several steps using standard nanofabrication techniques, and it is identical for device A and B. The substrate consists of 525~$\upmu$m-thick high-resistivity silicon, covered in \SI{100}{nm} of low pressure chemical vapor deposited $\rm{Si_3N_4}$. On top of this a \SI{20}{nm} thick NbTiN film is sputtered, into which the gate electrodes and circuit elements are patterned using an electron-beam lithography mask and $\rm{SF_6}$/$\rm{O_2}$ reactive ion etching. Subsequently, \SI{30}{nm} of $\rm{Si_3N_4}$ dielectric is deposited on top of the gate electrodes using plasma enhanced chemical vapor deposition and then etched with buffered oxide etchant. The nanowire is then deterministically placed on top of the dielectric using a nanomanipulator and an optical microscope. For this we use an approximately \SI{10}{um}-long vapour-liquid-solid (VLS) hexagonal InAs nanowire with a diameter of \SI{100}{nm} and a \SI{6}{nm}-thick epitaxial Al shell covering two facets \cite{Krogstrup2015}. After placement, two sections of the aluminium shell are removed by wet etching with MF-321 developer. These sections form the quantum dot junction and the reference junction, with lengths \SI{200}{nm} and \SI{110}{nm} respectively. A zoom-in of the the quantum dot junction is shown in Fig.~2(d) of the main text. The reference junction is controlled by a single \SI{110}{nm}-wide electrostatic gate, set at a DC voltage \Vj. The quantum dot junction is defined by three \SI{40}{nm}-wide gates separated from each other by \SI{40}{nm}, set at DC voltages  \Vl, \Vc~and \Vr. Note that in Fig.~2(d) the gates appear wider (and the gaps between gates appear smaller) than stated due to distortion by the $\rm{Si_3N_4}$ layer; the given dimensions are therefore determined from a scanning electron microscopy image taken before the deposition of the dielectric. After the junction etch the nanowire is contacted to the transmon island and to ground by an argon milling step followed by the deposition of \SI{150}{nm}-thick sputtered NbTiN. Finally, the chip is diced into 2 by 7 millimeters, glued onto a solid copper block with silver epoxy, and connected to a custom-made printed circuit board using aluminium wirebonds. 

\subsection{General chip overview}
\begin{figure}[h!]
    \center
    \includegraphics[scale=1]{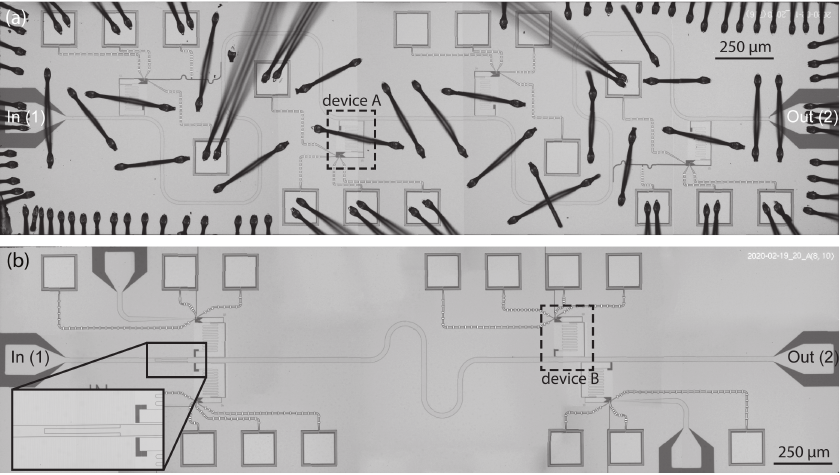}
    \caption{{\bf Chip design.} (a) The chip of device A, containing four nearly identical devices coupled to the same transmission line. The image is taken after wire-bonding onto a PCB. (b) The chip of device B, incorporating an input capacitor in the transmission line (enlarged in inset). The image is taken before wire-bonding onto a PCB.}
    \label{fig:full_chip}
\end{figure}

 Optical microscope images of the  chips containing devices A and B are shown in Figs.~\ref{fig:full_chip}(a) and (b), respectively. Each chip, 7~mm long and 2~mm wide, consists of four devices coupled to the same transmission line. For the chip containing device A, only one device was functional. Out of the other three, one did not have a nanowire, another contained three nanowires stuck together, and for the third device a gate electrode showed no response. The chip of device B includes an on-chip capacitor on the input port of the transmission line to increase the signal-to-noise ratio. For this chip only two of the devices were bonded: device B, which was functional, and another device that did not show any response to the electrostatic gates. The two unbonded devices were dismissed based on prior optical inspection, containing two and no nanowires respectively.

\subsection{Flux control with in-plane magnetic field}
\begin{figure}[h!]
    \center
    \includegraphics[scale=1]{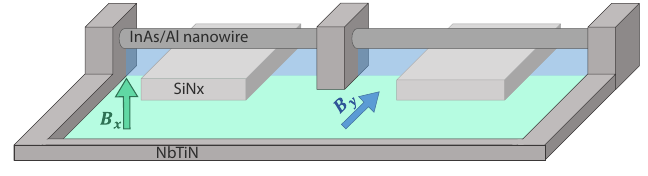}
    \caption{{\bf Flux control with $\mathbf{B_y}$.} The nanowire is elevated with respect to the NbTiN plane due to the gate dielectric. This defines a loop area perpendicular to $B_y$. $B_y$ can therefore be used to control the flux through the SQUID loop while keeping the out-of-plane field component ($B_x$) fixed, reducing the occurance of external flux jumps. }
    \label{fig:By_diagram}
\end{figure}

In all measurements we control the external flux \flux~with the in-plane component of the magnetic field perpendicular to the nanowire, $B_y$, as illustrated in Fig.~\ref{fig:By_diagram} \cite{Wesdorp2021b}. This is done since flux tuning with the out-of-plane magnetic field $B_x$ led to strong hysteric behaviour in the resonator as well as flux jumps in the SQUID loop. We attribute these effects to Abrikosov vortex generation and the presence of superconducting loops on the chip, causing screening currents. 

\subsection{Flux jumps in device A when $|B|<$~\SI{9}{mT}}
\begin{figure}[h!]
    \center
    \includegraphics[scale=1]{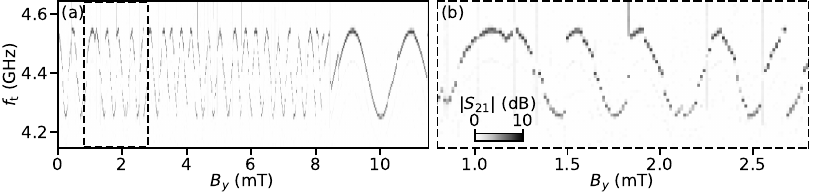}
    \caption{{\bf Flux jumps under $|\mathbf{B}|$~=~9~mT for device A.} Multiple flux jumps and a distorted periodicity observed at low magnetic fields disappear when $|B|>$~\SI{9}{mT}. Here, $B_z=B_x=0$}
    \label{fig:By_v1}
\end{figure}

For all measurements of device A, the value of the applied magnetic field is kept above \SI{10}{mT} to prevent flux jumps observed when $|B|<$~\SI{9}{mT}. In particular, for Figs.~3-5 in the main text, $B_z=$~\SI{10}{mT}. The reason for this is purely technical. Device A contains various on-chip aluminium wire-bonds connecting separate sections of the ground plane together. Below the critical magnetic field of aluminium ($\sim$\SI{10}{mT} \cite{Luthi2018}) these wire bonds create superconducting loops close to the device region, and have a significant cross-section perpendicular to the chip plane. In this regime, the application of an in-plane magnetic field $B_y$ generates unwanted currents across these superconducting loops, which in turn result in multiple jumps observed in the flux through the SQUID loop (Fig.~\ref{fig:By_v1}), making it impossible to reliably control \flux. Applying a field $|B|>$~\SI{9}{mT} turns the aluminium wire bonds normal and prevents the unwanted flux jumps, as shown in Fig.~\ref{fig:By_v1}(a). As this magnetic field is small compared to other energy scales involved, it should not affect the physics under study. We further note that the absence of superconducting loops containing wire-bonds in device B made it possible to measure this device at $B_z=$~0~mT without suffering from similar flux jumps.

\clearpage

\subsection{Cryogenic and room temperature measurement setup}
\begin{figure}[h!]
    \center
    \includegraphics[scale=0.5]{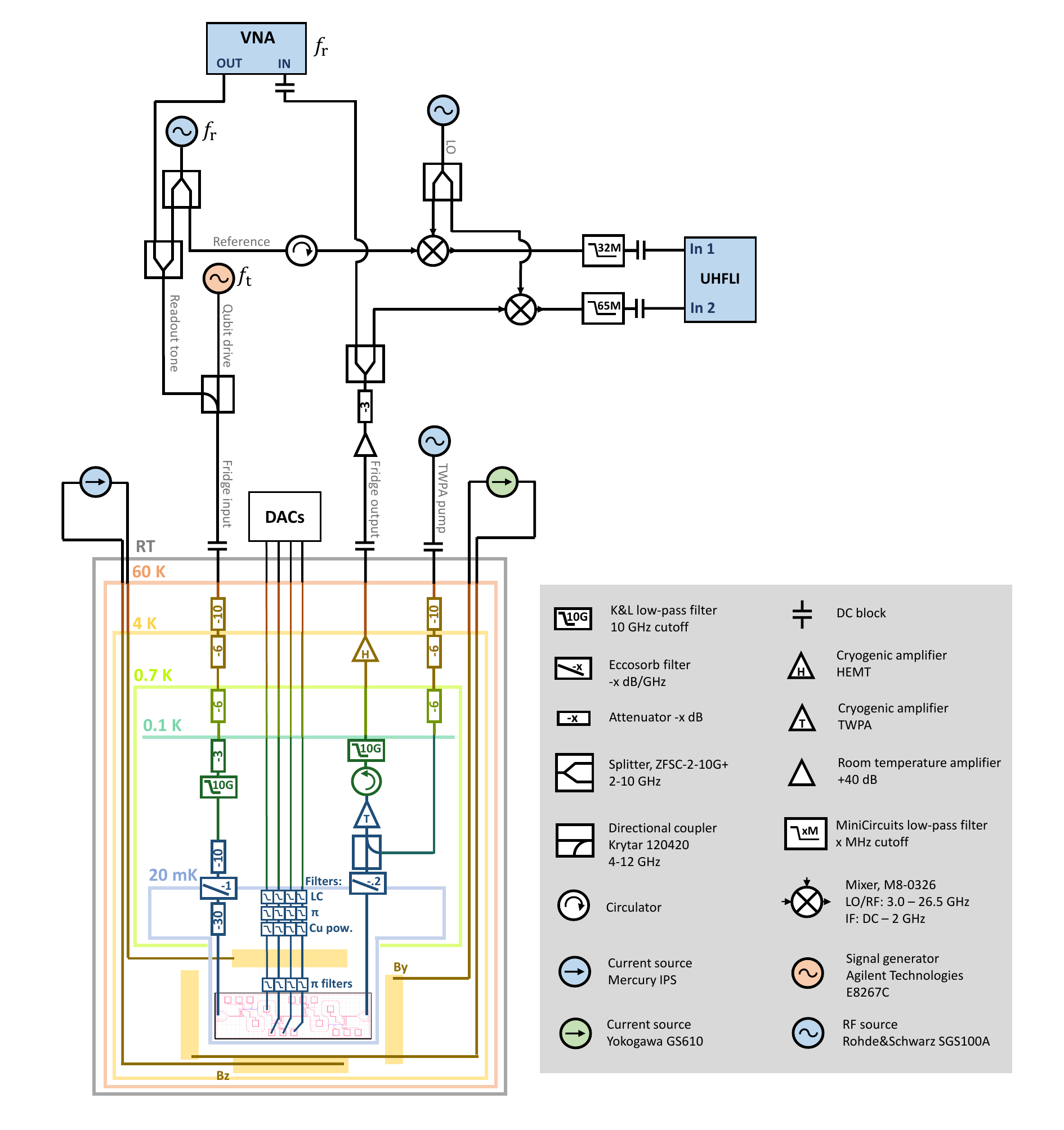}
    \caption{{\bf Measurement setup at cryogenic and room temperatures.} Both devices are measured in the same Triton dilution refrigerator with a base temperature of \SI{20}{mK}. It contains an input RF line, an output RF line and multiple DC gate lines. The DC gate lines are filtered at base temperature with multiple low-pass filters connected in series. The input RF line contains attenuators and low-pass filters at different temperature stages, as indicated. The output RF line contains a travelling wave parametric amplifier (TWPA) at the \SI{20}{mK} temperature stage,  a high-electron-mobility transistor (HEMT) amplifier at the \SI{4}{K} stage, and an additional amplifier at room temperature. A three-axis vector magnet (x-axis not shown) is thermally anchored to the \SI{4}{K} temperature stage, with the device under study mounted at its center. The $B_z$ component of the magnetic field is controlled with a MercuryiPS current source while the $B_x$ and $B_y$ axes are controlled with Yokogawa GS200 and GS610 current sources respectively. At room temperature a vector network analyzer (VNA) is connected to the input and output RF lines for spectroscopy at frequency $f_{\rm r}$. On the input line, this signal is then combined with the qubit drive tone at frequency $f_{\rm t}$ for two-tone spectroscopy. A separate tone at $f_{\rm r}$ only used for time-domain measurements is also combined onto this line. For time-domain measurements the output signal is additionally split off into a separate branch and down-converted to 25 MHz to be measured with a Zurich Instruments ultra-high frequency lock-in amplifier.}
    \label{fig:cryogenic_setup}
\end{figure}

\section{Basic characterization and tune up of device A}
\label{sec:basic_char}
\subsection{Reference junction characterization}
\begin{figure}[h!]
    \centering
    \includegraphics[scale=1.0]{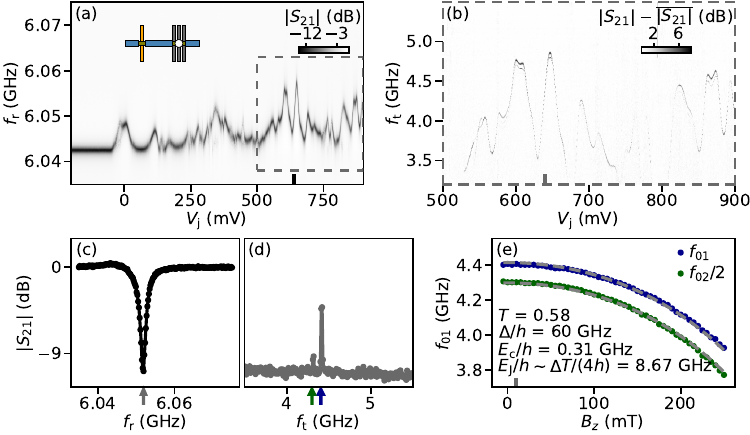}
    \caption{{\bf Reference junction characterization for device A.} (a) \Vj~dependence of single-tone spectroscopy when the quantum dot junction is pinched-off (\Vc~=~\SI{52.4}{mV}, \Vl~=~\SI{470}{mV}, \Vr~=~\SI{373}{mV}). At low \Vj~values the reference junction is pinched-off and $E_{\rm J} \sim 0$, thus the resonator is at its bare resonance frequency. As \Vj~increases, the resonator frequency increases non-monotonically due to mesoscopic fluctuations of the overall increasing transmission of different junction channels.  (b)  \Vj-dependence of two-tone spectroscopy for the \Vj~range indicated in (a) with a dashed line rectangle. The black lines in (a) and (b) indicate the \Vj~=~\SI{640}{mV} set-point which sets the transmon frequency to its set-point used for the main text figures, $f_{01} = f_{\rm 01}^0 = \SI{4.4}{GHz}$. (c) Line-cut of (a) at the \Vj~set-point, showing a resonance.  (d) Line-cut of (b) at the \Vj~set-point, showing two peaks. The highest peak, at higher frequency, appears when the second tone frequency matches the transmon frequency ($f_{\rm t}=f_{01}^0$). The lower peak corresponds to $f_{02}/2$ and shows the anharmonicity of the transmon.   For (d), the first tone frequency $f_{\rm r}$ is fixed at the bottom of the resonance, indicated with a grey arrow in (c). (e) $B_z$ evolution of $f_{01}^0$ and $f_{02}/2$ at \Vj~=~\SI{640}{mV}. }
    \label{fig:ref_junction}
\end{figure}

In this section we investigate the basic behaviour of the reference junction versus junction gate voltage \Vj~and magnetic field $B_z$ when the quantum dot junction is completely closed. This information is used to choose a \Vj~set-point, \Vj~=~\SI{640}{mV}, which maintains a good SQUID asymmetry in all regimes of interest. Figs.~\ref{fig:ref_junction}(a) and (b) show the \Vj~dependencies of the resonator and transmon frequencies, respectively. As \Vj~is varied, different junction channels open sequentially \cite{Spanton2017, Hart2019}, with transparencies that increase non-monotonically due to mesoscopic fluctuations at the junction. This in turn affects the transmon's $E_{\rm J}$ and results in the observed fluctuations of its frequency. 

The $B_z$ dependencies of $f_{01}$ and $f_{02}/2$ at \Vj~=~\SI{640}{mV} are shown in Fig.~\ref{fig:ref_junction}(e).
From this we estimate both the transmon island charging energy $E_{\rm{c}}$ (not to be confused with $U$, the charging energy of the quantum dot junction) and the parameters of reference junction potential used in Sec.~\ref{sec:matching} to match the measurements to the numerical calculations. Illustrated in this figure is a fit of the data with a Josephson potential governed by a single Andreev level at the junction $V(B, \delta) = -\Delta(B)\sqrt{1-T{\rm sin}^2{\frac{\delta}{2}}}$. Here $\Delta(B) = \Delta \sqrt{1-(B/B_{\rm c})^2}$ is the field dependent superconducting gap \cite{Luthi2018}, $\Delta$ is the superconducting gap at zero field, $B_{\rm c}$ is the critical magnetic field and $T$ is the transparency of the junction. As the fit is not constrained well enough to provide a unique solution, we fix $\Delta/h = 60$~GHz based on recent experiments on the same nanowires \cite{Splitthoff2022}. We obtain $E_{\rm c}/h=306$~MHz, $T=0.58$, and $B_{\rm c} = 413$~mT, resulting in an effective $E_{\rm J} \sim \Delta T/4= 8.7$~GHz. A similar procedure is then performed for $V_{\rm J}=E_{\rm J}(1-\cos\delta)$, resulting in $E_{\rm{c}}/h=\SI{210}{MHz}$ and $E_{\rm J}/h = 12.8$~GHz.

We can use these parameters to estimate the experimentally-realized SQUID asymmetry $\alpha_\mathrm{S} = E_{\rm J}/E_{\rm J, QD}$ where $E_{\rm J, QD}$ denotes the effective quantum dot junction Josephson energy. To do so we estimate $E_{\rm J, QD}$ from the calculated qubit frequencies of the singlet and doublet obtained in Sec.~\ref{sec:matching} through the relation $\hbar \omega_{01} \approx \sqrt{8E_{\rm J, QD} E_{\rm c}}-E_{\rm c}$ \cite{Koch2007}. We find that $\alpha_\mathrm{S}>10$ for almost all of the parameter range, exceeding 30 for low values of $V_\mathrm{t}$. The asymmetry is at its smallest for the upper values of $V_\mathrm{t}$ in the vicinity of $\xi=0$, where we find a minimum asymmetry $\alpha_S=4$. We note that the effects of these variations in asymmetry are fully captured by the numerical model; its effects are predominantly on the modulation of the qubit transition frequency with flux and not on the position of the singlet-doublet transition boundaries.

\subsection{Quantum dot junction characterization}
\begin{figure}[h!]
    \centering
    \includegraphics[scale=1.0]{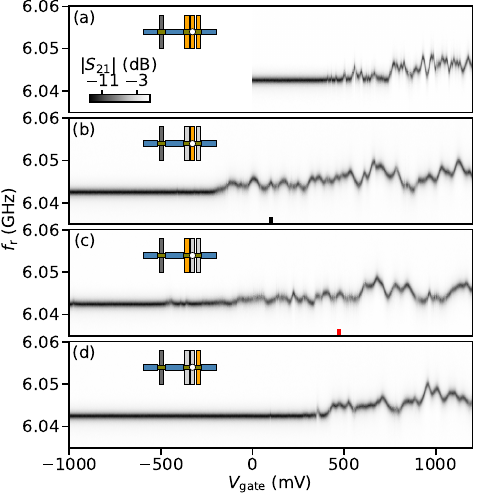}
    \caption{{\bf Quantum dot gates characterization for device A.} (a) Gate voltage dependence (\Vl~=~\Vc~=~\Vr~=~$V_{\rm gate}$) of single-tone spectroscopy, showing how the quantum dot junction is pinched off at $V_{\rm gate}$ values lower than \SI{300}{mV}.  (b-d) \Vc, \Vl~and \Vr~dependence, respectively of single-tone spectroscopy. In each panel, the two unused gates are kept at \SI{1250}{mV}. This shows how each of the three quantum dot gates can independently pinch off the quantum dot junction.  For all panels, the reference junction is closed (\Vj~=~\SI{-200}{mV}). The red line in (c) indicates the fixed value of \Vl~=~\SI{470}{mV} at which all main text figures are taken. }
    \label{fig:QD-gates}
\end{figure}

In this section we show the basic behaviour of the quantum dot gates when the reference junction is closed. Fig.~\ref{fig:QD-gates} shows effective pinch-off curves for all three quantum dot gates ramped together (a) and for each of them separately, when the other two are kept at \SI{1250}{mV} (b-d). This shows that each of the three quantum dot gates can independently pinch off the quantum dot junction even if the other gates are in the open regime, signifying strong lever arms and good gate alignment. We note that these are not pinch-off curves as encountered in conventional tunnel spectroscopy. They reflect the voltages at which there is no longer a measurable transmon transition frequency mediated by the quantum dot junction, which could either be due to low tunneling rates or a full depletion of the quantum dot. 

\subsection{\label{sec:tuneup} Device tune up}

\begin{figure}[h!]
    \centering
    \includegraphics[scale=1.0]{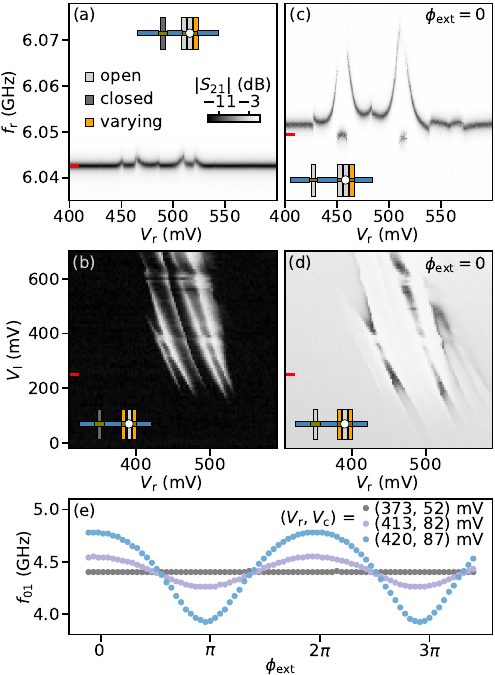}  \caption{{\bf Quantum dot tune up for device A.}  (a) Single-tone spectroscopy measured at \Vl~= 250 mV, exhibiting two small resonances. Here the reference junction is fully closed (\Vj~= -200 mV). The red line indicates the readout frequency used in panel (b). (b) Single frequency readout of the resonator. Bright colors indicate a shift in the resonance frequency, marking the onset of supercurrent through the dot. The red line indicates the \Vl~value of panel (a). (c) Same as panel (a) but with the reference junction opened to the \Vj~= 640 mV setpoint used throughout the manuscript. The two junctions in parallel form a SQUID, increasing the qubit frequency and in turn the resonance frequency. Measured at \flux~= 0. (d) Same as panel (b) but with the reference junction set to \Vj~= 640 mV and \flux~= 0, measured at the frequency indicated with a red line in (c). For (a-d), \Vc~=~100~mV (close to pinchoff), indicated with a black line in Fig.~\ref{fig:QD-gates}. (e) \ft~versus \flux~at fixed \Vj~=~640 mV, for three quantum dot gates setpoints corresponding to a quantum dot junction which is fully closed (grey), slightly open (violet) or very open (blue) showing the DC SQUID behaviour of the two parallel Josephson junctions. 
    }
    \label{fig:tuneup}
\end{figure}

This section describes the process of tuning up the quantum dot gates to the setpoint used for the main text figures. We start by closing the reference junction (\Vj~=~\SI{-200}{mV}) and going to a point in quantum dot gate voltages near pinchoff (\Vc~=~\SI{100}{mV}, \Vl~=~\SI{250}{mV} and \Vr~=~\SI{400}{mV}, see Fig.~\ref{fig:QD-gates}). Monitoring the frequency of the resonator while varying one of the gates reveals small shifts away from its bare frequency which resemble the shape expected for quantum dot resonances (Fig.~\ref{fig:tuneup}(a)). Fixing the readout frequency $f_{\rm r}$ at the bare frequency of the resonator, one can map out the regions where these shifts happen on a two-dimensional map versus the left and right gates (Fig.~\ref{fig:tuneup}(b)). In such maps, a pixel with  a dark color indicates the resonator is not shifted from its bare frequency while a bright pixel indicates a shift of the resonator frequency, which we can use to identify potential regions of interest. 

After identifying such a region in \Vl-\Vr~space, we open the reference junction to its set-point \Vj~=~\SI{640}{mV}, which lifts the reference transmon frequency to $f_{\rm 01}^0 = \SI{4.4}{GHz}$, closer to the bare resonator frequency. This magnifies the dispersive shift of the resonator and, furthermore, brings the external flux into the picture. As shown in Fig.~\ref{fig:tuneup}(e), the asymmetric SQUID behaves as expected for different quantum dot gate setpoints. The reference junction sets the reference value for the transmon frequency, $f_{\rm 01}^0 $, and the quantum dot contributes with small variations above or below this setpoint due to constructive or destructive interference, respectively. 

Fixing \flux~=~0 and repeating the initial measurement versus \Vr~with the reference junction open reveals much stronger deviations of the resonant frequency than before (Fig.~\ref{fig:tuneup}(c)). Importantly, the observed resonant frequency is now discontinuous, which, as detailed in the main text, is a signature of a singlet-doublet transition of the quantum dot junction. We tentatively identify the regions for which the resonator frequency is shifted to lower values as doublet regions and perform single frequency readout versus \Vr~and \Vl, now with $f_{\rm r}$ fixed at the resonator frequency corresponding to doublet regions (Fig.~\ref{fig:tuneup}(d)). The resulting two-dimensional map reveals regions for which the transmission amplitude signal is low (dark regions in Fig.~\ref{fig:tuneup}(d)) which we identify as potential regions with a doublet ground state.

\begin{figure}[h!]
    \centering
    \includegraphics[scale=1.0]{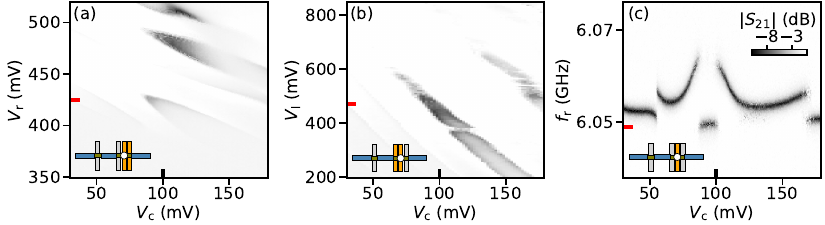}
    \caption{{\bf Quantum dot gate dependence for device A.} (a) Single frequency readout of the resonator at the frequency indicated in Fig.~\ref{fig:tuneup}(c) with a red line, performed versus \Vc~and \Vr~for fixed \Vl~=~\SI{470}{mV}.  (b) Same as (a) but versus \Vc~and \Vl~and for fixed \Vr~=~\SI{425}{mV}. (c) Single-tone spectroscopy versus \Vc, measured at \Vl~=~\SI{470}{mV} and \Vr~=~\SI{425}{mV}, revealing a quantum dot resonance.  For all panels \flux~=~0.
    }
    \label{fig:left-gate}
\end{figure}

The next step for tuning up is identifying an isolated region where the quantum dot is in a doublet ground state and exploring the behaviour versus the central quantum dot gate. This is shown in Fig.~\ref{fig:left-gate}. As \Vc~is varied at \flux~=~0 (Fig.~\ref{fig:left-gate}(c)), the resonator first shows a displacement towards higher frequencies 
to then abruptly drop to a lower frequency, to then finally go back to the higher frequencies once-more. As detailed in the main text, we identify this behaviour with a singlet-doublet transition as the relative level of the quantum dot $\xi$ is being varied. Figs.~\ref{fig:left-gate}(a) and (b) show how this central doublet ground state region varies with each of the two lateral quantum dot gates. In both cases we observe a dome shape, resembling the behaviour we would expect when varying the tunnel coupling between quantum dot and leads. However, these dome shapes are rotated in \Vc-\Vr~and \Vc-\Vl~space. This is understood as the result of cross coupling between the different quantum dot gates.

\begin{figure}[h!]
    \centering
    \includegraphics[scale=1.0]{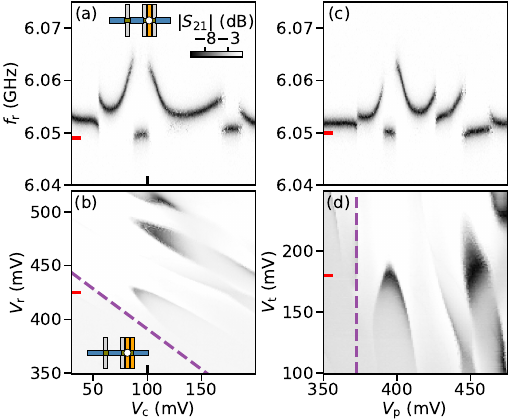} \caption{{\bf Gate compensation for device A.}  (a) Single-tone spectroscopy versus \Vc at \Vr~=~\SI{427}{mV}. (b) Single frequency readout of the resonator measured versus the central (\Vc) and right (\Vr) quantum dot gate voltages, performed at a at fixed \Vl~\SI{470}{mV}. The red line indicates the \Vr~value of panel (a). (c) Resonator spectroscopy versus \Vp~at \Vt~=~\SI{180}{mV}.  (d) Same as (b) but in the transformed coordinate frame, measured vs. the virtual plunger (\Vp) and right tunnel (\Vt) gate voltages. In (a) and (c), the red lines indicate the readout frequencies used in panels (b) and (d), respectively. For all panels \flux~=~0.  }
    \label{fig:compensation}
\end{figure}

After identifying the cross coupling effect between different quantum dot gates, we define a new set of virtual gates in an attempt to tune the model parameters independently. We fix \Vl~=470~mV (set-point kept for all results shown in the main text) and focus on \Vr-\Vc~space. Fig.~\ref{fig:compensation}(b) shows the dome shape previously identified in \Vr-\Vc~space. We identify a line along the dome (indicated with a dashed line) for which the quantum dot level appears to be fixed and define new plunger virtual gate (\Vp, perpendicular to this line) and right tunnel virtual gate (\Vt, along this line) (see Fig.~\ref{fig:compensation}(d)). This rotated gate frame is the one used for the main text. Note that this routine does not guarantee that \Vp~does not affect the tunneling rates. It rather ensures that \Vt~does not (strongly) affect the quantum dot level $\xi$.

\subsection{\label{sec:double_res} Larger tunnel voltage range}

\begin{figure}[h!]
    \centering
    \includegraphics[scale=1.0]{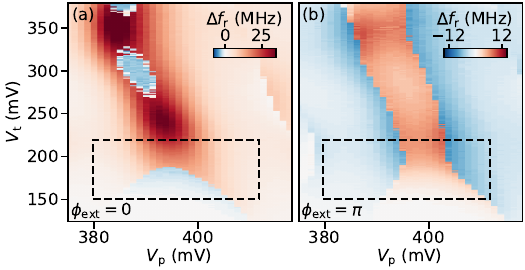}  \caption{{\bf Extended \Vt~dependence.}  (a)  $\Delta f_{\rm r}$ versus \Vp~and \Vt~ at \flux~=~0, revealing singlet (red) and doublet (blue) ground state regions separated by sharp transitions. (b) Same as (a) but for \flux~=~$\pi$. We note that the plunger gate axis is shifted by about \SI{5}{mV} with respect to (a) and the data shown in the main text, which we speculate is due to an irreproducible gate jump. Dashed rectangles indicate the gate ranges in which the measurements of Fig.~~5 of the main text are taken.}
    \label{fig:tunnel_long}
\end{figure}

In Fig.~\ref{fig:tunnel_long} we show the behaviour of the singlet and doublet regions beyond the \Vt~range investigated in Fig.~5 of the main text. At \flux~=~$\pi$ we do not observe the doublet phase boundary fully closing for any \Vt. According to theory, this should only occur if $\xi\approx0$ and $\Gamma_{\rm L}\approx\Gamma_{\rm R}$ are maintained at each gate setting in the experiment. That this condition would remain satisfied for any \Vt~is implausible given the cross-coupling present in the system.
We instead speculate that at higher gate voltages the tunnel rates cease to be a monotonically increasing function, which is substantiated by the tunnel gate dependence at \flux~=~0. Here we observe a temporary recovery of the doublet region at higher~\Vt, which should not occur for increasing values of $\Gamma$. We further speculate that in this regime of increasingly large $\Gamma/U$ the dot can eventually be tuned to a different charge configuration, involving energy levels not captured by the single-level model of main text Eq.~(6).

We note that for these measurements only single tone spectroscopy was performed. We therefore plot $\Delta f_{\rm r} = f_{\rm res} - f_{\rm res}^0$, where $f_{\rm res}^0$ denotes the resonator frequency with the quantum dot junction pinched off. Its qualitative interpretation is the same as that of $\Delta f_{\rm 01}$ used in the main text.

\subsection{\label{sec:state_selective} State selective spectroscopy}

\begin{figure}[h!]
    \centering
    \includegraphics[scale=1.0]{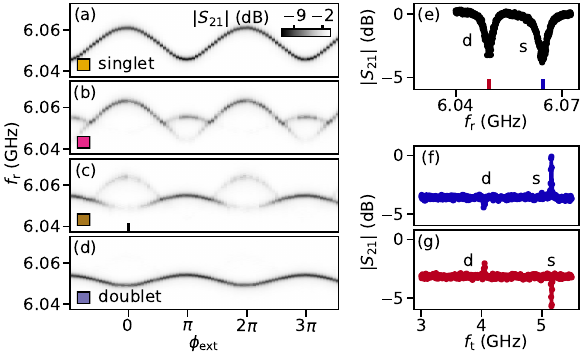}  \caption{{\bf State selective spectroscopy.}  (a-d) \flux~dependence of single-tone spectroscopy at four representative \Vp~values, indicated in Fig.~3 in the main text. Traces at intermediate \Vp~values show two resonances simultaneously due to switches on timescales faster than the integration time.  (e) Linecut of (c) at \flux~=~0, indicated by a black line in (c). (f) Two-tone spectroscopy at the same settings as in (e), with the first tone at the frequency of the singlet resonance. The measurement shows a peak at the transmon frequency of the singlet state. (g) Same as (f) but with the readout frequency corresponding to the doublet resonance, which shows a peak at the transmon frequency of the doublet state.
    }
    \label{fig:double-res}
\end{figure}
For the measurements performed close to singlet-doublet transitions, single-tone spectroscopy simultaneously shows two resonances whose relative depth varies with the distance from the transition. 
This is once more illustrated in Fig.~\ref{fig:double-res}, which shows single-tone spectroscopy at several different \Vp~regions while \flux~is varied. It corresponds to the measurements of Fig.~4 of the main text. In panels (a) and (d) we observe only a single resonance; at these plunger gate values the quantum dot junction is sufficiently deep in the singlet and doublet parity sector respectively that only one state is occupied. However, at the plunger gate values between these two regimes (panels (c-d)) the behaviour is more complex. We simultaneously observe two resonances and their depth becomes a function of flux. 

For the two-tone spectroscopy measurements in the main text we make use of the averaged occupation of the states captured in the single-tone spectroscopy measurement to identify most occupied state. This can be inferred from the relative depth of the resonances: for example in Fig.~\ref{fig:double-res}(e) the most occupied state is the singlet, albeit by a small margin. This in turn allows us to do state selective two-tone spectroscopy, revealing the transmon transition that corresponds to the most occupied state of the system. To do so we fix the frequency of the first tone $f_{\rm r}$ at the bottom of the deepest resonance, corresponding to the most populated sector of the system. We illustrate this in Figs.~\ref{fig:double-res}(f) and (g), where by fixing $f_{\rm r}$ at the bottom of the resonance corresponding to the singlet (doublet) state we observe a peak only when $f_{\rm t}$ is equal to the transmon frequency corresponding to the singlet (doublet) state. It is this peak position that we report as $f_{01}$.

\section{Magnetic field dependence of device A}

\begin{figure}[h!]
    \centering
    \includegraphics[scale=1.0]{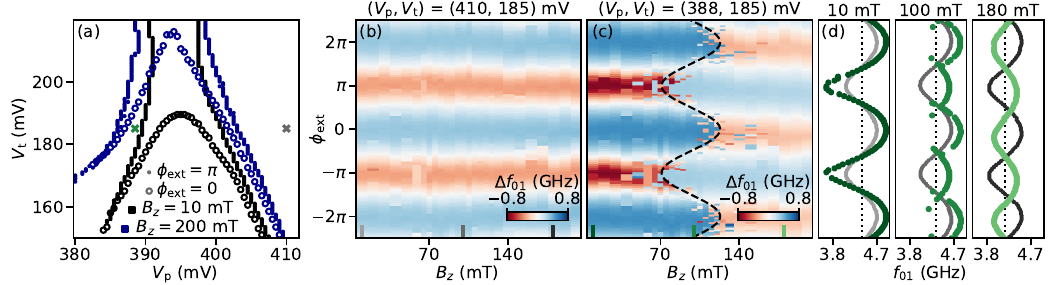}
    \caption{{\bf Data analysis for magnetic field dependence of device A.} (a)  Borders between singlet and doublet regions for $B_z$~=~\SI{10}{mT} (black) and $B_z$~=~\SI{200}{mT} (blue). Solid and empty markers correspond to \flux~=~$\pi$ and \flux~=~0, respectively. (b) and (c) show $\Delta f_{\rm 01}$ versus $B_z$ and \flux, measured at the two gate points indicated in (a) with grey and green markers, respectively. In (b), the singlet is the ground state for all $B_z$. This gate point is used to identify a flux reference for each $B_z$. For (c), there is a singlet-doublet ground state transition with $B_z$, where the sinusoidal dashed line serves as a guide for the eye. (d) \ft~versus \flux~for the three $B_z$ values indicated in (b) and (c). The dotted line indicates $f_{\rm 01}^0$, which decreases with $B_z$ as shown in Fig.~\ref{fig:ref_junction}(e).}
    \label{fig:field}
\end{figure}

In this section we elaborate on the analysis of the data shown in Fig.~6(c) in the main text. When varying both \flux~and $B_z$ in a measurement, one has to consider the possibility of an unwanted misalignment of the magnetic field with respect to the nanowire axis. This, in combination with the multiple orders of magnitude difference between the applied $B_z$ (hundreds of mT) and the $B_x$ (less than a $\upmu$T) or $B_y$ (several mT) needed to thread a flux quantum through the SQUID loop, can result in big changes of the \flux~=~0 point for different values of $B_z$. Therefore, one has to re-calibrate the value of $B_y$ that corresponds to \flux~=~0 for each $B_z$ value. To do so, we use the flux dependence of $f_{01}$ at a gate point for which the quantum dot junction ground state remains a singlet for the whole $B_z$ range as a reference for identifying \flux~=~0. This gate point is indicated with a grey cross in Fig.~\ref{fig:field}(a). 

The measurement shown in Fig.~6(c) is therefore performed as follows:
\begin{algorithmic}[0]
    \For{each $B_z$ value}
        \State{apply $B_z$}
        \For{each $B_y$ value}
            \State{apply $B_y$}
            \State{measure $f_{01}$ at the grey gate point}
            \State{measure $f_{01}$ at the green gate point}
        \EndFor
    \EndFor
\end{algorithmic}

For each $B_z$ value we then reconstruct the $B_y$ dependence of \flux~through the dependence of the reference gate point (grey). Furthermore, we use this method to identify points in $B_y$ where flux jumps happen and correct for them. While they almost never occur for small magnetic fields, and none of the other data required such a correction, we found that at increasing $B_z$ jumps would occur more often. We believe this is due to a small misalignment between $B_z$ and the plane of the chip. The resulting corrected \flux~reference is shown in Fig.~\ref{fig:field}(b), while Fig.~\ref{fig:field}(d) shows several linecuts.

\section{Parity lifetime extraction procedure}

\begin{figure}[h!]
    \centering
    \includegraphics[scale=1.0]{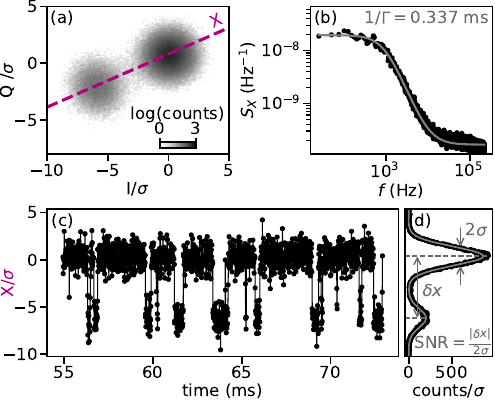}
    \caption{{\bf Parity lifetime analysis.}  (a) Logarithmic-scale histogram of the resonator response in the (I,Q)-plane after integrating a \SI{2}{s} time trace with time bins of $t_{\rm int} =$ 11.4~$\upmu$s. It exhibits two separate Gaussian distributions whose centers define an axis, X, indicated with a dashed line. (b) Power spectral density (black) of an unintegrated \SI{2}{s} time trace projected onto the X axis. In grey, best fit of a Lorentzian lineshape with a white noise background (Eq.~\eqref{eq:PSD}). (c)  \SI{18}{\milli s} cut of the integrated response projected onto the X axis, revealing jumps between two distinct states. (d) 1D histogram of the response in (a) projected onto the X axis (black) and the best fit of a double Gaussian line-shape (grey, Eq.~\eqref{eq:double_Gaussian}). For all panels \Vl~=~\SI{325}{mV}, \Vt~=~\SI{-60}{mV}, \Vp~=~\SI{551.4}{mV}, $B_z$~=~0 and \flux~=~0. 
    }
    \label{fig:QPP_analysis}
\end{figure}

In this section we elaborate on  the analysis method for extracting the characteristic lifetimes of the singlet and doublet states, $T_{\rm s}$ and $T_{\rm d}$. We start with a continuous measurement at a fixed readout frequency where we monitor the demodulated output signal integrated in time bins of $t_{\rm int} =$ 2.3~$\upmu$s. This reveals a complex random telegraph signal jumping between two states in the (I,Q)-plane. The histogram of the acquired (I,Q) points shows two states (Fig.~\ref{fig:QPP_analysis}(a))  whose centers define an axis X. A segment of the measured telegraph signal, projected onto this X axis, is shown in Fig.~\ref{fig:QPP_analysis}(c). Taking the histogram along this axis results in a double Gaussian distribution (Fig.~\ref{fig:QPP_analysis}(d))
that is well-described by
\begin{equation} \label{eq:double_Gaussian}
g(x) = \frac{A_1}{\sqrt{2\pi\sigma^2}}e^{\frac{-(x-x_1)^2}{2\sigma^2}} + \frac{A_2}{\sqrt{2\pi\sigma^2}} e^{\frac{-(x-x_2)^2}{2\sigma^2}}
\end{equation}
Here, $A_{1,2}$ are the relative populations of singlet and doublet occupation, $x_{1,2}$ are the centers of each Gaussian and $\sigma$ is their standard deviation.  For the data shown in Fig.~\ref{fig:QPP_analysis}, the fit results in $A_1 = 2169\sigma$, $A_2 = 506\sigma$, $x_1 = 0.37\sigma$ and $x_2 = -6.19\sigma$, from which we determine the ${\rm SNR} = |x_1 - x_2|/2\sigma =$~3.28.

From the time domain information of the signal we construct its power spectral density (PSD), which is its squared discrete Fourier transform  (Fig.~\ref{fig:QPP_analysis}(b))
\begin{equation} \label{eq:PSD}
S_X(f) = \frac{\Delta t}{N \pi} \abs{\sum_{n=1}^{N} X(n \Delta t)e^{-i2\pi f n \Delta t}}^2
\end{equation}
where $X(t)$ is the measured signal (as projected onto the previously defined X-axis), $\Delta t =$~2.3~$\upmu$s is the discrete time bin in which the data is measured, $N=\frac{T}{\Delta t} $ is the number of points and $T$ is the total signal length. In practice we use Welch's method with a Hanning window \cite{Welch1967} to calculate the power spectral density, dividing the trace into 50 sections of length \SI{40}{ms} that overlap by \SI{20}{ms} and averaging the power spectral density of all segments. This results in a spectrum that is well fit by a single Lorentzian of the form 
\begin{equation} \label{eq:PSD-fit}
S(f) = A \frac{4\Gamma}{(2\Gamma)^2 + (2\pi f)^2} + B,
\end{equation}
from which we obtain $1/\Gamma = 0.337$~ms, $A = 5.75 \cdot 10^{-5} $ and $B = 1.65 \cdot 10^{-10}$~${\rm Hz}^{-1}$.

Combining the amplitude ratio $R=A_1/A_2$ obtained from the Gaussian fit of the two quadratures and the $\Gamma$ value obtained from the Lorentzian fit of the PSD, we calculate 
\begin{equation} \label{eq:rates1}
T_{\rm s} = 1/\Gamma_{\rm s} = \frac{1+R}{2\Gamma R}
\end{equation}
\begin{equation} \label{eq:rates2}
T_{\rm d} = 1/\Gamma_{\rm d} = \frac{1+R}{2\Gamma}
\end{equation}
to obtain $T_{\rm s} =$~0.89~ms,  $T_{\rm d} =$~0.21~ms. 

\section{Extended parity lifetime data}

\subsection{Parity lifetimes linecut versus flux}
\begin{figure}[h!]
    \centering
    \includegraphics[scale=1.0]{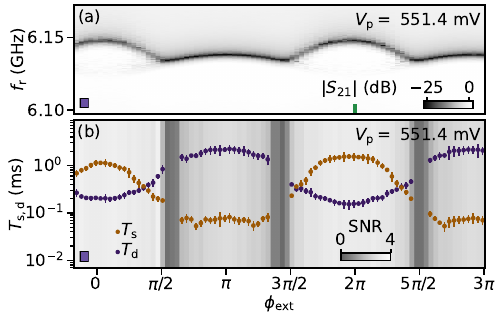}
     \caption{{\bf Flux dependence of parity lifetimes}  (a) \flux~dependence of single-tone spectroscopy at  \Vp~=~551.4~mV.  (b) \flux~dependence of the parity lifetimes extracted following the analysis in Fig.~\ref{fig:QPP_analysis} at \Vp~=~551.4~mV. Markers  indicate the mean  and error bars indicate the maximum and minimum values of 10 consecutive \SI{2}{s} time traces. SNR~$=\frac{|\delta x|}{2\sigma}$ is shown in greyscale in the background. For points where SNR~$<1$, the extracted parity lifetimes are not shown as we do not consider them reliable.  Measured at the same \Vt~, \Vl~ and $B_z$ as for Fig.~\ref{fig:QPP_analysis}. }
    \label{fig:QPP_flux}
\end{figure}
Fig.~\ref{fig:QPP_flux} shows the flux dependence of the lifetimes of the singlet and doublet states at \Vp~=~554.4~mV, which accompanies main text Fig.~7. We find that both singlet and doublet lifetimes show an approximate sinusoidal dependence on the applied flux. As discussed in the main text, this flux dependence most likely originates from the oscillation of the singlet-doublet energy gap with flux. However it could also be indicative of a coherent suppression of the tunneling rates \cite{Pop2014}. We further note that the sudden drops in SNR are due to crossings of the transmon frequencies of the singlet and doublet states. At these points both resonator frequencies become indistinguishable and their lifetimes can not be measured.

\subsection{Power and temperature dependence of parity lifetimes}
\begin{figure}[h!]
    \centering
    \includegraphics[scale=1.0]{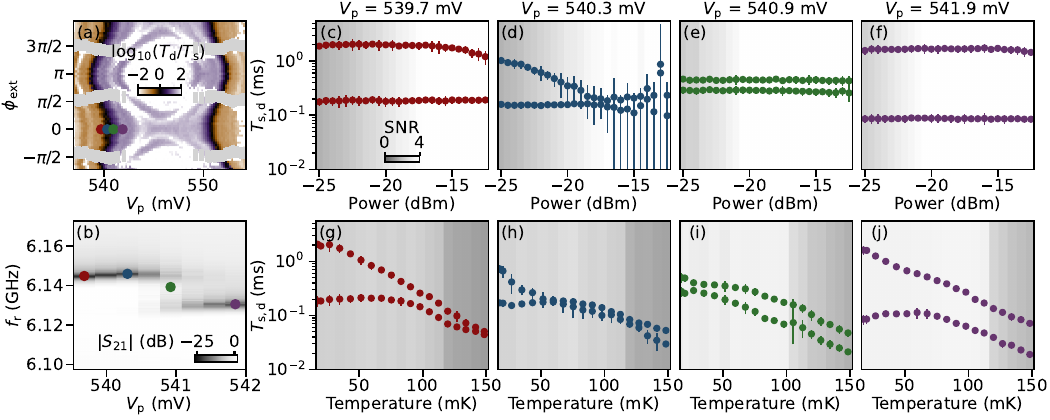}
    \caption{{\bf Power and temperature dependence of parity lifetimes across the singlet-doublet transition} (a) 2D map of ${\rm log_{10}}(T_{\rm d}/T_{\rm s})$ versus \Vp~and \flux, extracted from a 2 s time trace for each pixel. This is the same panel as Fig.~7(e) in the main text. (b) \Vp~dependence of single-tone spectroscopy at \flux~=~0, across a singlet/doublet transition. For (a) and (b), the mixing chamber temperature is \SI{18}{mK} and the readout power is \SI{-22}{dBm}. (c-f) Readout power dependence at \SI{18}{mK} of the extracted parity lifetimes at the plunger points indicated in (a) and (b). Markers  indicate the mean  and error bars indicate the maximum and minimum values of 10 consecutive \SI{2}{s} time traces.  The SNR is shown in greyscale in the background. For points where ${\rm SNR}<1$, the extracted parity lifetimes are discarded. (g-j) Same as (c-d) but versus temperature and at a power of \SI{-22}{dBm}. All powers are given at the fridge input.}
    \label{fig:QPP_power_and_temp}
\end{figure}

Here we present additional data on the readout power and temperature dependence of the parity lifetimes shown in Fig.~7 of the main text. The power dependence at four selected points across a phase boundary is shown in Figs.~\ref{fig:QPP_power_and_temp}(c-f). Away from the transition (purple) and right on top of the transition (green) the readout power does not have a strong effect on the extracted lifetimes in the investigated range. For plunger gate values \Vp~ closer to the transition, however, the asymmetry of the lifetimes decreases with power (blue). Although the origin of this dependence is not clear, we conjecture it is related to parity pumping effects \cite{Wesdorp2021}.

Temperature dependencies at the same gate points, measured at a readout power of -22 dBm at the fridge input, are shown in Figs.~\ref{fig:QPP_power_and_temp}(g-j). Here the mixing chamber temperature of the dilution refrigerator is measured with a ruthenium oxide resistance thermometer and increased in a controlled step-wise fashion with a variable-output heater mounted on the mixing chamber plate. We observe different effects of temperature for each of the gate points. In general, there is a temperature independent regime at low temperatures, followed by a temperature dependent drop above a certain characteristic temperature, which varies over tens of mK for different gate points.  For some of the gate points, however, the temperature independent contribution is absent and the effect of increased mixing chamber temperature starts immediately at base temperature (Fig.~\ref{fig:QPP_power_and_temp}(i)). These results are indicative of non-equilibrium effects playing a role in the physics of the devices under study, their exact behaviour dependent on the energy level configuration of the quantum dot junction.

\begin{figure}[h!]
    \centering
    \includegraphics[scale=1.0]{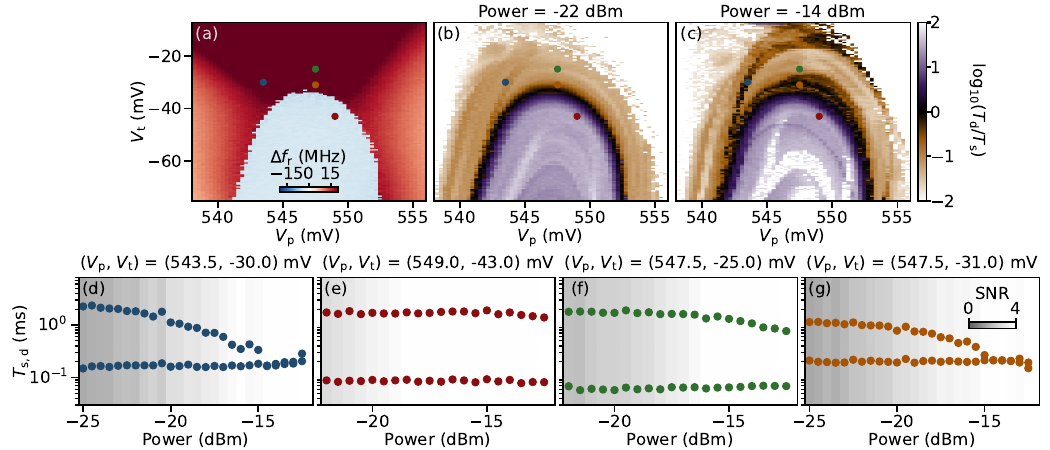}
    \caption{{\bf Tunnel, plunger and power dependence of parity lifetimes} (a)  $\Delta f_{\rm r} = f_{\rm res} - f_{\rm res}^0$ versus \Vp~and \Vt~measured at \flux~=~0. It shows a regions of constructive and destructive interference, separated by sharp dome-like boundary. (b) Two-dimensional map of ${\rm log_{10}}(T_{\rm d}/T_{\rm s})$ versus \Vp~and \Vt, measured at a power of \SI{-22}{dBm}.  (c) Same as (b) but for a power of \SI{-14}{dBm}.  (d-e) Power dependence of the extracted parity lifetimes at the gate points indicated in (a-c). All powers are given at the fridge input.}
    \label{fig:QPP_plunger_tunnel_powers}
\end{figure}

\subsection{Parity lifetimes versus tunnel gate}
To complement the data shown in Fig.~7 of the main text, taken at \Vt~=~\SI{-60}{mV}, we also show the \Vt~dependence of the parity lifetimes at \flux~=~0 in Fig.~\ref{fig:QPP_plunger_tunnel_powers}. As for device A, the doublet ground state region exhibits a dome shape in \Vp~and \Vt~space, and at the transition between singlet and doublet ground states the lifetimes for both states become equal. Away from the transition, the lifetime asymmetry increases and the lifetimes differ by more than one order of magnitude. We note that the gate compensation of device B was not ideal, resulting in a small tilt of the dome.

Similarly to the behaviour shown in the main text for \flux~and \Vp, in this case we also observe contours of equal ratio where the lifetime asymmetry abruptly increases or decreases. For higher readout power these contours become accentuated, as shown in Fig.~\ref{fig:QPP_plunger_tunnel_powers}(c). Furthermore, for higher power the region with similar lifetimes around the ground state transition becomes wider. This is due to the parity lifetimes having a different dependence on power for different regions in gate space. For most regions in gate space there is again almost no dependence on readout power in the range explored (Fig.~\ref{fig:QPP_plunger_tunnel_powers}(e,f)). However, on special gate regions, such as close to ground state transitions (Fig.~\ref{fig:QPP_plunger_tunnel_powers}(g)) and on top of the observed contours (Fig.~\ref{fig:QPP_plunger_tunnel_powers}(d)), the lifetime asymmetry decreases rapidly with power, similar to the effect shown in Fig.~\ref{fig:QPP_power_and_temp}.

\bibliography{supplement.bib}